\documentclass[twocolumn,nofootinbib, preprintnumbers, superscriptaddress]{revtex4-1}

\usepackage{amsmath,amssymb,amscd,simplewick}
\usepackage{amsmath,bm}
\usepackage{listings}
\usepackage{multirow}
\usepackage{dsfont}
\usepackage{slashed}
\usepackage{textcomp}
\usepackage{color}
\usepackage{ulem}
\usepackage{float}
\usepackage{makecell}
\usepackage{accents}
\usepackage{graphicx}
\usepackage{epstopdf}
\usepackage{subfigure}
\usepackage{epsfig}

\usepackage{xcolor}
\usepackage[colorlinks=true,
            linkcolor=blue,
            urlcolor=blue,
            citecolor=green,          
            bookmarks=true,
            bookmarksnumbered=true,
            breaklinks=true,
            pdfpagemode=Fullscreen,
            pdfstartview=FitBH]{hyperref}

\hypersetup{pdfauthor = {Shao-Feng Ge},
	     pdftitle = {}, 
	     pdfsubject = {}, 
             pdfkeywords = {}, 
	     pdfcreator = {LaTeX with hyperref package},
	     pdfproducer = {dvips + ps2pdf} }

\definecolor{gesfpurple}{rgb}{0.47,0.19,0.42}

\definecolor{gesflanse}{rgb}{0.00,0.50,0.50}

\definecolor{gesfblue}{rgb}{0.08,0.42,0.76}

\definecolor{gesfred}{rgb}{1,0,0}

\definecolor{gesfwhite}{rgb}{1,1,1}

\definecolor{gesfblack}{rgb}{0,0,0}

\definecolor{gesfxd}{rgb}{0,0,1}

\definecolor{gesfgreen}{rgb}{0.99,0,0.52} 

\newcommand{\gsec} [1]{{\hypersetup{linkcolor=red}Sec.~\ref{#1}\hypersetup{linkcolor=blue}}}
\newcommand{\gapp} [1]{{\hypersetup{linkcolor=red}App.~\ref{#1}\hypersetup{linkcolor=blue}}}
\newcommand{\geqn} [1]{\hypersetup{linkcolor=blue}(\ref{#1})\hypersetup{linkcolor=blue}}
\newcommand{\gfig} [1]{{\hypersetup{linkcolor=violet}Fig.~\ref{#1}\hypersetup{linkcolor=blue}}}
\newcommand{\gtab} [1]{{\hypersetup{linkcolor=gesflanse}Tab.~\ref{#1}\hypersetup{linkcolor=blue}}}

\definecolor{Orange}{cmyk}{0,0.61,0.87,0}
\definecolor{JungleGreen}{cmyk}{0.99,0,0.52,0}
\definecolor{OliveGreen}{cmyk}{0.64,0,0.95,0.40}
\definecolor{Brown}{cmyk}{0,0.81,1,0.60}
\definecolor{RoyalBlue}{cmyk}{0.71,0.53,0,0.12}
\definecolor{Gray}{cmyk}{0,0,0,0.40}
\definecolor{LightPink}{cmyk}{0.0,0.25,0,0}
\definecolor{LLightPink}{cmyk}{0.0,0.10,0,0}
\definecolor{LightBlue}{cmyk}{0.25,0,0,0}
\definecolor{LightGray}{cmyk}{0,0,0,0.2}

\graphicspath{{FinalFigures/}}

\begin{document}

\title{\Large Improving CP Measurement with THEIA and Muon Decay at Rest}
\author{Shao-Feng Ge}
\email{gesf@sjtu.edu.cn}
\affiliation{Tsung-Dao Lee Institute \& School of Physics and Astronomy, Shanghai Jiao Tong University, Shanghai 200240, China}
\affiliation{Key Laboratory for Particle Astrophysics and Cosmology (MOE) \& Shanghai Key Laboratory for Particle Physics and Cosmology, Shanghai Jiao Tong University, Shanghai 200240, China}

\author{Chui-Fan Kong}
\email{kongcf@sjtu.edu.cn}
\affiliation{Tsung-Dao Lee Institute \& School of Physics and Astronomy, Shanghai Jiao Tong University, Shanghai 200240, China}
\affiliation{Key Laboratory for Particle Astrophysics and Cosmology (MOE) \& Shanghai Key Laboratory for Particle Physics and Cosmology, Shanghai Jiao Tong University, Shanghai 200240, China}

\author{Pedro Pasquini}
\email{ppasquini@sjtu.edu.cn}
\affiliation{Tsung-Dao Lee Institute \& School of Physics and Astronomy, Shanghai Jiao Tong University, Shanghai 200240, China}
\affiliation{Key Laboratory for Particle Astrophysics and Cosmology (MOE) \& Shanghai Key Laboratory for Particle Physics and Cosmology, Shanghai Jiao Tong University, Shanghai 200240, China}

\begin{abstract}
We explore the possibility of using the recently proposed
THEIA detector to measure the
$\bar \nu_\mu \rightarrow \bar \nu_e$ oscillation with
neutrinos from a muon decay at rest ($\mu$DAR) source to improve
the leptonic CP phase measurement. Due to its intrinsic
low-energy beam, this $\mu$THEIA configuration
($\mu$DAR neutrinos at THEIA) is only sensitive to the
genuine leptonic CP phase $\delta_D$ and not contaminated
by the matter effect. With detailed study of neutrino energy
reconstruction and backgrounds at the THEIA detector,
we find that the combination with
the high-energy DUNE can significantly reduce the CP
uncertainty, especially around the maximal CP violation cases
$\delta_D = \pm 90^\circ$. Both the $\mu$THEIA-25 with 17\,kt
and $\mu$THEIA-100 with 70\,kt fiducial volumes are considered.
For DUNE + $\mu$THEIA-100, the CP uncertainty can be better
than $8^\circ$.
\end{abstract}

\maketitle 

\section{Introduction}
\label{intro}

The charge-parity (CP) symmetry violation is a key to understand 
the existence of baryon asymmetry in the Universe, namely,
why there are more matter than anti-matter \cite{Branco:2011zb,
Canetti:2012zc,Balazs:2014eba,Garbrecht:2018mrp,Bodeker:2020ghk}.
There are at least two possible sources of CP violation in the
Standard Model (SM) of particle physics: the CP phase in the
quark mixing matrix \cite{CPbook,PDG20-QuarkCP} and the leptonic
CP phases in the neutrino mixing matrix \cite{PDG20-NuCP}.
Especially, the leptonic CP phases at low energy play an important
role \cite{Rebelo:2007rv, Granelli:2021fyc} in the leptogenesis
mechanism \cite{Fukugita:1986hr,Buchmuller:2005eh,Davidson:2008bu}. 
Both the Dirac and Majorana CP phases can contribute to the
leptogenesis mechanism. However, only the Dirac CP phase manifests
itself in neutrino oscillation and can be measured by oscillation
experiments \cite{Bilenky:1998dt}.

The nonzero reactor mixing angle ($\theta_r \equiv \theta_{13}$)
measured  by Daya Bay \cite{DayaBay:2012fng} and RENO
\cite{RENO:2012mkc} heralds the precision era
of neutrino oscillation experiments. A nonzero
$\theta_r$ allows the Dirac CP phase $\delta_D$ to have
physical effect since these two variables always appear together
as $\sin \theta_r e^{\pm i \delta_D}$ in the standard parametrization
\cite{PDG20-NuCP} of the
Pontecorvo-Maki-Nakagawa-Sakata (PMNS) matrix 
\cite{Pontecorvo:1957qd,Maki:1962mu}. Typically,
the neutrino oscillations from the muon flavor to the
electron flavor ($\nu_\mu \rightarrow \nu_e$ and
$\bar \nu_\mu \rightarrow \bar \nu_e$) are used by
the long-baseline accelerator experiments to measure $\delta_D$
\cite{Feldman:2012jdx}.

The current long-baseline experiments T2K 
\cite{T2K:2011qtm} and NO$\nu$A \cite{NOvA:2004blv}
are approaching the discovery threshold.
The 2019 T2K result \cite{T2K:2019bcf} with
$\delta_D = 252^\circ{}^{+40^\circ}_{-33^\circ}$
and $281^\circ{}^{+28^\circ}_{-31^\circ}$
for the normal and inverted orderings (NO and IO),
respectively, has excluded almost half of the
parameter space except [$165^\circ, 358^\circ$] 
and [$-214^\circ, 342^\circ$] at $3\sigma$ confidence level (C.L.).
It is interesting to see that the maximal CP phase
$\delta_D = - 90^\circ$ is around the best-fit point.
However, the 2019 NO$\nu$A result is
$\delta_D=0^\circ{}^{+74^\circ}_{-23^\circ}$
\cite{NOvA:2019cyt} for NO with best-fit value at vanishing
CP phase, $\delta_D = 0^\circ$. 
In 2021, T2K and NO$\nu$A updated their results
with the best-fit value from T2K remaining the same,
$\delta_D = 252^\circ{}^{+40^\circ}_{-33^\circ}$ (NO)
and $281^\circ{}^{+28^\circ}_{-31^\circ}$ (IO)
\cite{T2K:2021xwb} while the NO$\nu$A best fit changes to
$\delta_D=148^\circ{}^{+49^\circ}_{-157^\circ}$ for NO
\cite{NOvA:2021nfi}. 

Although not significant,
there is a tension between the T2K and NO$\nu$A data
that they exclude each other at $1\sigma$\,C.L.
\cite{Rahaman:2021zzm,Rahaman:2022rfp}.
New physics can explain the tension.
Both belonging to accelerator neutrino experiments,
T2K and NO$\nu$A have very different configurations.
While the T2K baseline is 295\,km and the peak energy is at 0.6\,GeV, 
the NO$\nu$A baseline is 810\,km and peak energy at 2\,GeV.
These differences in baseline and beam energy leave room for
new physics. For example, the
non-standard interaction (NSI) contributes extra matter potential
\cite{Wolfenstein:1977ue} and hence its
effect on oscillation probabilities is energy dependent
\cite{Miranda:2015dra,Farzan:2017xzy} to provide a possible
solution \cite{Denton:2020uda,Chatterjee:2020kkm}.
The tension is reduced when the data
are analyzed in the context of Lorentz invariance violation 
(LIV). However, it is accompanied by a new mild tension between the best-fit values of $\sin^2\theta_{23}$ \cite{Rahaman:2021leu}.
Besides, the non-unitarity mixing due to heavy neutrinos
\cite{Fernandez-Martinez:2007iaa,Antusch:2009gn,Martinez-Soler:2018lcy} 
allows extra CP phases to fake the genuine CP effect which can
also explain the tension \cite{Miranda:2019ynh}.
However, a more recent work \cite{Forero:2021azc} points out the 
non-unitarity cannot explain the tension with the
bounds on non-unitarity parameters from the combination of short- and long-baseline data.
Similar thing happens for the light sterile neutrino scheme
\cite{Chatterjee:2020yak}.
Whether this tension is truly new physics or not needs further 
investigation at current and future experiments.

Even if the current tension between T2K and NO$\nu$A measurements
vanishes with more data, correct interpretation of CP
measurements still faces intrinsic issues including event rate
inefficiency, $\delta_D \leftrightarrow \pi - \delta_D$ degeneracy,
and large CP uncertainty around the maximal values
\cite{Ge:2017qqv,Ge:2020xkm,Ge:2020ffj}. These issues still remain
for the next-generation experiments like T2HK \cite{Hyper-Kamiokande:2018ofw}
and DUNE \cite{DUNE:2015lol}. Although its wide spectrum
can help to reduce the $\delta_D$ degeneracy, DUNE has much
larger matter effect than T2K and NO$\nu$A due to higher
energy peaking around 2.5\,GeV \cite{Kelly:2018kmb}.
At long-baseline experiments, the genuine
CP effect can be faked by the ubiquitous matter effect, reducing
the experimental sensitivity to $\delta_D$
\cite{Barger:2001yr,Mena:2004sa,Minakata:1998bf}.
In addition, the uncertainties in the matter effect can
also reduce the CP sensitivity at DUNE \cite{Kelly:2018kmb}.

In this paper, we propose $\mu$THEIA as combination of the
THEIA detector
\cite{Theia:2017xtk,Fischer:2018zsr,Theia:2019non,Guffanti:2020mui} 
and a $\mu$DAR neutrino flux to
improve the Dirac leptonic CP phase measurement together with DUNE.
\gsec{sec:muTHEIA} summarizes the contamination of matter effect
in the CP measurement and explains why the $\mu$DAR neutrino flux
with lower energy can help. Then in \gsec{sec:modes}, we
describe the low-energy mode at $\mu$THEIA and the high-energy
mode at DUNE, including selection criteria, energy 
reconstruction, smearing, and backgrounds. A combination of $\mu$THEIA
and DUNE can significantly improve the CP sensitivity as
illustrated in \gsec{sec:CP}. Therein, we also give the details 
of simulation and $\chi^2$ analysis. Our $\mu$THEIA proposal 
is compared with the existing configurations/proposals in
\gsec{sec:results} and summarized in \gsec{sec:conclu}.

\section{Improving CP Measurement with Multiple Baselines and Beam Energies}
\label{sec:muTHEIA}

The current T2K and NO$\nu$A experiment aims for the discovery of
leptonic CP violation, namely, excluding $\delta_D = 0$ and $\pi$.
Once a nontrivial $\delta_D$ is measured, the next step is
precision measurement of its value.
Several experimental configurations have been proposed to 
improve the CP measurement after the reactor mixing angle
$\theta_r$ was measured by Daya Bay and RENO.
The upgrade from existing experiments includes:
(1) Intensity Upgrade: the beam intensity is
significantly enhanced, such as T2K-II \cite{Abe:2016tii};
(2) Detector Upgrade: T2HK \cite{Hyper-Kamiokande:2018ofw}
has a much larger detector Hyper-K than Super-K;
(3) Spectrum Upgrade: DUNE \cite{DUNE:2015lol} adopts
wider on-axis spectrum than the off-axis one of NO$\nu$A; and
(4) Baseline Upgrade with longer baseline such as
T2HKK \cite{Hagiwara:2005pe,Hagiwara:2006vn,
Hyper-Kamiokande:2016srs}
and DUNE. In addition, there are also several new proposals:
(5) the accelerator experiments such as P2O \cite{Choubey:2018rnl},
ESS$\nu$SB \cite{ESSnuSB:2021azq}, and MOMENT \cite{Cao:2014bea};
(6) CP measurement with sub-GeV atmospheric neutrino oscillation such
as Super-PINGU \cite{Razzaque:2014vba,Razzaque:2015fea}, Super-ORCA
\cite{Hofestadt:2019whx}, and even at JUNO \cite{JUNO:2015zny}
or DUNE \cite{Kelly:2019itm}. Comparison among various
experimental configurations can be found in
\cite{Ballett:2016daj,Raut:2017dbh,Chakraborty:2017ccm,Ghosh:2019sfi}.

\subsection{Matter Contamination on CP Measurement}

However, the neutrino energy for all these designs is not
low enough to avoid contamination from the ubiquitous matter
effect \cite{Arafune:1997hd,Koike:1998hy,Mocioiu:2000st,
Brahmachari:2003bk,Ge:2016dlx,Kelly:2018kmb,Datta:2019uwv}. 
We will try to describe how the CP
measurement is contaminated by the matter effect. Based on
this, one can see the possible solutions.

The neutrino propagation through matter is described by
the following Hamiltonian \cite{Wolfenstein:1977ue,Mikheyev:1985zog},
\begin{eqnarray}
        \mathcal{H}_M
        =
        \frac{1}{2 E_\nu}U\begin{pmatrix}
        0 & 0 & 0\\
        0 & \Delta m^2_s & 0 \\
        0 & 0 & \Delta m^2_a
        \end{pmatrix}U^\dagger
        +
     \begin{pmatrix}
     V & 0 & 0 \\
     0 & 0 & 0\\
     0 & 0 & 0
     \end{pmatrix}.
\quad
\label{eq:HM}
\end{eqnarray}
The first term is the vacuum Hamiltonian that is a product of
the PMNS matrix $U$ \cite{Pontecorvo:1957qd,Maki:1962mu} and
the diagonal mass matrix with solar
$\Delta m^2_s \equiv \Delta m^2_{21}$ and atmospheric 
$\Delta m^2_a \equiv \Delta m^2_{31}$ mass squared differences.
The neutrino energy $E_\nu$ in the denominator contributes
as an overall factor. So it is not a suppression in the
vacuum term but actually an enhancement of the matter
potential in the second term.

Induced by the SM weak interaction, the matter potential
$V\equiv \sqrt 2 G_F n_e$ is proportional to the Fermi constant
$G_F$ and the electron number density $n_e$,
\begin{equation}
  V({\bf x})
\approx
  \frac{\rho({\bf x})}{\mbox{g/cm}^3} Y_e({\bf x})
\times
  7.56 \times 10^{-14}\mbox{eV},
\end{equation}
and only appears in the first element of the potential matrix
in \geqn{eq:HM} for the electron flavor \cite{Wolfenstein:1977ue}.
The matter potential $V({\bf x})$ is not just proportional to
the matter density $\rho({\bf x})$ but also the number of electrons
per nucleon $Y_e({\bf x})$. Both the matter density and chemical
composition vary with position. Consequently, the matter potential
is in general also a function of $\bf x$.
In the Earth crust, the electron fraction is approximately $Y_e\sim 0.5$
and the average matter density is $\bar \rho 
\equiv \langle \rho({\bf x}) \rangle \approx 
2.845$ g/cm$^3$ \cite{Kelly:2018kmb}.
For simplicity, we ignore the density variation and adopt
the averaged matter density $\bar \rho$ as constant along
the baseline of DUNE.

To make the CP and matter effect explicit, we perform a series
expansion of the $\nu_\mu\rightarrow\nu_e$
($\bar{\nu}_\mu\rightarrow\bar{\nu}_e$) oscillation
probability  in terms of the ratio between the two mass squared 
differences $\alpha$ ($\equiv \Delta m_s^2/\Delta m_a^2 \approx 3\%$)
and  the reactor mixing angle $s_r$ ($\equiv \sin \theta_r \approx 0.15$)
\cite{Freund:2001pn, Akhmedov:2004ny, Coloma:2012wq},
\begin{eqnarray} \nonumber 
  P_{\overset{{\nu_\mu \rightarrow\nu_e}}{{\overline \nu_\mu \rightarrow \overline \nu_e}}}
& \approx &
  \alpha^2 \sin^2 2 \theta_s c^2_a
  \frac{\sin^2 (A \Delta_a)}{A^2}
+
  4 s^2_{r}s^2_{a}
  \frac{\sin^2 [(1 \mp A) \Delta_{a}]}{(1 \mp A)^2}
  \nonumber \\
& + &
  2 \alpha s_{r}\sin 2\theta_{s}\sin 2\theta_{a} \cos (\Delta_{a}\pm\delta_{D})
 \nonumber \\
& \times &
  \frac{\sin(A \Delta_{a})} A
  \frac{\sin [(1 \mp A)\Delta_{a}]}{(1 \mp A)}.
\label{eq:Pnumu_nue_approx}
\end{eqnarray}
The sign $\pm$ ($\mp$) is for neutrino (the upper one) and
anti-neutrino (the lower one), respectively. For convenience,
we have used
$(s_{a,r}, c_{a,r}) \equiv (\sin \theta_{a,r}, \cos \theta_{a,r})$
to denote the sine and cosine functions of the atmospheric
($\theta_a \equiv \theta_{23}$) and reactor ($\theta_r$)
mixing angles while $\theta_s \equiv \theta_{12}$ is the
solar mixing angle. The matter term
$A \equiv 2 E_\nu V / \Delta m^2_a$ appears in two combinations,
$\sin (A \Delta_a) / A$ and $\sin [(1 \mp A) \Delta_a] / (1 \mp A)$.
In the limit of tiny matter effect, $A \rightarrow 0$, the two
combinations reduce to approximately $\Delta_a$ and $\sin \Delta_a$,
respectively. In addition, the CP phase $\delta_D$ appears as a linear
combination with the atmospheric oscillation phase
$\Delta_a \equiv |\Delta m^2_a| L / 4 E_\nu$
where $L$ is the oscillation baseline \cite{He:2016dco}.

For neutrino CP measurement, the essential observable is
the difference between the neutrino and anti-neutrino
oscillation probabilities,
$P_{\nu_\mu \rightarrow\nu_e} - P_{\bar \nu_\mu \rightarrow \bar \nu_e}
\propto \sin \Delta_a \sin \delta_D$, that is proportional to
$\sin \delta_D$ in the absence of matter potential.
However, a realistic measurement has sign difference
in not just the Dirac CP phase $\delta_D$ but also
the matter term $A$.
With a typical size of $A \approx (0.05$, 0.13, 0.21)
estimated with the peak neutrino energies (0.55, 2, 2.5)\,GeV
at T2K/T2HK, NO$\nu$A, and DUNE, respectively,
the matter effect on neutrino CP measurement cannot be ignored.
The matter potential can fake the genuine CP violation and
blur the CP measurement.

At a single long-baseline neutrino oscillation experiment,
there is only one independent CP observable but two parameters
($A$ and $\delta_D$). To disentangle the matter contamination
($A$) from the genuine CP effect ($\delta_D$), a combination
of two different baselines is a promising choice. For example,
the T2HKK with a 295\,km baseline to the Kamioka site and a
much longer baseline around 1100\,km to the Korea site can
effectively remove the faked CP by matter potential and
achieve a better CP uncertainty 
\cite{Raut:2017dbh,Chakraborty:2017ccm,Cho:2019ctv,King:2020ydu}.
In addition, the atmospheric measurement with neutrinos produced
around the Earth intrinsically has multiple baselines
\cite{Razzaque:2015fea,Razzaque:2014vba,Hofestadt:2019whx,
Kelly:2019itm,JUNO:2015zny}.
Nevertheless, the matter effect at accelerator and atmospheric
neutrino oscillation experiments is not negligible in the first
place. Even with
multiple baselines, intrinsic uncertainty from the matter effect can
still reduce the sensitivity of $\delta_{\rm D}$.

\subsection{Improvement with $\mu$DAR Neutrinos}
\label{sec:improvement}

A better way is significantly reducing the matter effect with
low-energy neutrino beam to make $A \ll 1$
\cite{Minakata:2000ee}. One possibility is using the
muon decay at rest \cite{Agarwalla:2010nn}, such as
DEA$\delta$ALUS \cite{Alonso:2010fs}, JUNO supplemented with
$\mu$DAR sources \cite{Ciuffoli:2014ika, Smirnov:2018ywm},
TNT2K/TNT2HK \cite{Evslin:2015pya,Ge:2016xya,Ge:2016dlx,
Agarwalla:2017nld,Soumya:2019kto},
and C-ADS \cite{Ciuffoli:2015uta}. 
All these designs share the feature of multiple baselines.
Especially, the TNT2K/TNT2HK configuration incorporates
both low- and high-energy beams by supplementing the existing T2K/T2HK
with $\mu$SK/$\mu$HK ($\mu$DAR source together with the Super-K/Hyper-K
detectors) to significantly improve the CP sensitivity.
The possibility of detecting $\mu$DAR neutrinos at long
baseline has also been studied but is unfortunately
diluted too much over such a long baseline \cite{Harnik:2019iwv}.

As mentioned above, the effect of matter potential on the neutrino
oscillation is modulated by the neutrino energy. The higher neutrino
energy, the larger contamination on the CP measurement. Comparing with
the 30\% effect at T2K for $E_\nu \approx 550$\,MeV \cite{Ge:2016dlx},
DUNE with peak energy around $2.5$\,GeV suffers from larger
matter effect.
So it is more urgent for the DUNE experiment to have a complementary 
baseline with low-energy beam to further improve the CP measurement.

However, it is impossible to simply add a $\mu$DAR source and share the
same liquid Argon detectors of DUNE in a similar way as TNT2K. This is
because there are no free protons to provide inverse beta decay
(IBD) for unique probe of the electron anti-neutrino and hence the
$\bar \nu_\mu \rightarrow \bar \nu_e$ oscillation. Besides, the
$\bar \nu_e -$Ar cross section is too small to detect the $\mu$DAR flux 
at DUNE \cite{DUNE:2015lol}.

A new THEIA detector at the same site of SURF was
recently proposed
\cite{OrebiGann:2015gus,Theia:2017xtk,Fischer:2018zsr,Theia:2019non,Guffanti:2020mui}.
With a new technique of water-based liquid scintillator (WbLS),
it is possible to use both scintillation and Cherenkov lights
\cite{Wei:2016vjd,Sawatzki:2020mpb,Guo:2017nnr,Caravaca:2020lfs,
Land:2020oiz}.
This opens the possibility of detecting the low-energy
$\mu$DAR neutrino oscillation to supplement the high-energy mode
at DUNE. For convenience, we call the combination of $\mu$DAR
and THEIA as $\mu$THEIA.

\begin{figure}[t]
\centering
\includegraphics[width=\linewidth]{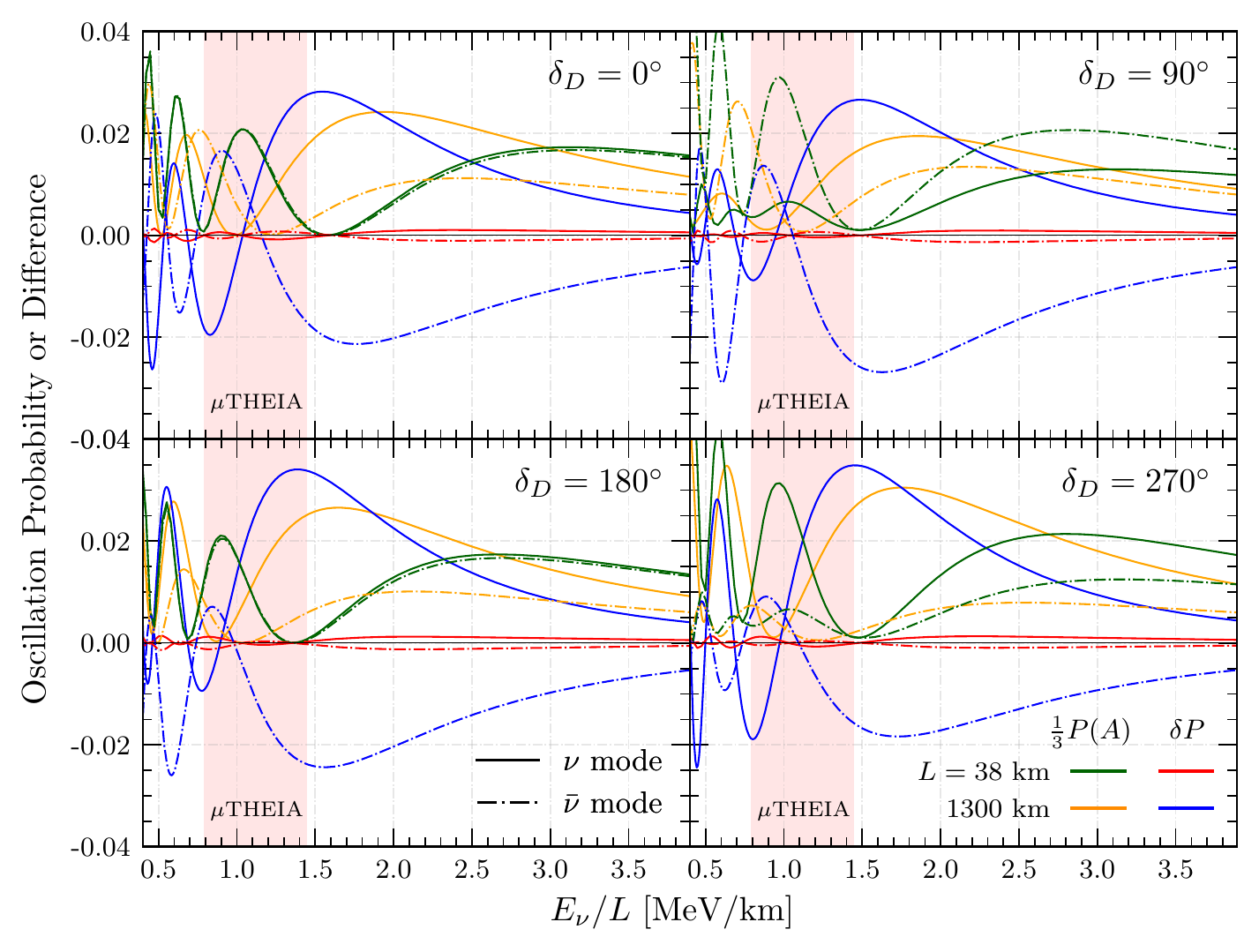}
\caption{The oscillation probability $\frac 1 3 P(A)$ and
oscillation probability difference
$\delta P_{\mu e} \equiv P_{\mu e}(A) - P_{\mu e}(A=0)$
between the matter-induced and vacuum cases
as a function of $E_\nu/L$ for both neutrino (solid)
and anti-neutrino (dashed) modes. Both $\mu$THEIA ($L = 38$\,km
with green line for $P_{\mu e}(A) / 3$ and red one for
$\delta P_{\mu e}$) and DUNE ($L = 1300$\,km with yellow
line for $P_{\mu e}(A) / 3$ and blue one for $\delta P_{\mu e}$)
are illustrated. The pink band indicates the $\mu$THEIA
energy range [30, 50]\,MeV while the DUNE energy window
spans the whole range.}
\label{fig:dunedar}
\end{figure}

The difference between the oscillation probabilities with and
without matter,
$\delta P_{\mu e} \equiv P_{\mu e}(A) - P_{\mu e}(A = 0)$,
in \gfig{fig:dunedar} shows explicitly the matter
effect at the DUNE and $\mu$THEIA configurations.
For DUNE ($L = 1300$\,km with blue and yellow lines),
the difference can be as large as $\delta P_{\mu e} \approx 0.03$ for
$E_\nu / L \approx 1.5\,\mbox{MeV/km}$ which is roughly
52\% (62\%) of the
CP-violating oscillation probability $P_{\mu e} = 0.058$
($|P^\nu_{\mu e} - P^{\bar \nu}_{\mu e}| = 0.048$).
Even for the neutrino energy
peak at $E_\nu / L \approx 1.9\,\mbox{MeV/km}$, the size of the
matter effect is still as large as 0.025 and 33\% (52\%)
of $P_{\mu e} = 0.075$
($|P^\nu_{\mu e} - P^{\bar \nu}_{\mu e}| = 0.048$).
It is interesting to see that the probability difference
is almost independent of the Dirac CP phase $\delta_D$.
This is because the major matter effect,
$\delta P_{\mu e} \approx 8 s^2_r s^2_a A \approx 0.02$,
comes from the second term of \geqn{eq:Pnumu_nue_approx} at
the oscillation peak without involving $\delta_D$.
Although the matter effect also appears
through the third term of \geqn{eq:Pnumu_nue_approx},
the effect is further suppressed by a prefactor of
$\Delta_a \sin 2 \theta_s \times (\alpha / s_r) \approx 0.3$
and hence is a minor effect. Altogether, the matter effect at
DUNE is at the same order as the genuine CP effect.

In contrast, the matter effect at $\mu$THEIA ($L = 38$\,km with
red and green lines) is negligibly small.
Being essentially insensitive to the matter potential,
$\mu$THEIA can focus on the genuine CP phase while DUNE
probes both. Their combination can significantly
improve the CP sensitivity.
We will further discuss the details of $\mu$THEIA and its
interplay with DUNE in \gsec{sec:modes} for neutrino
detection and \gsec{sec:CP} for CP sensitivity.
\section{Neutrino Detection at THEIA and DUNE Detectors}
\label{sec:modes}

As proposed above, the essential feature of the DUNE and $\mu$THEIA
complex is a combination of different baselines and neutrino beams.
In addition, the DUNE and THEIA detectors are also quite different with
liquid Argon and WbLS targets, respectively. Each combination of neutrino
beam and detector has its own characteristics in neutrino detection.
For clarity, we elaborate separately the details of the
``low-energy mode'' (LEM) that the $\mu$DAR beam is detected by
the THEIA detector in \gsec{sec:lowEnergyMode}
as well as the the ``high-energy mode'' (HEM) that the LBNF beam
is detected by both the DUNE and THEIA detectors in
\gsec{sec:highEnergyMode}. We study in detail the event
reconstruction for both signal and background to obtain their normalized
transfer tables. The event rate including the information of flux, running time,
cross section, and detector size can be found in the following \gsec{sec:CP}
when estimating the CP sensitivity.

\subsection{The Low-Energy Mode}
\label{sec:lowEnergyMode}

The $\mu$DAR neutrinos are produced by a cyclotron complex.
For example, a typical 800\,MeV proton beam hits a thick target to first
generate pions. Although both $\pi^\pm$ can be produced,
$\pi^-$ is mostly absorbed by the positively charged nuclei while
$\pi^+$ decays at rest via $\pi^+ \rightarrow \mu^+ \nu_\mu$.
The decay product $\mu^+$ also loses its energy
and decays at rest via $\mu^+ \rightarrow e^+ +\nu_e + \bar \nu_\mu$.
During this process, three neutrinos ($\nu_\mu$, $\nu_e$, and
$\bar \nu_\mu$) are produced. Of them, $\bar \nu_\mu$ experiences
the $\bar \nu_\mu \rightarrow \bar \nu_e$ oscillation that is of
interest to CP measurement. Since $\mu^+$ decays at rest, 
$\bar \nu_\mu$ has a well-understood
spectrum with maximum energy of 53\,MeV \cite{LSND:2001aii}.

\subsubsection{Signal at THEIA Detector}
\label{sec:IBDsignal}

As mentioned in the previous section,
the low-energy $\bar \nu_\mu$ cannot be detected by the DUNE
detector. But the THEIA detector is an ideal equipment.
For the major oscillation channel $\bar \nu_\mu \rightarrow \bar \nu_e$
for CP measurement with $\mu$DAR neutrinos, the IBD process
($\bar \nu_e + p \rightarrow e^+ + n$) is an ideal
detection method. With a large volume of WbLS,
THEIA has a significant fraction of free protons
(hydrogen) to allow IBD. Both positron and neutron in the final
state are detectable by THEIA.

The WbLS allows THEIA to detect both scintillation and Cherenkov lights.
There are several differences between the Cherenkov and scintillation
lights that can be used for separation \cite{Wei:2016vjd,Sawatzki:2020mpb}:
(1) the arrival time, i.e, 
Cherenkov light arrives nanoseconds earlier 
than the delayed scintillation light; 
(2) the angular topology, i.e, 
the Cherenkov light emission will cause a local 
enhancement on top of the isotropic 
scintillation signal; (3) the wave-length, i.e, the scintillation light
has shorter wavelength while the Cherenkov
light has relatively longer one.

For particle identification,
the Cherenkov ring from $e^\pm$ and $\gamma$ has a smeared pattern.
In contrast, the heavier muon or charged pion has much
sharper Cherenkov ring. In addition, the production rates
of Cherenkov and scintillation lights are different for
different particles \cite{Wei:2016vjd}. We will give more
details once needed in later discussions.

\begin{figure}[t]
\centering
\includegraphics[width=\linewidth]{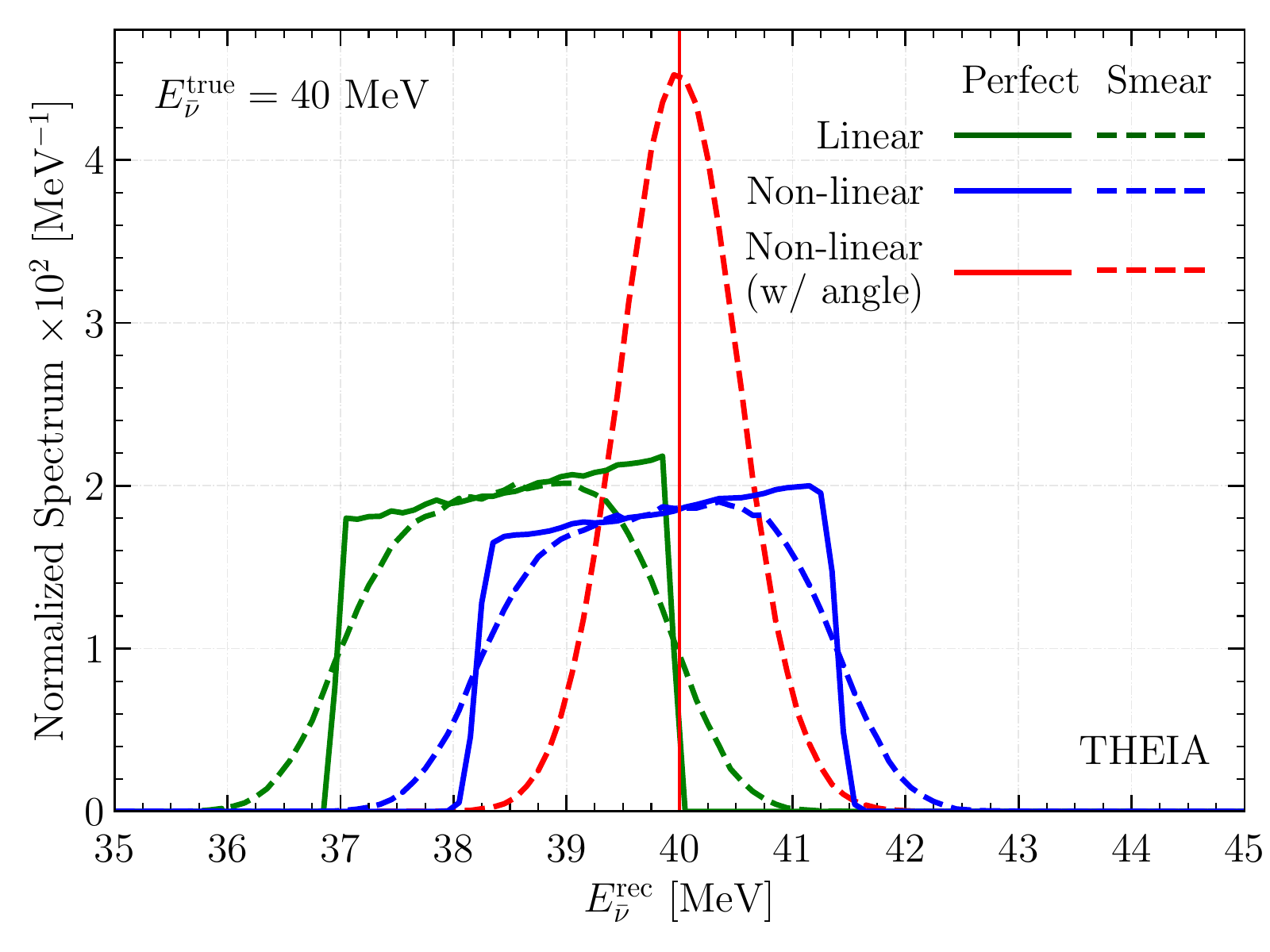}
\caption{The reconstructed neutrino energy $E^{\rm rec}_{\bar \nu}$
spectrum of IBD events at the THEIA detector for a true energy
$E^{\rm true}_{\bar \nu_e} = 40$\,MeV with linear (green)
and non-linear reconstruction methods with (red) or without
(blue) angular information. For comparison, both results with
(dashed) and without (solid) energy smearing are shown.}
\label{fig:IBDeRec}
\end{figure}

Both Cherenkov and scintillation
lights can be used for energy reconstruction.
For low energy reactor and supernova $\bar \nu_e$ neutrinos,
the energy is typically reconstructed with linear form,
$E_{\bar\nu} = E_e + m_n - m_p$ \cite{Ge:2012wj} where
$m_n$ ($m_p$, $m_e$) is the neutron (proton, electron) mass
and $E_e$ the positron energy.
Nevertheless, this linear formula cannot reconstruct the
neutrino energy very precisely. To see the deviation clearly,
we simulate the IBD events with GENIE \cite{Andreopoulos:2009rq,
Andreopoulos:2015wxa}
and reconstruct the neutrino energy according to the linear formula.
As shown in \gfig{fig:IBDeRec}, the reconstructed energy spectrum
for neutrinos with $E^{\rm true}_{\bar \nu} = 40$\,MeV spreads
into a trapezoid (green solid line)
and the central value shifts leftward by almost 2\,MeV.
In other words, for the $\mu$DAR neutrinos with $\mathcal O(10$\,MeV)
energy, the neutrino energy reconstruction is no longer a
linear dependence on the positron energy with a constant shift.

Another possibility is adding a correction term with nonlinear dependence
\cite{IBD},
\begin{equation}
  E^{\rm rec}_{\bar{\nu}}
=
  E_e + \Delta
+ \frac{2E_e(E_e+\Delta)+\Delta^2-m^2_e}{2m_p},
\label{eq:EIBD}
\end{equation}
where $\Delta \equiv m_n-m_p$ is the difference between the
neutron ($m_n$) and proton ($m_p$) masses.
The blue solid line clearly shows that
the reconstructed trapezoid spectrum shifts back to center around
the true neutrino energy at $E^{\rm true}_{\bar \nu} = 40$\,MeV.
Nevertheless, the reconstructed
spectrum is still a trapezoid even without smearing due to the
detector resolution. With a half width being almost 1.3\,MeV,
the corresponding energy uncertainty can be as large as 3.25\%
which is even larger than the statistical uncertainty as elaborated
below.

To get a precise reconstruction of the neutrino energy,
we adapt the reconstruction formula from long-baseline experiments
\cite{Abe:2017vif} to incorporate the scattering angle,
\begin{equation}
  E^{\rm rec}_{\bar{\nu}}
=
  \frac{m^2_n-m_p^2-m^2_e+2m_p
  E_e}{2(m_p-E_e + |{\bf p}_e| \cos\theta_e)},
\label{eq:EIBD-angle}
\end{equation}
where ${\bf p}_e$ is the 
electron momentum while $\theta_e$ is the angle between
the neutrino beam and positron direction. The directional
information is a benefit due to the capability of detecting
Cherenkov light for a WbLS detector.
In the absence of detector resolution, \geqn{eq:EIBD-angle}
can obtain exactly the true neutrino energy and the
reconstructed spectrum becomes a $\delta$-function
(red solid line) in \gfig{fig:IBDeRec}.
Note that the positron energy $E_e$ cannot be directly measured.
The positron not just loses its kinetic energy but also annihilates
with an environmental electron to produce two 511\,keV $\gamma$'s.
The total deposited energy that is visible in the detector
receives an extra electron mass in addition to the positron energy,
$E_{\rm vis} \equiv E_e + m_e$. In other words, the positron
energy in the linear as well as non-linear energy
reconstruction formula \geqn{eq:EIBD} and \geqn{eq:EIBD-angle} 
is reconstructed as $E_e = E_{\rm vis} - m_e$.

With systematical uncertainty from the reconstruction formula
eliminated, the remaining uncertainty mainly
comes from the detector resolutions of the visible
energy $E_{\rm vis}$ and scattering angle $\theta_e$.
Since the light yield is proportional to the deposited energy,
the visible energy $E_{\rm vis}$ can be
reconstructed from the number of photoelectrons (p.e.) which
is roughly 80 (130) p.e./MeV for the Cherenkov
(scintillation) light \cite{Sawatzki:2020mpb}. 
The relative statistical uncertainties are
$(\sigma_{E_{\rm vis}} / E_{\rm vis})_{\rm Ch} = 1 / \sqrt{80 \times E_{\rm vis} / \mbox{MeV}}$
for the Cherenkov light and
$(\sigma_{E_{\rm vis}} / E_{\rm vis})_{\rm sc} = 1 / \sqrt{130 \times E_{\rm vis} / \mbox{MeV}}$
for the scintillation light, respectively. The combined uncertainty
is
\begin{equation}
\label{eq:sigma_over_Evis}
  \frac {\sigma_{E_{\rm vis}}} {E_{\rm vis}}
=
  \frac 1 {\sqrt{210 \times E_{\rm vis} / \mbox{MeV}}}
\approx
  \frac {7\%}{\sqrt{E_{\rm vis} / \mbox{MeV}}}.
\end{equation}
Since the $\mu$DAR neutrino energy is typically $\mathcal O(10$\,MeV),
the relative error $\sigma_{E_{\rm vis}} / {E_{\rm vis}} \lesssim 2\%$
is actually quite good.
Note that the scintillation light yield is more efficient than the
Cherenkov one and both can play important roles in energy 
reconstruction. At $E^{\rm true}_{\bar\nu} = 40$\,MeV, the 
positron energy resolution leads to about $1.17\%$
uncertainty on the reconstructed neutrino energy.

In addition, the scattering angle $\theta_e$ in \geqn{eq:EIBD-angle}
also receives an imperfect detector resolution and blur the reconstructed
neutrino energy. Although the actual resolution for the $\mu$DAR flux
that has higher energy should be better, we take a conservative value
$\sigma_\theta=10^\circ$ \cite{Theia:2019non} estimated for
supernova neutrino.
The angular uncertainty leads to a
$0.44\%$ uncertainty on the neutrino energy
reconstruction which is slightly less than
the one from positron energy resolution.
In \gfig{fig:IBDeRec}, the dashed lines show the combined
neutrino energy uncertainty, which is 1.28\% for
$E^{\rm true}_{\bar \nu} = 40$\,MeV assuming Gaussian smearing. 

Once produced, the neutron from IBD would experience thermalization
and lose energy before being captured. The scintillation
light emitted during this process can mix with the
light from positron.
Consequently, the visible energy receives an extra
contribution from the neutron kinetic energy $T_n$,
$E_{\rm{vis}}=E_e+m_e+Q_F\times T_n$ where $Q_F$ is the neutron
quenching factor.
For the $\mu$DAR neutrino at 40\,MeV, the neutron kinetic
energy $T_n$ is typically 3\,MeV according to GENIE simulation.
With WbLS, it is possible to reconstruct the neutron
kinetic energy $T_n$ by measuring the positron scattering angle
\cite{Wei:2020yfs}. The neutron quenching factor
$Q_F$ keeps decreasing with $T_n$ and is 15\%
at $T_n = 0.25$\,MeV \cite{Wei:2020yfs}.
Then from the measured visible energy $E_{\rm vis}$, one can
first solve the positron energy
$E_e = E_{\rm vis} - m_e - Q_F \times T_n$ and put it into
\geqn{eq:EIBD-angle} to reconstruct $E_{\bar \nu}$.
In principle, the neutron kinetic energy is not a problem
for the IBD energy reconstruction. But the neutron quenching factor
$Q_F$ still remains to be measured in a WbLS as far as we know.
So for simplicity, we omit the neutron kinetic energy in
our current phenomenological study.

\subsubsection{Background Suppression with WbLS}
\label{sec:mudarbkg}

As illustrated in \gsec{sec:IBDsignal}, the $\bar \nu_e$ signal
contains not just a positron but also a neutron in the final state.
Although a single positron can already allow precise energy
reconstruction, it receives various backgrounds. The final-state
neutron is extremely important for selecting out the IBD signal
with double coincidence.

The neutron is produced at the same time as the positron.
But the neutron signal comes out much later with a delay
of 250\,$\mu$s \cite{talk}. Since the time resolution required
to separate the Cherenkov and scintillation lights (typically 
0.1\,ns) is much smaller \cite{Theia:2019non}, the delayed
neutron signal can be well separated from the positron
one. When neutron is captured by nuclei, the delayed $\gamma$s
also produce $e$-like Cherenkov rings and scintillation lights
to allow neutron tagging at THEIA \cite{talk}.
The neutron tagging efficiency can reach almost 90\%
\cite{Askins:2019oqj}.

The double coincidence requiring both positron and neutron
signals can remove those events that contain only one faked
positron. But there are still various backgrounds that can survive
\cite{Evslin:2015pya,Shaevitz:2015uar} including the beam neutrinos
from the complementary high-energy beam neutrinos, the low-energy
reactor, solar and supernova neutrinos, the intrinsic beam neutrinos from
the same $\mu$DAR flux, as well as the atmospheric
neutrinos. The invisible muon decay from atmospheric neutrinos
is particularly difficult to remove since both positron and
neutron can be produced to fake the signal. Fortunately, the
new technique of WbLS can be very efficient in suppressing these
backgrounds as we elaborate below.
\\

\noindent
{\bf Reactor, Solar, Geo-, Supernova, and DUNE Beam Neutrinos} --
At low energy, there are four neutrino sources
from reactor \cite{Mueller:2011nm,Huber:2011wv},
solar \cite{Bahcall:1987jc}, geo-radioactivity \cite{Mantovani:2003yd,Sramek:2012nk},
and supernova \cite{Dighe:1999bi,Keil:2002in}.
While the solar neutrinos are not anti-neutrinos and hence
cannot fake the $\bar \nu_e$ IBD signal, the other three
may be incorrectly identified as the signal.
Fortunately, all these neutrinos typically have much lower
energy than majority of the $\mu$DAR neutrinos.
With an energy cut, $E_{\bar \nu} > 30$\,MeV, they
can be safely removed \cite{Evslin:2015pya}.
On the other side, the LBNF beam
can also contribute $\bar \nu_e$ flux. With pulsed beam at
much higher energy, a combined cut on the arrival time and
energy, $E_{\bar \nu} < 55$\,MeV, will eliminate the contamination
from the complementary DUNE experiment.
\\

\noindent
{\bf Intrinsic $\mu$DAR Beam} --
During the production of $\mu$DAR with cyclotron,
various neutrinos can be produced via
$\pi^\pm \rightarrow \mu^\pm + \nu_\mu (\bar \nu_\mu)$
and $\mu^\pm \rightarrow e^\pm + \bar \nu_\mu (\nu_\mu) + \nu_e (\bar \nu_e)$.
One needs to examine all possible intrinsic backgrounds.
The $\bar \nu_e$ flux can produce
exactly the same IBD signal and becomes an intrinsic background
from the same $\mu$DAR source. There is no way to remove or
suppress this background on the detector side. Fortunately,
the negatively charged $\pi^-$ is efficiently absorbed by the
positively charged nuclei inside the target before decay.
So the $\mu^-$ production is much smaller than $\mu^+$ with
a suppression as large as $10^{-4}$ \cite{Evslin:2015pya}. Since the
$\bar \nu_\mu \rightarrow \bar \nu_e$ oscillation probability 
is typically $\mathcal O(1\%)$ and hence two orders larger,
the $\mu^-$ contamination can be safely neglected.

The second beam background is the electron neutrino
$\nu_e$ from the same $\mu$DAR. After oscillation, most
$\nu_e$ neutrinos can survive and scatter with
oxygen ($\nu_e+ {}^{16}O \rightarrow e^- + {}^{16}F$) or
carbon ($\nu_e+ {}^{12}C \rightarrow e^- + {}^{12}N$)
\footnote{Since the scintillator composition and fraction
for THEIA are not decided yet and only a possible fraction
$1\% \sim 10\%$ is mentioned \cite{Theia:2019non},
we leave the carbon contribution
open and focus on the major water target.}
to produce an electron in the final state.
Fortunately, the scattering cross section with oxygen
is much smaller than the IBD
one in the $\mu$DAR energy range
\cite{Vogel:1999zy, Formaggio:2012cpf} and
hence can be neglected. Even if some events can still be
produced, the double
coincidence of neutron capture can remove this background.

The muon neutrino and anti-neutrino cannot experience
charged-current scattering with not enough energy to
produce a $\mu^\pm$ in the final state.
However, all neutrinos and anti-neutrinos can elastically
scatter with electron to produce an energetic electron in the
final state. But these backgrounds can also be removed by
requiring neutron capture \cite{Wei:2016vjd,Sawatzki:2020mpb}.
\\

\noindent
\textbf{Atmospheric Neutrinos} --
The atmospheric neutrino flux \cite{SajjadAthar:2012dji,Honda}
contains all neutrino flavors
($\nu_e$, $\bar{\nu}_e$, $\nu_\mu $, $\bar{\nu}_\mu$,
$\nu_\tau$, and $\bar{\nu}_\tau$) to
contribute as background \cite{Evslin:2015pya}.
With much higher energy, the
atmospheric neutrino backgrounds can experience all
types of neutrino-nucleus scattering and hence need
much more dedicated treatment. Fortunately, the
excellent performance of the WbLS with both Cherenkov
and scintillation detection can effective suppress
these backgrounds. Except $\bar \nu_e$ and $\bar \nu_\mu$,
the other components can be removed at the THEIA detector
as we elaborate below.

$\boldsymbol \nu_{\bf e}$:
The charged-current quasi-elastic scattering (CC-QES) process
$\nu_e + {}^{16} O \rightarrow e^- + {}^{16}F$ cannot produce
a neutron. Then the neutron tagging with WbLS can efficiently
remove this background. For the atmospheric $\nu_e$ with 
higher energies, 30\% of the CC-QES scattering process can kick
out a neutron and a proton from the nuclei via
$\nu_e + {}^{16}O \rightarrow e^- +n + {}^{15}O^* + p$.
Since energy is needed to kick out $n$ and $p$,
the primary electron energy is much smaller than the neutrino
energy. Mainly $E_\nu \in [80, 130]$\,MeV
will produce electrons with energy reconstructed in the
$\mu$DAR range $[30, 55]$\,MeV.
In addition, roughly half of the single neutron events
are also accompanied by a monoenergetic photon 
($\sim$6.2\,MeV) due to the de-excitation of $^{15}O^*$,
which can also serve as background veto.
Both electron ($T_e$) and proton ($T_p$) kinetic
energies deposit as visible energy $E_{\rm vis}$.
Requiring $E_{\rm vis} \in [30,55]$\,MeV, only 
0.07 events per year at THEIA-25 can survive,
which is a negligible amount.

Moreover, those neutrinos with even higher
energy can also have resonant (RES) and deep inelastic (DIS)
scatterings with at least a pion ($\pi^0$ or $\pi^+$) starting
around $E_\nu \approx$ 200\,MeV, and possibly a neutron in the
final state. In addition, the interaction can also produce
protons in the final state or kick off protons from the nucleus 
\cite{SajjadAthar:2020nvy}. The neutron capture process
emits a single 2.2\,MeV photon \cite{Sawatzki:2020mpb} that
can be used to separate RES and DIS events. In the energy range
$E_\nu \in [200, 600]$\,MeV that can contribute to the
$\mu$DAR IBD energy window, roughly 30\% of CC-RES
can have a single neutron.
The events containing $\pi^0$ can
be vetoed by energy cut $E_{\rm vis} < 55$\,MeV,
since the $\pi^0$ decays immediately into a pair of photons
each with energy $E_\gamma \geq m_\pi/2$. This reduces the
remaining atmospheric $\nu_e$ background by another 50\% to
only 15\% of the total CC-RES. The
charged $\pi^+$ first deposits its kinetic energy as
scintillation light and then decays at rest to produce
a $\mu^+$, since its decay length
is typically at least one order longer than the radiation length in
\begin{widetext}
\centering
\begin{minipage}{\linewidth}
\begin{figure}[H]
\centering
\includegraphics[width=0.49\linewidth]{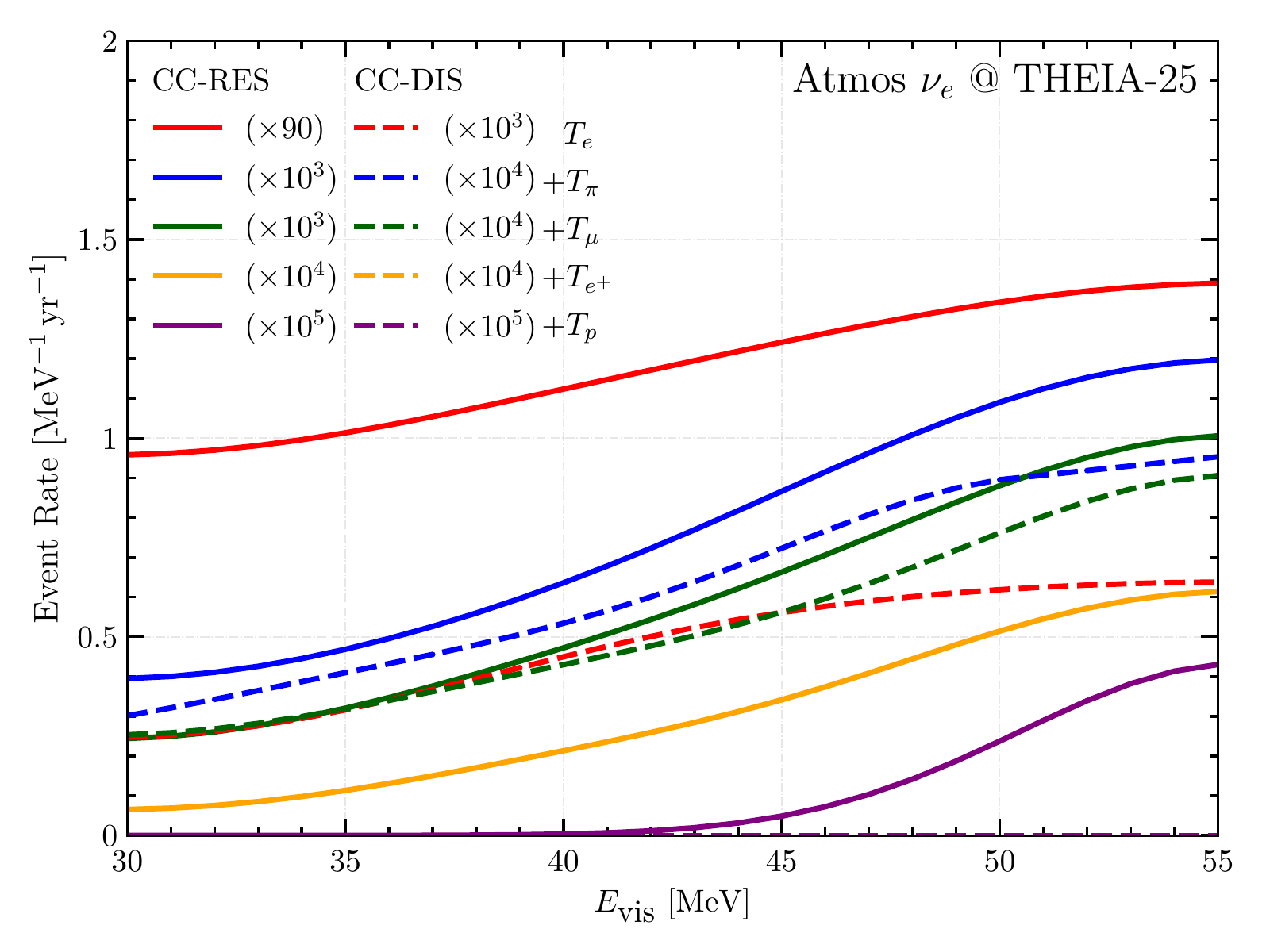}
\hfill
\includegraphics[width=0.49\linewidth]{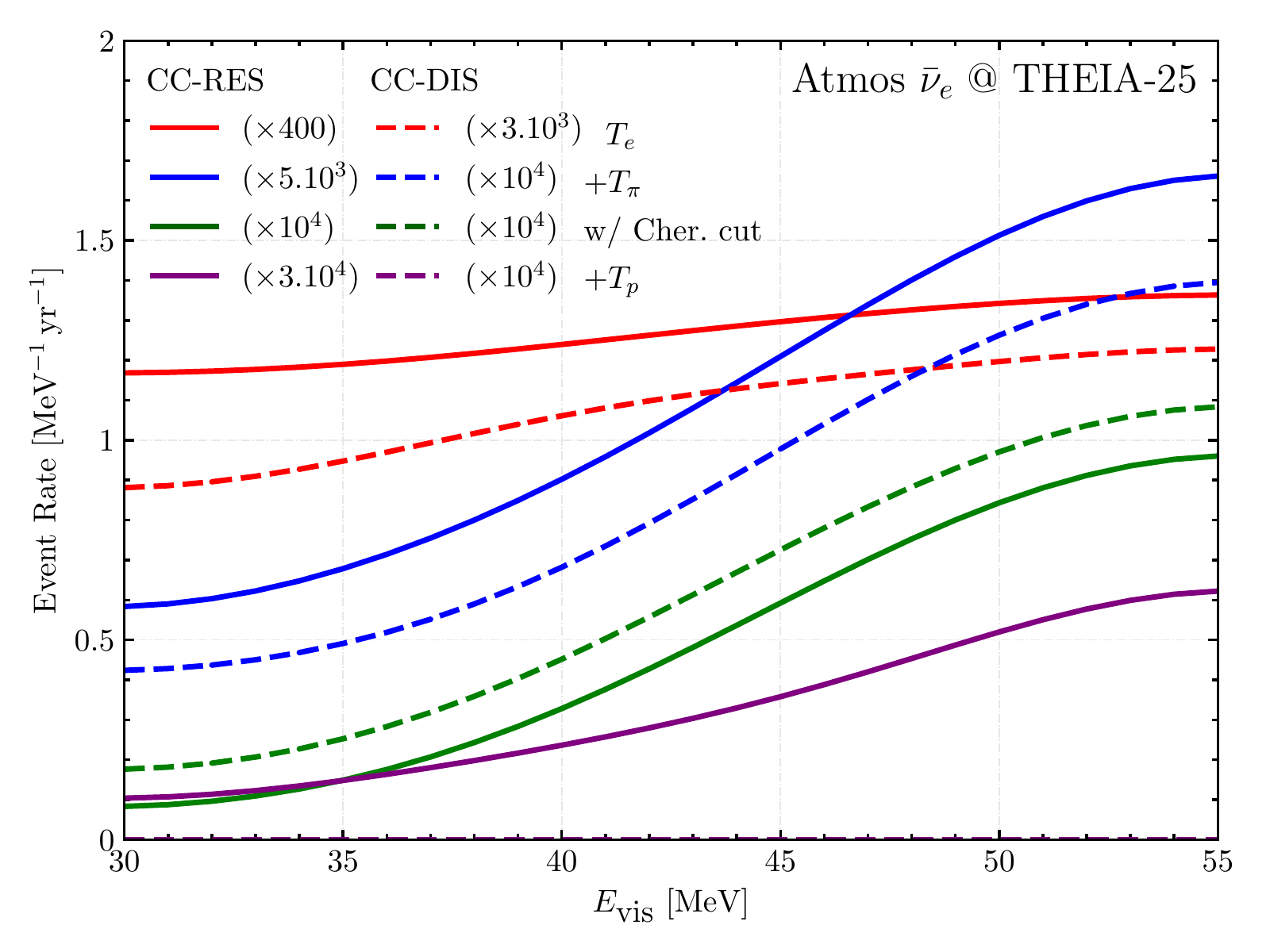}
\caption{
The atmospheric $\nu_e$ (left) and $\bar \nu_e$ (right) background
event rate spectrum from CC-RES (solid) and 
CC-DIS (dashed) interactions as a function of the visible energy
$E_{\rm{vis}}$ at THEIA-25. With energy deposits
from the primary electron ($T_e$, red), charged $\pi^+$ ($+T_\pi$,
blue), $\mu^+$ ($+T_\mu$, green), the Michel electron ($+T_{e^+}$,
yellow), and proton ($+T_p$, purple) gradually added up,
the $\nu_e$ event spectrum moves out of the IBD energy
window [30, 55]\,MeV to higher energy.
Similarly, the $\bar \nu_e$ spectrum has the same feature with
the primary electron ($T_e$, red), charged $\pi^-$ ($+T_\pi$,
blue) or with Cherenkov cut (w/ Cher. cut, green),
and proton ($+T_p$, purple).
Multiplication factors have been used to make the
highly suppressed spectrum visible.}
\label{fig:atmCC_bkg}
\end{figure}
\end{minipage}
\end{widetext}
\noindent
water \cite{PDG20-Matter}. Then the energy 
deposition is composed of five parts: 1) the kinetic energy $T_e$ of the primary electron;
2) the proton kinetic energy $T_p \equiv E_p - m_p$;
3) the $\pi^+$ kinetic energy $T_\pi \equiv E_\pi - m_\pi$;
4) the $\mu^+$ kinetic energy
$T_\mu = (m_\pi - m_\mu)^2 / 2 m_\pi \approx 4$\,MeV
that is uniquely determined by the pion decay at rest; and
5) the $e^+$ energy $T_{e^+}$ ranging from 0 to $m_\mu / 2$.
Since the positron always annihilation with an environmental
electron, we also include the released $2 m_e$ photon energy
in $T_{e^+}$ for convenience. For a $\pi^+$ with energy
$E_\pi$, the minimal total energy deposit is
$E_\pi - m_\pi + T_\mu + 2 m_e \approx E_\pi - 134$\,MeV if
both the primary and Michel electrons have vanishing momentum.
Those events with $E_\pi \gtrsim 189$\,MeV can be vetoed by
$E_{\rm vis} \geq 55$\,MeV. Requiring the primary and Michel
electrons to be produced at rest is actually very stringent.
Even with less energetic $\pi^+$, the atmospheric $\nu_e$
background can be easily vetoed. As shown in the left figure of \gfig{fig:atmCC_bkg},
once the $\pi^+$, $\mu^+$, and $e^+$ energies are considered,
the $\nu_e$ CC-RES event rate is suppressed to negligible level
with just around $8 \times 10^{-4}$ events/year for
$E_{\rm vis} \in [30, 55]$\,MeV at the THEIA-25 detector.
\gfig{fig:atmCC_bkg} also shows the CC-DIS case which has
only 20\% for single neutron events and its contribution
is only around $10^{-5}$ event per year. The typical value
of the proton kinetic energy is around
$\mathcal{O}(10\,\mbox{MeV})$ for sub-GeV neutrinos.
It can further suppress the background by one
order for RES and almost down to zero for DIS, 
shown as the purple curves of \gfig{fig:atmCC_bkg}.

\begin{figure}[t]
\centering
\includegraphics[width=\linewidth]{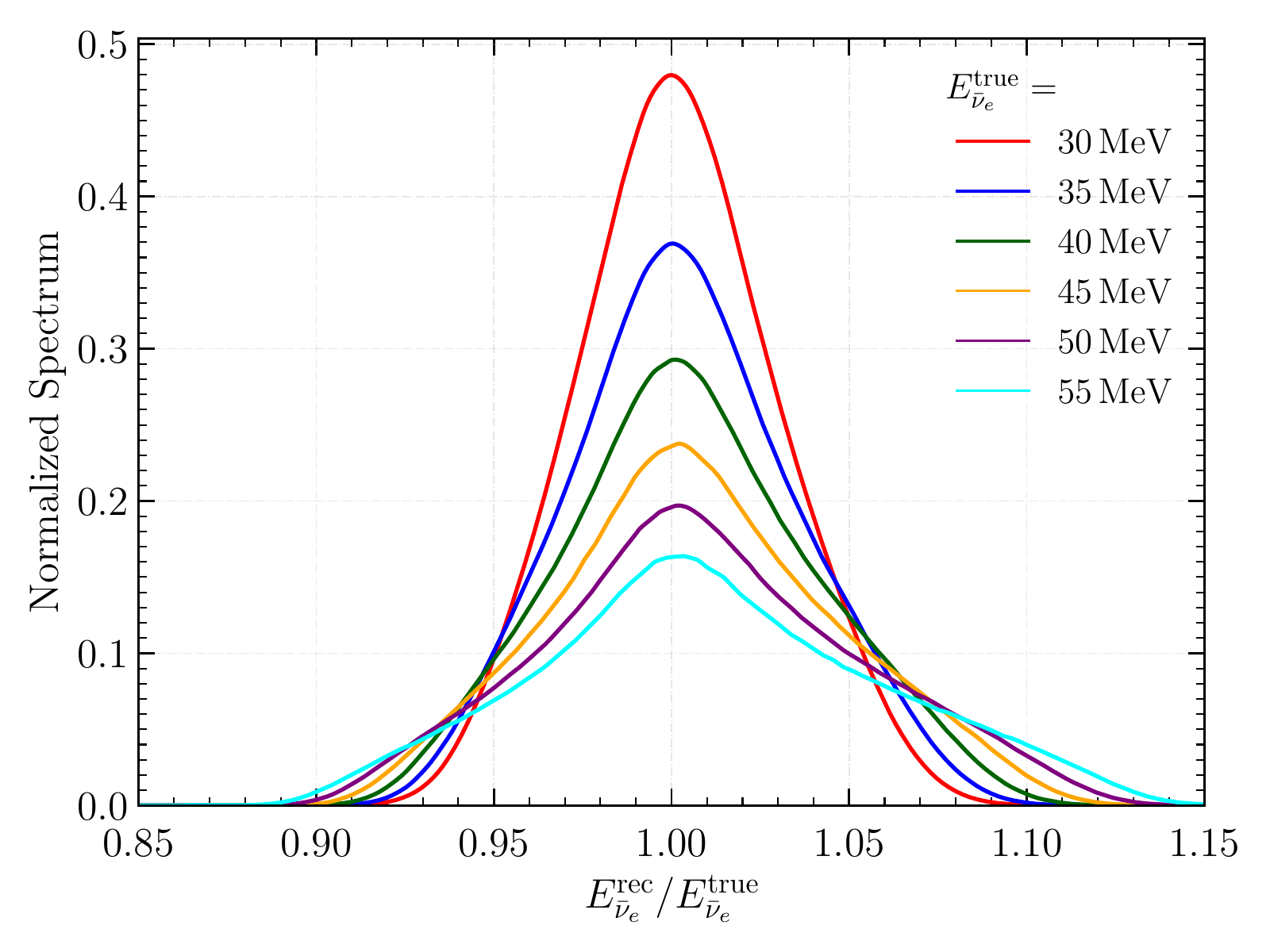}
\caption{The wrong scattering angle effect on the reconstructed
neutrino energy $E^{\rm rec}_\nu$ for the atmospheric
$\bar \nu_e$ background in the unit of the corresponding true
neutrino energy $E^{\rm true}_{\nu_e}$.
Several typical neutrino energies,
$E^{\rm true}_{\nu_e}$ = 30\,MeV (red), 35\,MeV (blue),
40\,MeV (green), 45\,MeV (yellow), 50\,MeV (purple),
and 55\,MeV (cyan), are shown for comparison.
}
\label{fig:atm-erec}
\end{figure}

$\bar{\boldsymbol{\nu}}_{\bf e}$:
The electron anti-neutrino $\bar{\nu}_e$ that scatters with
Hydrogen via CC-QES (or precisely IBD) process is an
irreducible background to the IBD signal and there is
no experimental solution. It contributes the major background
for $\mu$THEIA. Since the direction of the incoming
$\bar \nu_e$ is unknown, the scattering angle $\theta_e$ in
\geqn{eq:EIBD-angle} cannot be correctly measured but only inferred
from the $\mu$DAR source direction. This {\it wrong scattering
angle effect} introduces significant smearing for the
reconstructed neutrino energy $E_{\bar \nu}$. With more details
elaborated in \gapp{sec:angle}, \gfig{fig:atm-erec} shows that
the smearing can be as large as $3.4\%\sim5.5\%$ at half height for
$E_\nu \in [30, 55]$\,MeV. The wrong scattering angle effect
is much larger than the detector resolution in
\gfig{fig:IBDeRec} with known direction.
With higher neutrino energy, the wrong scattering angle effect
becomes more severe.
The atmospheric $\bar\nu_e$ CC-QES background
is estimated as 1.1 event per year at THEIA-25.

The $\bar{\nu}_e$ CC-RES and CC-DIS events mainly consist of
a $\pi^0$ or $\pi^-$ in addition to the primary positron.
Slightly higher than the $\nu_e$ mode, the single neutron events
contribute around 50\% (60\%) of the total CC-RES (CC-DIS) events.
The $\pi^0$ can be vetoed by the energetic photon with $E_\gamma \geq m_\pi/2$.
Since the pion decay length is much longer than its radiation length
in water, $\pi^-$ first loses its kinetic energy and
then is absorbed by the positively charged nuclei.
The minimal energy deposit is then only its kinetic
energy $T_\pi = E_\pi - m_\pi$ and no $\mu^-$ or subsequent
electron from $\mu^-$ decay. For comparison with the $\nu_e$
mode, the deposited $E_{\rm vis}$ is shown in the
right panel of \gfig{fig:atmCC_bkg}.
As mentioned earlier, 80 (130) photoelectrons can be
produced for each MeV energy deposit \cite{Sawatzki:2020mpb}
as Cherenkov (scintillation) lights. So the required $\mu$DAR
energy window [30, 55]\,MeV corresponds to [2400, 4400]
Chereknov photons and [3900,7150] scintillation ones, respectively.
The atmospheric $\bar\nu_e$ CC-RES (CC-DIS) background only
contributes $1.2 \times 10^{-3}$ ($1.5 \times 10^{-3}$)
event per year at THEIA-25. This is negligibly small
compared with the atmospheric $\bar\nu_e$ CC-QES background
with 1.1 events per year. This background is further
suppressed to 1/4 for the RES and almost 0 for the DIS
processes if the proton kinetic energy is considered.

$\boldsymbol{\nu_\mu} / \bar{\boldsymbol{\nu}}_{\boldsymbol{\mu}}$:
For the atmospheric muon neutrinos $\nu_\mu/\bar\nu_\mu$,
the charged-current scattering produces a $\mu^\mp$ in the
final state. If the muon energy is above the Cherenkov
threshold, the $\mu$-like ring has different pattern to be
distinguished from the $e$-like one. In addition,
the ratio between the Cherenkov ($N_{\rm ch}$) and
scintillation ($N_{\rm sc}$) lights is also quite different
between electron ($N_{\rm ch}/N_{\rm sc} \approx 0.2$) and
muon ($N_{\rm ch}/N_{\rm sc} \lesssim 0.08$) rings \cite{Wei:2016vjd}.

On the other hand, the Cherenkov light alone cannot see an
``invisible muon'' below the Cherenkov threshold but only its
decay product $e^\pm$ \cite{Evslin:2015pya}. Fortunately,
with a muon lifetime of 2.2\,$\mu$s, the time resolution
($\sim 0.1$\,ns) of WbLS is good enough to separate
the muon scintillation light from the $e^\pm$ Cherenkov and
scintillation lights \cite{Wei:2016vjd}. In other words,
the WbLS can identify the invisible muon with triple coincidence
($\mu$ scintillation light, $e^\pm$ Cherenkov and scintillation
lights, and the 2.2 MeV delayed $\gamma$ from neutron capture). The
triple coincidence can essentially remove all the invisible muon
background. Even for a conservative study by requiring the
Cherenkov photon and the combined scintillation photon to satisfy
$N_{\rm ch}/N_{\rm sc} > 0.4$, the background rate is suppressed
to only 2\% \cite{Wei:2016vjd}.

$\boldsymbol{\nu_\tau} / \bar{\boldsymbol{\nu}}_{\boldsymbol{\tau}}$:
Although tau neutrinos also exist in the atmospheric flux,
the primary $\tau^\pm$ from the CC
scattering can decay into either electron or muon with roughly 17\%
branching ratio each. However, the $\tau^\pm$ lepton is very heavy
with mass at 1.78\,GeV. The electron and muon from tau decay is then
much more energetic than the IBD signal. A simple cut on the visible
energy can effectively veto the atmospheric tau neutrino backgrounds.

\begin{figure}[t]
\centering
\includegraphics[width=1\linewidth]{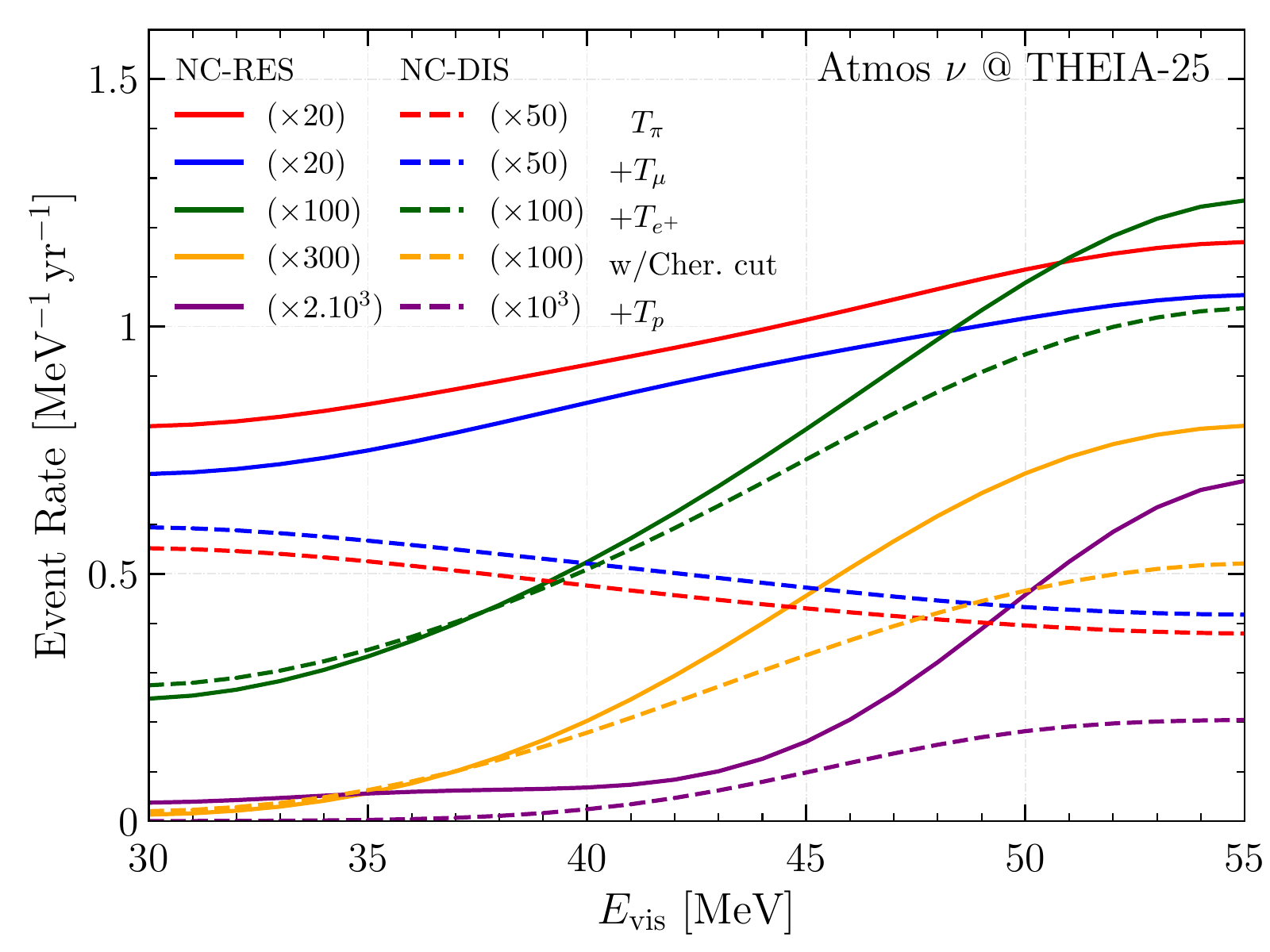}
\caption{The background event rate spectrum of the atmospheric NC-RES
(solid) and NC-DIS (dashed) events as a function of the
visible energy $E_{\rm{vis}}$ with one year running at THEIA-25.
For comparison, the deposited energies of pion ($T_\pi$, red),
muon ($+T_\mu$, blue), positron ($+T_{e^+}$) without (green)
or with (yellow) Cherenkov cut, and proton ($+T_p$, purple)
are added up step by step. To put all curves at similar heights,
multiplication factors are implemented accordingly.}
\label{fig:atm_NC_bkg}
\end{figure}

{\bf Neutral Current}:
In addition to the CC events, the neutral-current (NC)
scattering is also a potential background.
Since the neutral current process is flavor blind,
all flavors can contribute. First, the NC-QES scattering can
not contribute as background since there is no positron in the
final state. The NC-RES and NC-DIS processes both allow
a single $\pi^\pm$ and $\pi^0$ production in the final state. 
The interaction can also produce neutron and proton.
Since $\pi^0$ decays into a pair of photons each with energy larger
than $m_\pi / 2 \approx 70$\,MeV, the deposit energy is already
beyond the IBD window of $\mu$DAR $\bar \nu_e$. Among the remaining
$\pi^\pm$, $\pi^-$ is absorbed by nuclei and only $\pi^+$ below
Chereknov threshold can decay through
$\pi^+ \rightarrow \mu^+ \rightarrow e^+$
to fake the IBD positron.
For NC-RES (NC-DIS) events, around 43\% (45\%) have
a single neutron while 16\% (22\%) have a single charged
$\pi^+$ in the final state. The fraction reduces to
11\% (18\%) if requiring both neutron and $\pi^+$.
It has a probability for $E_{\rm vis}$ being within the energy
window of interest as shown in \gfig{fig:atm_NC_bkg}.  
Requiring $E_{\rm vis} \in [30, 55]$\,MeV, 
the integrated event number gives around 0.03 (0.06) events
per year for NC-RES (NC-DIS) at THEIA-25 which is also a
negligible amount. Since there is also proton, the background
can be further reduced by at least one order if the proton
kinetic energy is also taken into account.

\begin{figure}[t]
\centering
\includegraphics[width=1\linewidth]{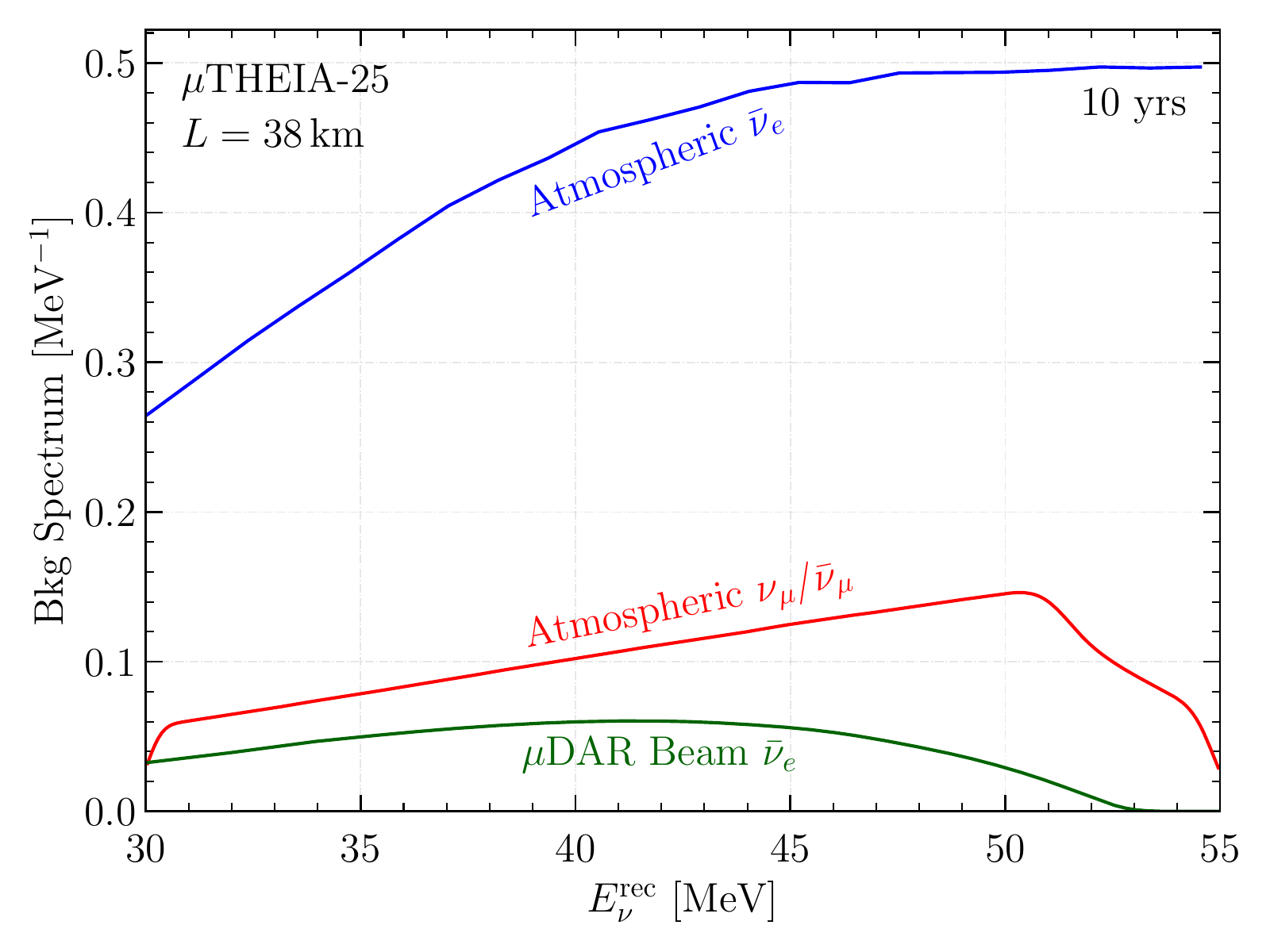}\\
\caption{The major background spectrum to the $\mu$DAR IBD
signal as a function of the reconstructed neutrino energy
$E^{\rm rec}_\nu$ at THEIA-25 (17\,kt) with 10 years running
and a baseline of $L = 38$\,km.
The remaining backgrounds are atmospheric
$\bar\nu_e$ (blue), invisible muon (red), and the
$\mu$DAR $\bar\nu_e$ (green).
}
\label{fig:bkg-spec}
\end{figure}

\gfig{fig:bkg-spec} shows the three survival background
spectra as discussed above, including the beam $\bar \nu_e$
with $1.83\times 10^{25}$ protons on target (POT) as
well as the atmospheric $\bar\nu_e$ and $\nu_\mu / \bar\nu_\mu$.
While the atmospheric invisible muon background dominates at a
Cherenkov detector such as Super-K and Hyper-K \cite{Evslin:2015pya},
it is a small minor contribution at the WbLS detector THEIA even
with a conservative selection procedure. The intrinsic $\mu$DAR
beam background is even smaller. THEIA is an ideal detector for
the CP measurement with $\mu$DAR source.

\subsubsection{Event Selection}

As elaborated above, the IBD signal is quite distinctive at
the THEIA detector. All backgrounds can be suppressed to negligibly
small amount. Here we summarize the selection criteria for
the IBD events,
\begin{itemize}
\item[(1)] Only one $e$-like Cherenkov ring;

\item[(2)] For the total visible energy $E_{\rm vis}$,
the number of scintillation photons $N_{\rm sc}$ is within
the range of [3900, 7150] and Cherenkov photons
$N_{\rm ch} \in [2400, 4400]$
that correspond to a $(30 \sim 55)$\,MeV positron;

\item[(3)] The ratio of Cherenkov and scintillation photons
$N_{\rm ch} / N_{\rm sc}$ is larger than $0.4$;

\item[(4)] Existence of delayed $\gamma$s from neutron capture;

\end{itemize}
For the above requirements, the energy window used
in the criterion (2) can remove reactor, solar, geo-,
supernova and DUNE beam neutrino backgrounds.
The criteria (2) and (4) as double coincidence is 
efficient in removing the $\mu$DAR flux background
except the intrinsic $\bar\nu_e$. The combination of
all criteria forms the triple coincidence which is even more
powerful to reduce the invisible muon background.
Finally, the criteria (1) and (2) can remove the atmospheric
CC-RES, CC-DIS, and NC backgrounds.

\begin{figure}[t]
\centering
\includegraphics[width=1\linewidth]{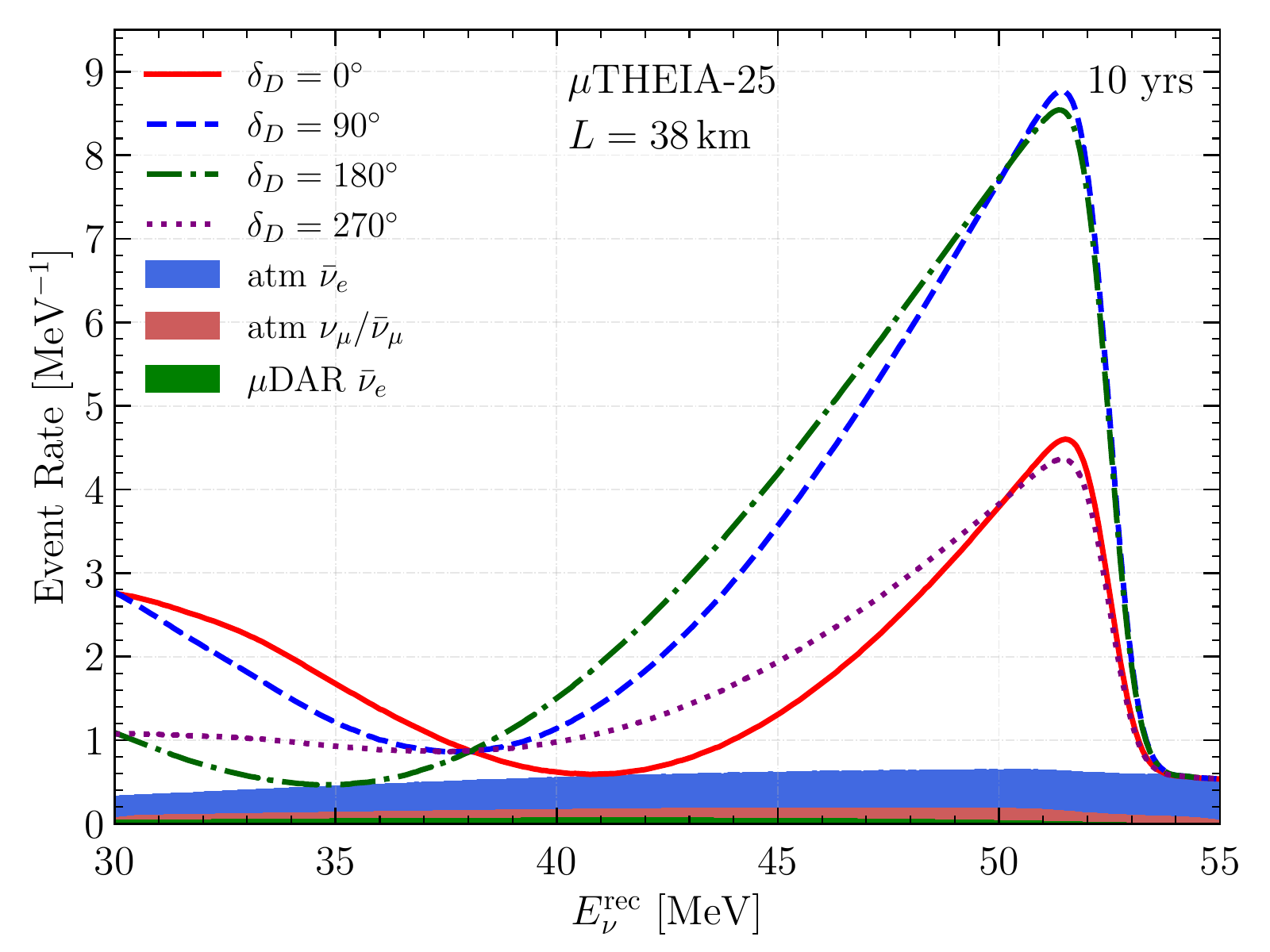}
\caption{
The event spectra of $\mu$DAR source at THEIA detector
at baseline $L = 38$\,km. The Four solid lines are
$\delta_D = 0^\circ$ (red), $90^\circ$
(blue), $180^\circ$ (green), and $270^\circ$ (purple). 
The filled regions stand for the backgrounds
the atmospheric $\overline \nu_e$ (blue), the 
atmospheric $\nu_\mu/\bar\nu_\mu$ (red) and the intrinsic
$\overline \nu_e$ from the $\mu$DAR beam (green).
}
\label{fig:muDAR}
\end{figure}

Note that these criteria are already quite conservative. For
the WbLS technique to be used by THEIA, more information can
help to distinguish signal from background. Especially, the
time information and pulse shape can be used to distinguish
the Cherenkov and scintillation lights \cite{Wei:2016vjd}. 
This could be extremely useful to further suppress the atmospheric
invisible muon background with triple coincidence fully
implemented.

In \gfig{fig:muDAR}, we show the signal and background event
rates for the low-energy mode at $\mu$THEIA. The 
WbLS can highly suppress the background especially
for the ``{\it invisible muon}'' that is reduced to a negligible
amount. With CP dependence dominating the event rate,
we can expect improvement of the CP sensitivity, which is
elaborated later in \gsec{sec:CP}.

\subsection{The High-Energy Mode}
\label{sec:highEnergyMode}

The high-energy LBNF neutrinos from Fermilab to SURF is
a wide beam spanning from 0.5\,GeV to 5\,GeV with peak
at 2.5\,GeV \cite{DUNE:2015lol} for both neutrino and
anti-neutrino modes. Each mode contains four different
flavor components: $\nu_e$, $\bar \nu_e$, $\nu_\mu$
and $\bar\nu_\mu$. Since the THEIA detector is at the same
SURF experimental site as the DUNE far detectors, it
can also probe the LBNF neutrinos. While the event reconstruction
of high energy neutrinos at the DUNE liquid Argon detectors
has already been studied carefully \cite{Abi:2021arg},
we focus on the detection at THEIA.

The key element for the neutrino CP measurement is
neutrino flavor reconstruction for
$\nu_\mu \rightarrow \nu_e$
($\bar \nu_\mu \rightarrow \bar \nu_e$).
It is interesting to see that the low-energy
$\mu$DAR neutrinos outside the energy
window [0.5\,GeV, 5\,GeV] cannot contribute as background.
Although atmospheric neutrinos can overlap in
energy, the pulse shape of the LBNF beam provides an
efficient way to suppress the atmospheric backgrounds.
Both signal and the remaining background actually come
from the same LBNF beam.

With broad energy range, several types of CC scatterings with a
target nuclei $N$ can happen.
In addition to a charged lepton $\ell_\alpha$, the final state
is either a single nuclei $N'$ for the quasi-elastic
($\nu_\alpha+N\rightarrow \ell_\alpha+N'$),
nuclei plus mesons for the resonant
($\nu_\alpha+N\rightarrow \ell_\alpha+N' + meson$),
or nuclei plus hadrons for the deep-inelastic
($ \nu_\alpha + N\rightarrow \ell_\alpha+N'
+hadrons$) CC scatterings \cite{Formaggio:2012cpf}.
Typically CC-QES dominates below 1\,GeV and CC-RES
between $1.2$\,GeV and $5 \sim 7$\,GeV, while
CC-DIS takes over above 7\,GeV. With the LBNF neutrino beam 
being below 5\,GeV, most of the interactions are CC-QES
and CC-RES. In order to make better neutrino reconstruction,
it is desirable to distinguish these different CC
scattering events.

The following discussions focus on the
signal of appearance channels
$\nu_\mu \rightarrow \nu_e$ and
$\bar \nu_\mu \rightarrow \bar \nu_e$.
Similar procedures shall also apply for the
disappearance channels $\nu_\mu \rightarrow \nu_\mu$ 
and $\bar \nu_\mu \rightarrow \bar \nu_\mu$.
Although the disappearance channels do not
contribute significantly to the leptonic CP
measurement, they can serve as supplementary
probe of the other oscillation parameters and
the neutrino flux. Both appearance and
disappearance channels are taken into account
in our GLoBES simulation in \gsec{sec:CP}.

\subsubsection{The CC-QES Category}
\label{sec:ccqe}

\begin{figure}[t]
\centering
\includegraphics[width=1\linewidth]{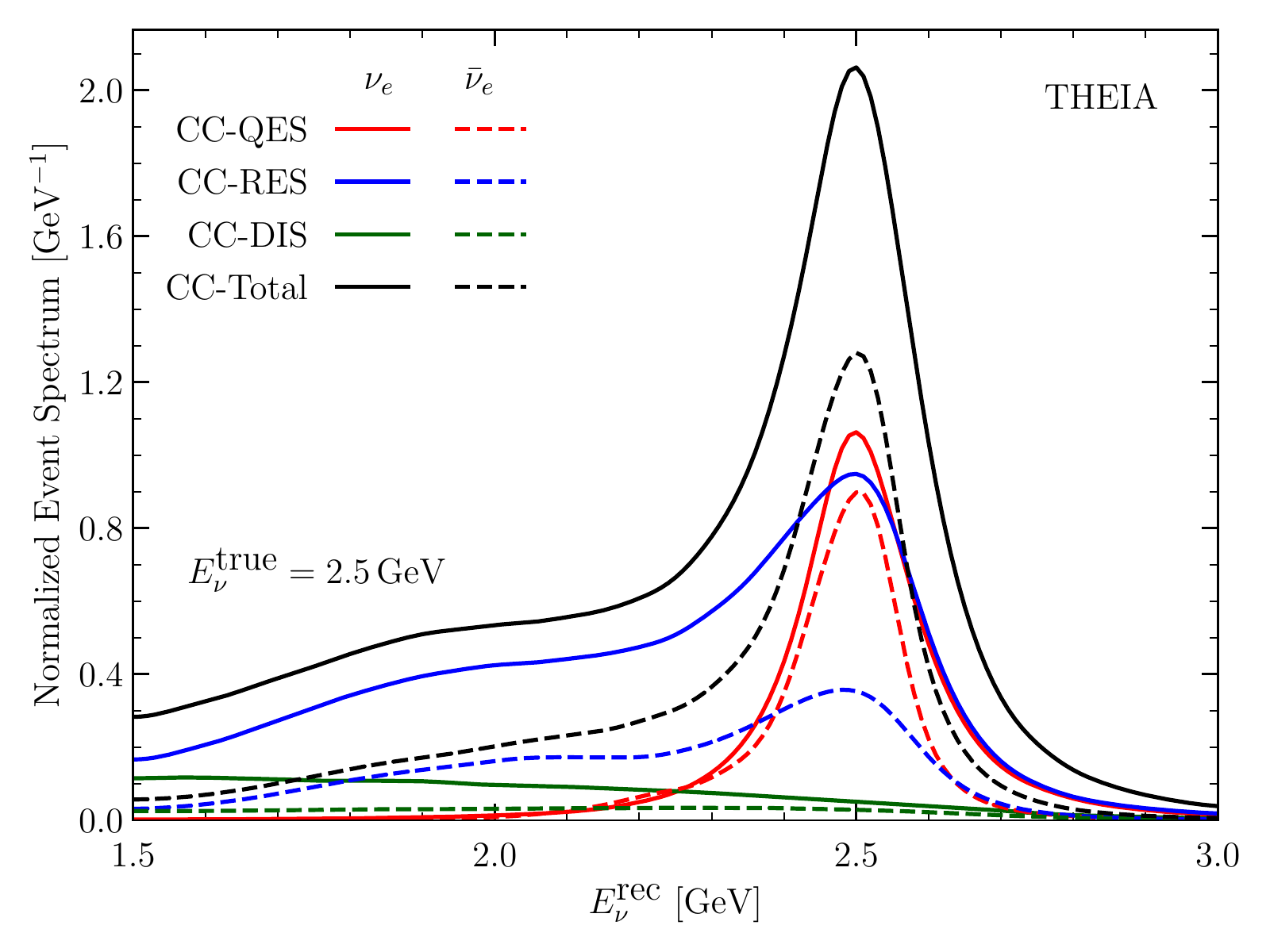}\\
\caption{
The reconstructed energies of a 2.5\,GeV $\nu_e$/$\bar \nu_e$
neutrino at the THEIA
detector for different CC scattering processes: QES (red),
RES (blue), and DIS (green). The combined total spectrum is 
plotted in black color.
For comparison, both $\nu_e$ (solid) and $\bar{\nu}_e$ (dashed)
modes are plotted. Note that the total spectrum for neutrino
mode is normalized and the others weighted by their
corresponding cross sections.}
\label{fig:DUNETHEIA}
\end{figure}

\noindent
\textbf{Signals} -- The CC-QES process has a two-body final state
with a primary lepton and a nuclei. A combination of Cherenkov
and scintillation lights can achieve outstanding
lepton identification as discussed in \gsec{sec:lowEnergyMode}.
Similar to the IBD case \geqn{eq:EIBD-angle},
the neutrino energy can be reconstructed
from the charged lepton energy $E_\ell$ (or momentum
$|\bf{p}_\ell|$) and scattering angle $\theta_\ell$
\cite{Abe:2017vif},
\begin{equation}
E^{\rm{rec}}_\nu 
  =
\frac{m^2_f-(m'_i)^2-m^2_\ell+2m'_iE_\ell}
{2(m'_i-E_\ell+|\bf{p}_\ell|\cos\theta_\ell)},
\label{eq:ccqe}
\end{equation}
where $m_f$ is the final-state nucleon mass.
On the other hand, the initial nucleon mass $m_i$ always
appears together with the binding energy $E_b$ for a nucleon
inside $^{16}O$ nuclei as $m'_i \equiv m_i-E_b$. The binding
energy $E_b = 42$\,MeV ($19$\,MeV) for the neutrino (anti-neutrino)
mode is adopted to make the reconstructed energy peak at
the true value for $E_\nu = 2.5$\,GeV as shown in \gfig{fig:DUNETHEIA}.
This energy reconstruction formula is similar to the IBD one
\geqn{eq:EIBD-angle} with the only difference that the
initial free proton (Hydrogen) mass $m_p$ is replaced by
$m'_i$ to account for the binding energy.
Experimentally, there is no efficient way to distinguish
whether it is the Oxygen or Hydrogen nuclei that is scattered.
Consequently, the neutrino energy reconstruction \geqn{eq:ccqe}
for Oxygen target is used universally since the mass fraction
of Oxygen is eight times larger than Hydrogen in the water
target. Similar to the low energy mode, the same Gaussian
smearing $7\%/\sqrt{E_{\rm vis}/ \mbox{MeV}}$
is used for simulating the detector resolution
of the deposit visible energy. In addition, we take the angular
resolutions $1.48^\circ$ for $e^\pm$ and $1.00^\circ$ for
$\mu^\pm$ from SK-IV \cite{Jiang:2019xwn}.

\gfig{fig:DUNETHEIA} shows the reconstructed spectrum for a
2.5\,GeV $\nu_e$ ($\bar \nu_e$) at the THEIA detector. For
the CC-QES events, the reconstructed neutrino energy spectrum
is the narrowest whose width at half height is roughly 80\,MeV.
The high-energy CC-RES or CC-DIS event always has at least one
$\pi^\pm / \pi^0$ in the final state which can be used to
distinguish from the CC-QES signal. That means the CC-QES
events can be separately from the CC-RES and CC-DIS counterparts.
For illustration, \gfig{fig:DUNETHEIA} also shows the
individual features of CC-RES and CC-DIS events
separately. The fact that CC-RES and CC-DIS partially overlap
with each other 
will be elaborated in later discussions.

\vspace{3mm}
\noindent
\textbf{Backgrounds} --
The LBNF beam can contribute intrinsic $\nu_e$ ($\bar \nu_e$) 
background $\nu_e\rightarrow\nu_e$ ($\bar\nu_e\rightarrow\bar\nu_e$)
which is irreducible. For the neutrino (anti-neutrino) mode,
the $\nu_e$ ($\bar \nu_e$) flux is two orders smaller than the 
dominant $\nu_\mu$ ($\bar \nu_\mu$) \cite{Abi:2021arg}.
Considering the
$\nu_\mu \rightarrow \nu_e$ ($\bar \nu_\mu \rightarrow \bar \nu_e$)
oscillation probability that is typically $\lesssim 5\%$,
the intrinsic background can be comparable or roughly one
order smaller 
\begin{widetext}
\centering
\begin{minipage}{\linewidth}
\begin{figure}[H]
\centering
\includegraphics[width=0.49\linewidth]{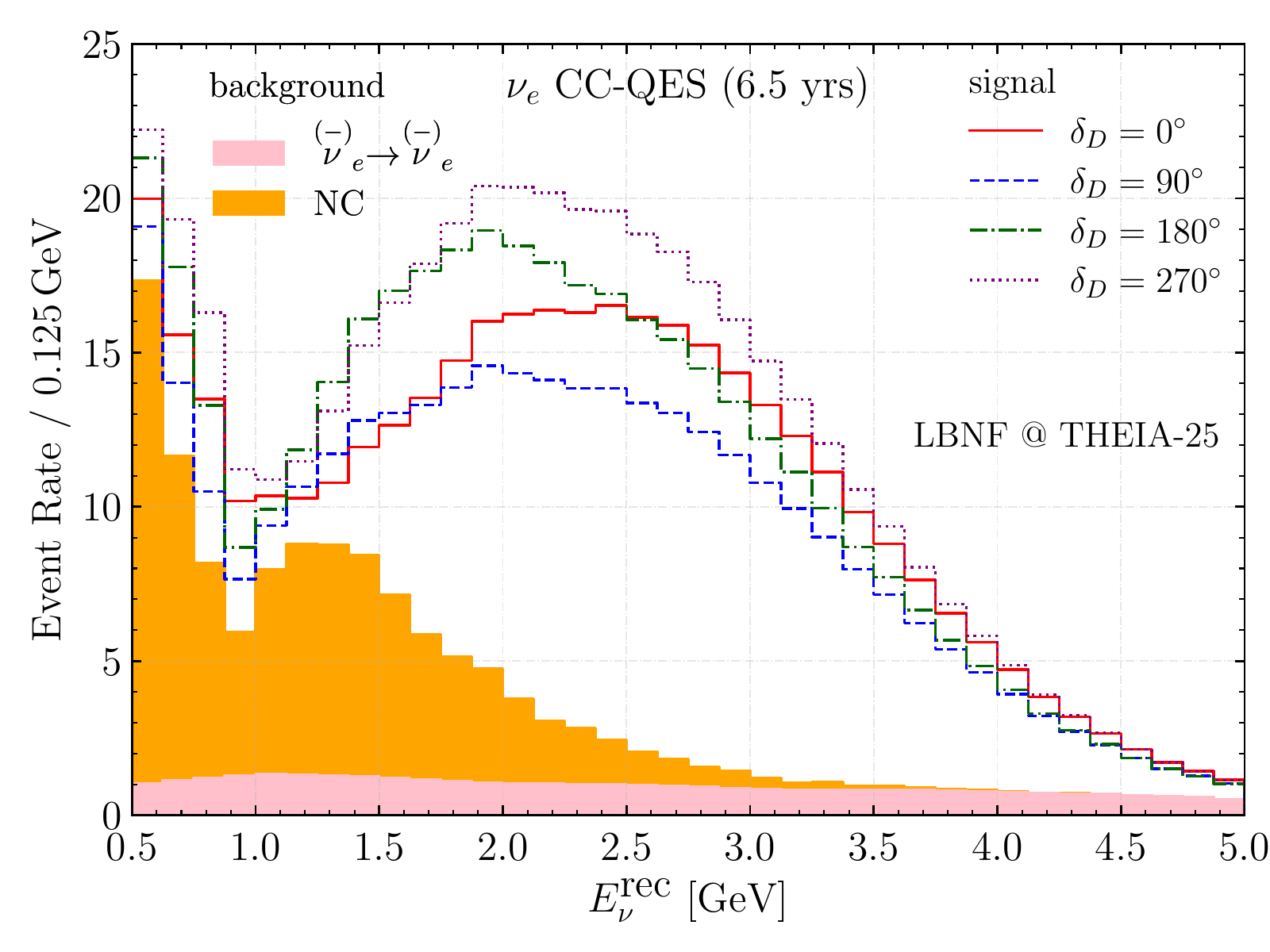}
\hfill
\includegraphics[width=0.49\linewidth]{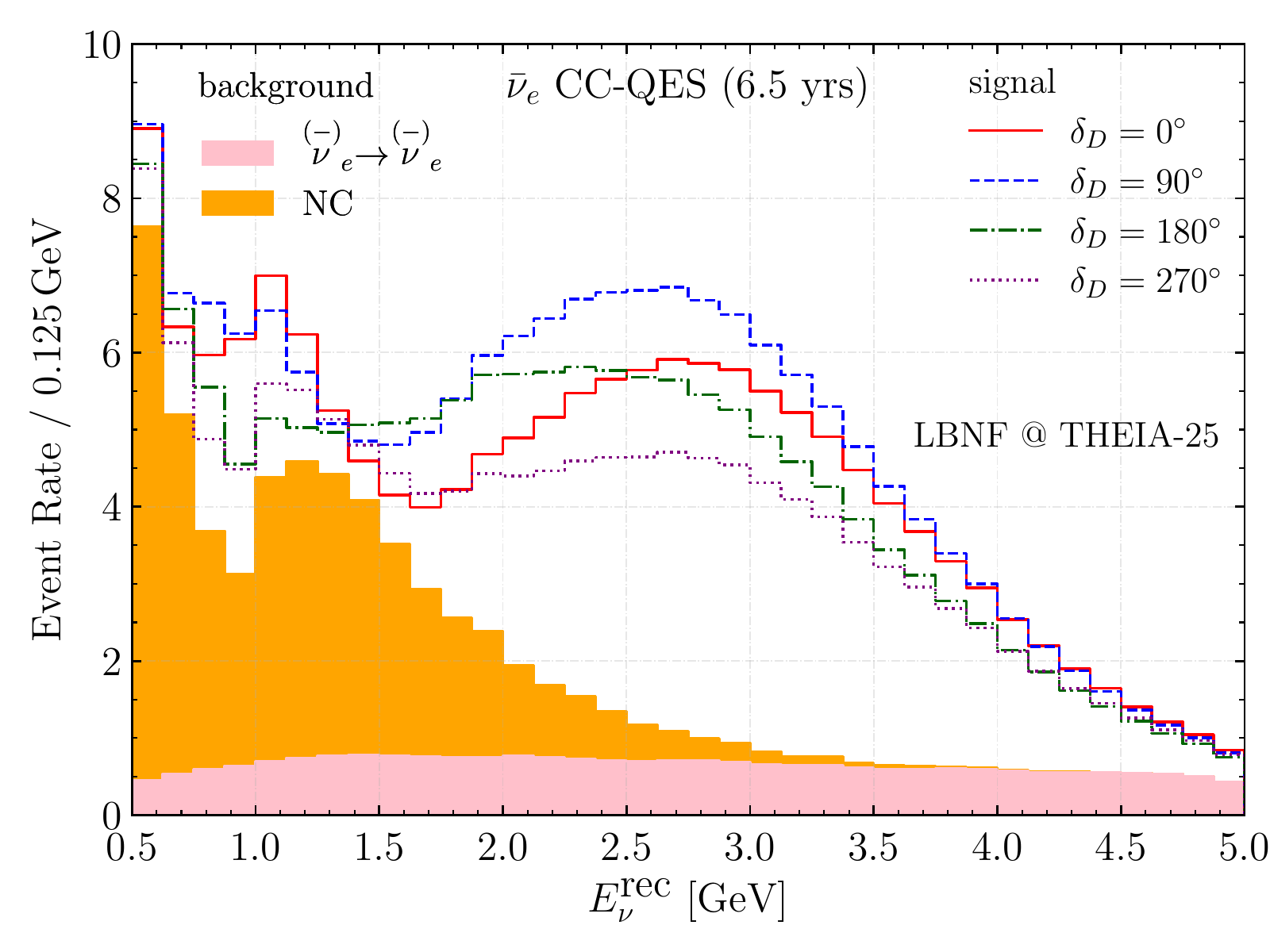}
\caption{The LBNF CC-QES signal (lines) and background (filled
regions) event spectra as functions of the reconstructed neutrino
energy $E^{\rm rec}_\nu$ at the THEIA-25 detector. Both neutrino
(left) and anti-neutrino (right) modes are shown, each with
6.5 years running. The signal curves adopts four different
values of the Dirac CP phase $\delta_D = 0^\circ$ (red),
$90^\circ$ (blue), $180^\circ$ (green), and $270^\circ$
(purple) to make comparison.
On the other hand, the background is divided into CC (pink) and neutral
current (NC, yellow) components.}
\label{fig:DUNEQES}
\end{figure}
\end{minipage}
\end{widetext}
\noindent
than the signal. In other words, the intrinsic
beam background is sizable and hence cannot be neglected.
Moreover, the beam also has intrinsic $\bar \nu_e$ ($\nu_e$)
fluxes which are smaller by another order. Although the positron
annihilation at the THEIA detector can in principle 
be used to distinguish the electric charge of the final-state
leptons, we assume the $\bar \nu_e$ ($\nu_e$) also contributes
as background to be conservative. The intrinsic beam backgrounds
are shown as the pink filled region in \gfig{fig:DUNEQES}.

For the disappearance channel of $\nu_\mu$ ($\bar \nu_\mu$),
the misidentification of muon as electron is also a potential
background. Nevertheless, the misidentification rate is 0.05\%
for a muon
misidentified as an electron and 0.02\% for the converse one 
\cite{Jiang:2019xwn}. This contribution is much smaller than
the intrinsic $\nu_e$/$\bar \nu_e$ background and
hence can be neglected for simplicity.

As for the neutral-current events, the NC-QES does not have
charged lepton in the final state which can serve as an effective veto.
However, the NC-RES and NC-DIS scatterings have both single and
multi-pion final states. The multi-pion background can be removed
by simply imposing a single-ring cut. Single pion production has
$\pi^\pm$ or $\pi^0$ in the final state.
The $\pi^\pm$ above the Cherenkov threshold can produce a $\mu$-like
ring, since both are heavy particles with similar masses. For $\pi^+$
below the Cherenkov threshold, it can experience a chain decay
$\pi^+ \rightarrow \mu^+ \rightarrow e^+$ with both $\pi^+$ and
$\mu^+$ decaying at rest. Although the final $e^+$ can produce an
$e-$like ring, its maximal energy is only $m_\mu / 2 = 53$\,MeV
and hence the reconstructed energy is far below the signal
energy window $[0.5, 5]$\,GeV. Moreover, the invisible $\pi^+$
and $\mu^+$ can actually been seen by the THEIA detector with
scintillation lights before the $e^+$ Cherenkov ring, which
provides an extra way to veto the invisible $\pi^+$ background.
Furthermore, the invisible $\pi^-$ is absorbed in the detector
and hence cannot produce an electron to fake signal. Neither
$\pi^+$ nor $\pi^-$ can become background. However, the photon
pair from $\pi^0$ decay can fake an electron. This happens if
(1) the two photons have a small opening angle
$\theta_{\gamma\gamma}\leq 17^\circ$ and
overlap with each other or (2) one photon is soft enough to be
invisible \cite{Hagiwara:2009bb}.
Roughly 3.3\% (16.2\%) of NC-RES (NC-DIS) single  
$\pi^0$ events at the peak energy $E_\nu = 2.5$\,GeV have
overlapping photons. For the soft photon in case (2),
we take a conservative $E_\gamma < 30$\,MeV Cherenkov detection
threshold \cite{Hagiwara:2009bb}. The soft photon events with
$\theta_{\gamma \gamma} > 17^\circ$ contribute 7.6\% (6.8\%)
of the NC-RES (NC-DIS) scatterings. In total, 10.9\% (23.0\%)
of the NC-RES (NC-DIS) events have a $\pi^0$ to fake the electron.
The $\pi^0$ direction and energy are then used as
the electron information to reconstruct the neutrino
energy via \geqn{eq:ccqe} for small opening angle case.
Since the soft photon direction is difficult to reconstruct,
the direction of the hard photon is used instead.
\\

\noindent
\textbf{Event Selection} --
As elaborated above, the CC-QES category has only one primary
lepton in the final state. Being different from the IBD signal
of the low-energy mode, there is no neutral for double coincidence
to significantly suppress the background.
The selection criteria for the CC-QES signal events 
are summarized below,
\begin{itemize}
\item[(1)] Only one primary $e-$like Cherenkov ring
    with $|{\bf p}_e|> 100$\,MeV. For the disappearance channel
    the primary muon is selected with $\mu$-like Cherenkov
    ring with $|{\bf p}_\mu| > 200$\,MeV.
    
\item[(2)] Reconstructed neutrino energy inside the range [0.25, 5]\,GeV.

\item[(3)] No extra Cherenkov light from mesons.
\end{itemize}

The event spectra of the CC-QES category after selection
and the corresponding backgrounds are shown as functions
of the reconstructed neutrino energy $E^{\rm rec}_\nu$
in \gfig{fig:DUNEQES}. For both neutrino and anti-neutrino
modes, the signal lines clearly show the CP dependence. Taking
$\delta_D = 90^\circ$ for illustration, the corresponding
blue line is at the bottom for the neutrino mode while
it is at the top for the anti-neutrino one. The opposite
feature happens for the other maximal CP phase
$\delta_D = 270^\circ$. This is a
reflection of the fact that the CP-violating term $\sin \delta_D$
in \geqn{eq:Pnumu_nue_approx} differs by a sign.
The NC background is contributed by
all flavors and the CC one by the intrinsic $\nu_e / \bar \nu_e$
beam background in addition to the misidentified
$\nu_\mu / \bar \nu_\mu$ comes from the disappearance channels.
While the signal and CC backgrounds can extend to 5\,GeV,
the NC background mainly contributes below 3.5\,GeV since
typically more particles are produced to split the energy.
The two NC background peaks here are produced due to the
soft photon contribution (lower energy peak) and the small
opening angle contribution (higher energy peak), respectively.
For CC-QES, the CC background spectrum is quite flat and
the NC backgrounds dominate below 2.5\,GeV. Around the peak
energy, $E^{\rm rec}_\nu = 2.5$\,GeV, the signals dominate
to provide a good CP sensitivity. Between neutrino and
anti-neutrino modes, the typical event rate differs by
a factor around 2.5.

\subsubsection{The CC-RES Category}

\noindent
\textbf{Signals} --
The CC-RES process for the appearance signal $\nu_\mu \rightarrow \nu_e$
($\bar \nu_\mu \rightarrow \bar \nu_e$) has 1 electron
(positron) plus 1 pion ($\pi^\pm$ or $\pi^0$).
Therefore, the CC-RES event can be distinguished
from the CC-QES one by ring counting if the final-state
pion is a charged one $\pi^\pm$. For $\pi^0$, the major
contribution to its decay final state is two resolved photons.
Nevertheless, both CC-RES and CC-DIS can
produce pion in the final state to have some
overlap with each other. But it is still possible
to partially distinguish CC-RES CC-DIS events by the number of pions.
Our simulation shows that for the neutrino (anti-neutrino)
mode at the peak energy of 2.5\,GeV, only 15\% (11\%)
of the CC-RES events have multiple pions, while CC-DIS
reaches 64\% (48\%). Hence the CC single-pion event can
categorized as CC-RES while the multi-pion one as CC-DIS.

The energy reconstruction formula \geqn{eq:ccqe} for CC-QES
can no longer apply due to different particles in the final
state. Instead of having nucleons in the initial and final
states, CC-RES has heavy $\Delta$ baryons as intermediate
resonance. Most of the pions are produced from $\Delta$ decays
$\Delta^{++}\rightarrow p+\pi^+$,  $\Delta^+\rightarrow p+\pi^0$,
and $\Delta^-\rightarrow n+\pi^-$ \cite{SajjadAthar:2020nvy}.
Usually the proton is not energetic enough to produce a
Cherenkov ring and hence cannot be uniquely identified but
only leaves some scintillation lights. Observationally,
the $\nu_e/\bar \nu_e$ CC-RES has a primary lepton and
a pion ($\pi^0$ or $\pi^\pm$). A reasonable neutrino energy
reconstruction for CC-RES is \cite{Abe:2017vif},
\begin{equation}
  E^{\rm rec}_\nu 
=
  \frac {m^2_\Delta - m^2_p - m^2_\ell + 2 m_p E_\ell}
        {2 (m_p-E_\ell+|\bf{p_\ell}|\cos\theta_\ell)}.
\label{eq:ccres}
\end{equation}
Comparing with \geqn{eq:ccqe}, the final-state nucleon mass
is replaced by the $\Delta$ baryon mass, $m_\Delta = 1.232$\,GeV.
As demonstrated with blue lines in \gfig{fig:DUNETHEIA}, the
CC-RES energy reconstruction formula
\geqn{eq:ccres} gives a correct peak position. However, for the
high-energy scattering process, not only $\Delta$ is produced
but also other resonant particles like $N(1440)$ which can
also produce pion particles. So \geqn{eq:ccres} cannot describe
all the resonant processes exactly and gives a wider distribution
than CC-QES.

\vspace{3mm}
\noindent
\textbf{Backgrounds} -- 
The beam background for the CC-RES category has three major
contributions. First, the beam electron-flavor neutrinos
(anti-neutrinos) of the disappearance channel
$\nu_e \rightarrow \nu_e$ ($\bar \nu_e \rightarrow \bar \nu_e$)
via either CC-RES or CC-DIS scatterings contributes as
irreducible background with the same $1e1\pi$ final state.
Being depicted as the
pink regions in \gfig{fig:DUNERES}, this type of background
for neutrino (left) and anti-neutrino (right) modes
contribute roughly 13.9\% and 14.9\% of the total
detected events for $\delta_D = -90^\circ$.

The second beam background comes from the muon neutrinos $\nu_\mu$
interacting via CC-RES or CC-DIS to produce $\mu^-$ and $\pi^0$
in the final state. The $\pi^0$ can fake electron if one of
the decay photons is soft ($E_\gamma < 30$\,MeV) or the two
photons are almost collinear ($\theta_{\gamma\gamma} < 17^\circ$).
For the peak energy $E_\nu = 2.5$\,GeV, only 7.6\% (13\%)
of such $\pi^0$ can fake an electron. In addition, the muon
needs to be misidentified as a charged pion. With both
misidentification, the beam $\nu_\mu$ can fake the CC-RES signal.
Although the new reconstruction algorithm of fitQun \cite{Jiang:2019xwn}
shows some capability of separating pion from muon at T2K
using the hadronic kinks in the pion propagation
\cite{Tobayama:2016dsi}, we assume $\mu^-$ and $\pi^+$ can
not be separated to be conservative. Not to say,
$\mu^-$ has an electron but $\pi^+$ has a positron in
their decay products, respectively, and the $e^+ e^-$ annihilation
photons from $\pi^+$ can also provide a distinguishable feature.
Note that there is no beam $\bar \nu_\mu$ CC background for the
anti-neutrino mode. This is because the $\pi^-$ in the $\bar \nu_e$
CC-RES signal is absorbed in the detector and does not produce
a delayed Michel electron which is different from the beam
$\bar \nu_\mu$ that can decay to $\mu^+$ and finally a Michel $e^+$.
For both neutrino and anti-neutrino modes,
muon misidentified as electron/positron might
happen together with the same $\pi^\pm$ to fake the CC-RES category.
But this background can be neglected since the misidentification
rate is negligibly small ($< 0.05\%$) as mentioned in
\gsec{sec:ccqe}. The beam $\nu_\mu$ CC background
is shown as green region in the left panel of \gfig{fig:DUNERES}
only for the neutrino mode. 
\begin{widetext}
\centering
\begin{minipage}{\linewidth}
\begin{figure}[H]
\centering
\includegraphics[width=0.49\linewidth]{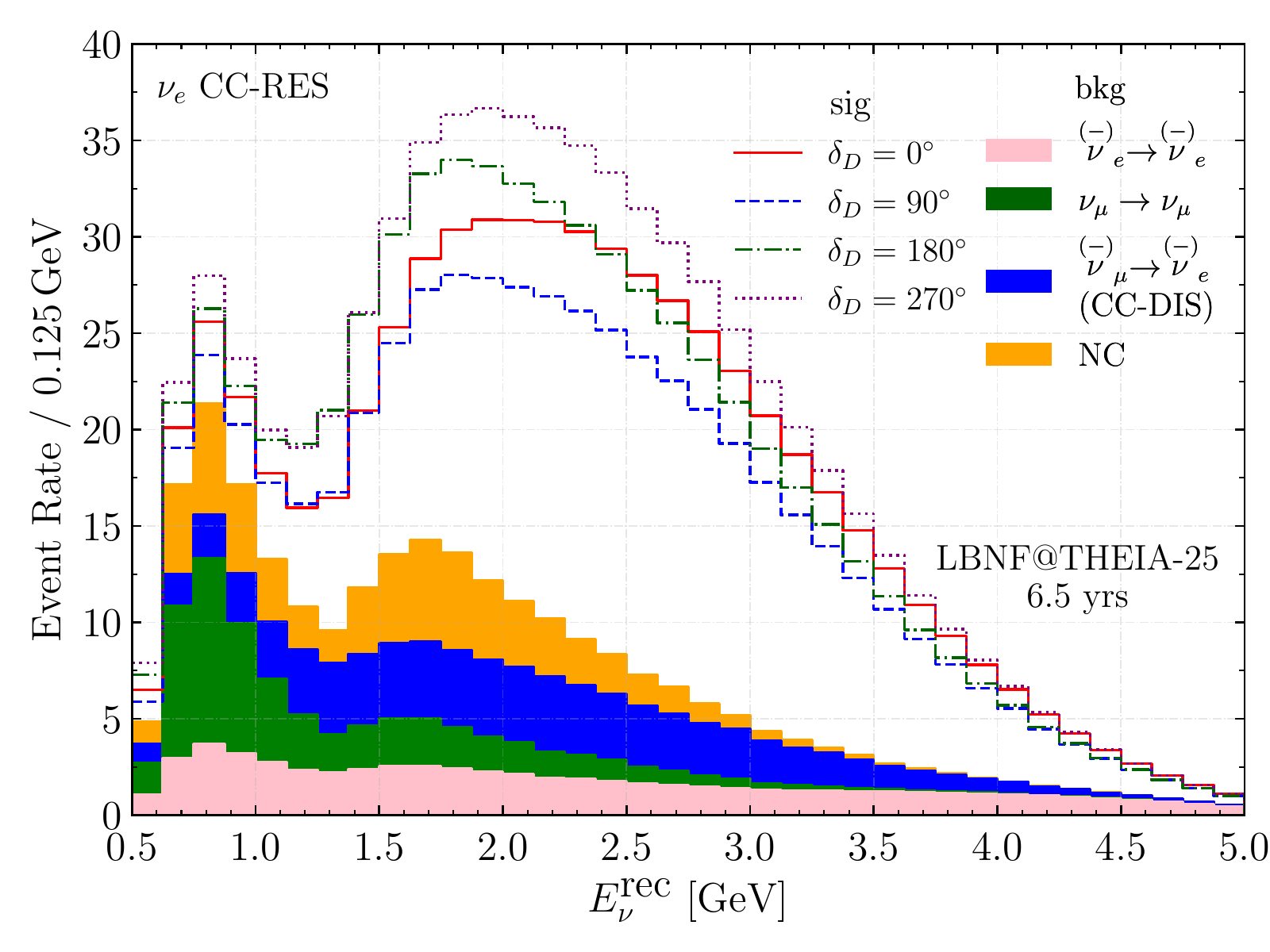}
\hfill
\includegraphics[width=0.49\linewidth]{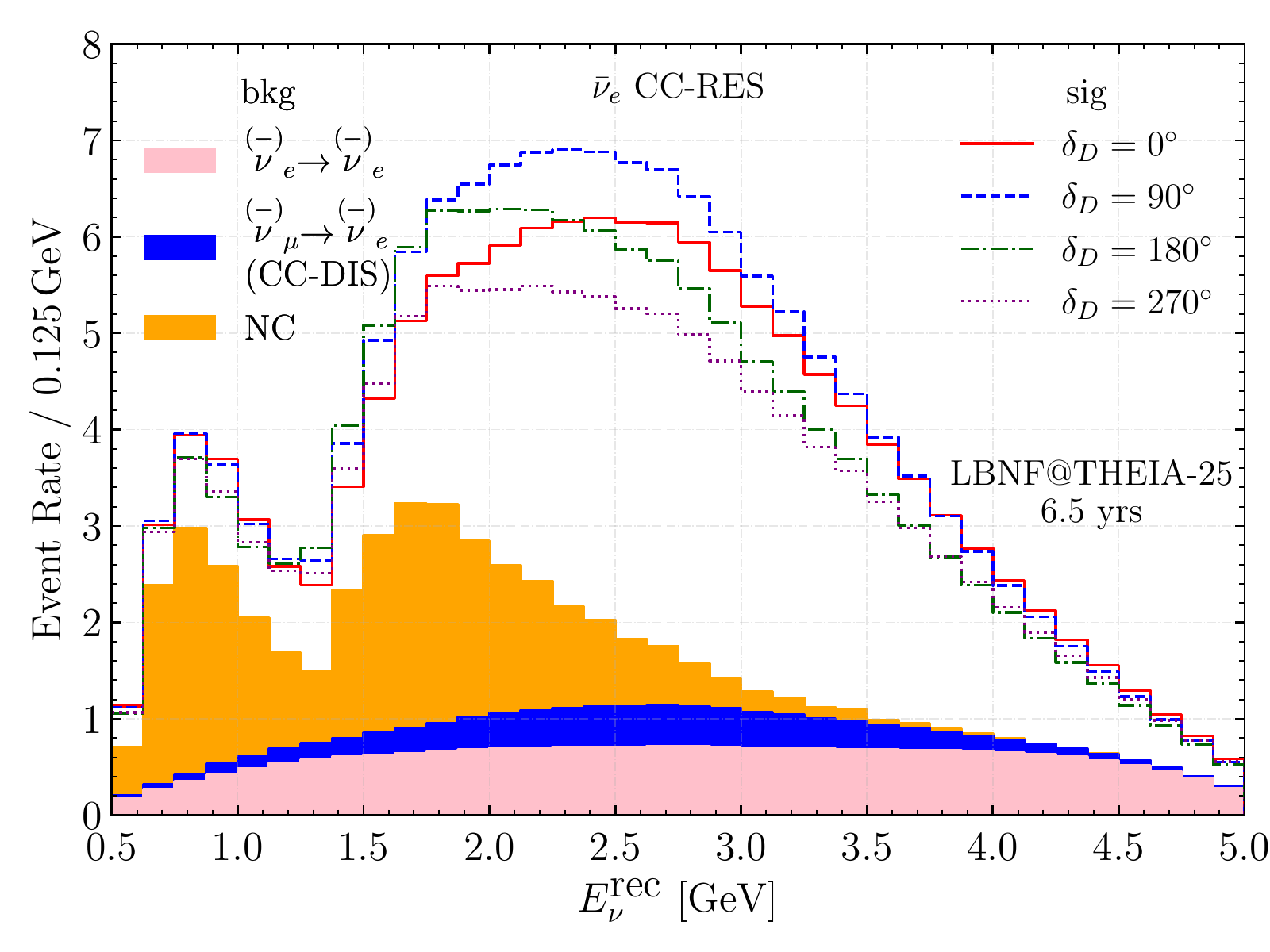}
\caption{The event spectra of CC-RES signal and its background
at the THEIA-25 detector with LBNF neutrino (left) and anti-neutrino
(right) beams. Four lines are used for
$\delta_D = 0^\circ$ (solid 
red line), $\delta_D = 90^\circ$ (dashed blue line), 
$\delta_D = 180^\circ$ (dot-dashed green line) and 
$\delta_D = 270^\circ$ (dotted purple line). For
comparison, the filled regions stand for the backgrounds of the intrinsic 
$\nu_e / \bar \nu_e$ beam CC (pink), the $\nu_\mu$ CC (green),  the $\nu_e / \bar \nu_e$ CC-DIS (blue), and NC (yellow). Both neutrino and anti-neutrino
modes have 6.5 years running.
}
\label{fig:DUNERES}
\end{figure}
\end{minipage}
\end{widetext}
The third part of the background comes from NC-RES and NC-DIS with
two pions ($2 \pi^0$ or $\pi^0 + \pi^\pm$)
in the final state. In such case, one $\pi^0$ decays into 
unresolved photons to fake the primary electron signal
while the other pion coincides with the one in the signal. 
The estimated NC background is shown as the
orange region of \gfig{fig:DUNERES}. We can see that
the reconstructed NC background mostly occur at
$E_{\nu}^{\rm rec} < 3$ GeV.

\vspace{3mm}
\textbf{Event Selection} --
In summary, the selection criteria for CC-RES signal is,
\begin{itemize}
\item[(1)] Only one primary $e$-like Cherenkov ring with $|p_{e}|> 100$ MeV.

\item[(2)] Reconstructed neutrino energy inside the range [0.25, 5]\,GeV.

\item[(3)] Single pion particle in the final state.
To be conservative, the events with a charged $\pi^\pm$
or a neutral $\pi^0$ are not mixed into a single CC-RES
category. More detailed study is necessary for maximizing
the CP sensitivity.
\end{itemize}

\gfig{fig:DUNERES} depicts the $\nu_e$ (left) and $\bar \nu_e$
(right) CC-RES signal and the corresponding backgrounds.
Comparing the two
panels, we can see that the neutrino event rates are
typically $4 \sim 5$ times larger than its anti-neutrino
counterparts. This is a direct consequence of the relatively
larger cross section for neutrino than anti-neutrino.
For backgrounds, the neutrino mode has
four components while the anti-neutrino mode has
only three. This is because the disappearance channel
$\overline \nu_\mu \rightarrow \bar\nu_\mu$ has $\mu^+$ and
Michel $e^+$ to be distinguished from $\pi^-$ of the
$\bar \nu_e$ CC-RES signal. For neutrino mode, all 
backgrounds contribute roughly the same size while for
anti-neutrino the NC and intrinsic $\nu_e$ background
dominate. Around the major peak, $E^{\rm rec}_\nu \approx 2$\,GeV,
the signal is slightly larger than the background for
neutrino and anti-neutrino modes.

\subsubsection{The CC-DIS Category}
\noindent
\textbf{Signals} --
In order to enhance the statistics, we also
include CC-DIS events as a separate category. As mentioned in
the previous CC-RES section, the events identified as CC-DIS
are those with an $e$-like ring accompanied with more than one
pion in the final state. The majority of events contains two pions.
The 2-pion  events include any of the following combination
$(\pi^\pm, \pi^\pm),~(\pi^\pm, \pi^0)$ and $(\pi^0, \pi^0)$.
Similar to the CC-RES category, only a pair of resolved
photons are identified as $\pi^0$. For events with one
or more $\pi^0$ and consequently 3 or more  $e$-like rings
in the final state, we take the most energetic $e$-like ring
as the primary electron/positron. Then the neutrino energy can
be reconstructed using the same \geqn{eq:ccres}. Nevertheless,
the scattering process of CC-DIS differs a lot from CC-RES
and the reconstructed neutrino energy is almost flat as
shown with green curves in \gfig{fig:DUNETHEIA}.
\\

\noindent
\textbf{Backgrounds} --
The background of the CC-DIS process also 
has three major contributions. The first is the irreducible 
one from the electron-flavor neutrinos via the CC-RES or
CC-DIS with the $1eN\pi$ final state with $N > 1$.
Being depicted as the pink regions in \gfig{fig:DUNEDIS},
this background is the largest contribution for
$E^{\rm rec}_\nu > 1.5$\,GeV and can be as large as 45\%
of the total events at 2.5\,GeV for $\delta_{\rm D} = - 90^\circ$.

\begin{widetext}
\centering
\begin{minipage}{\linewidth}
\begin{figure}[H]
\centering
\includegraphics[width=0.49\linewidth]{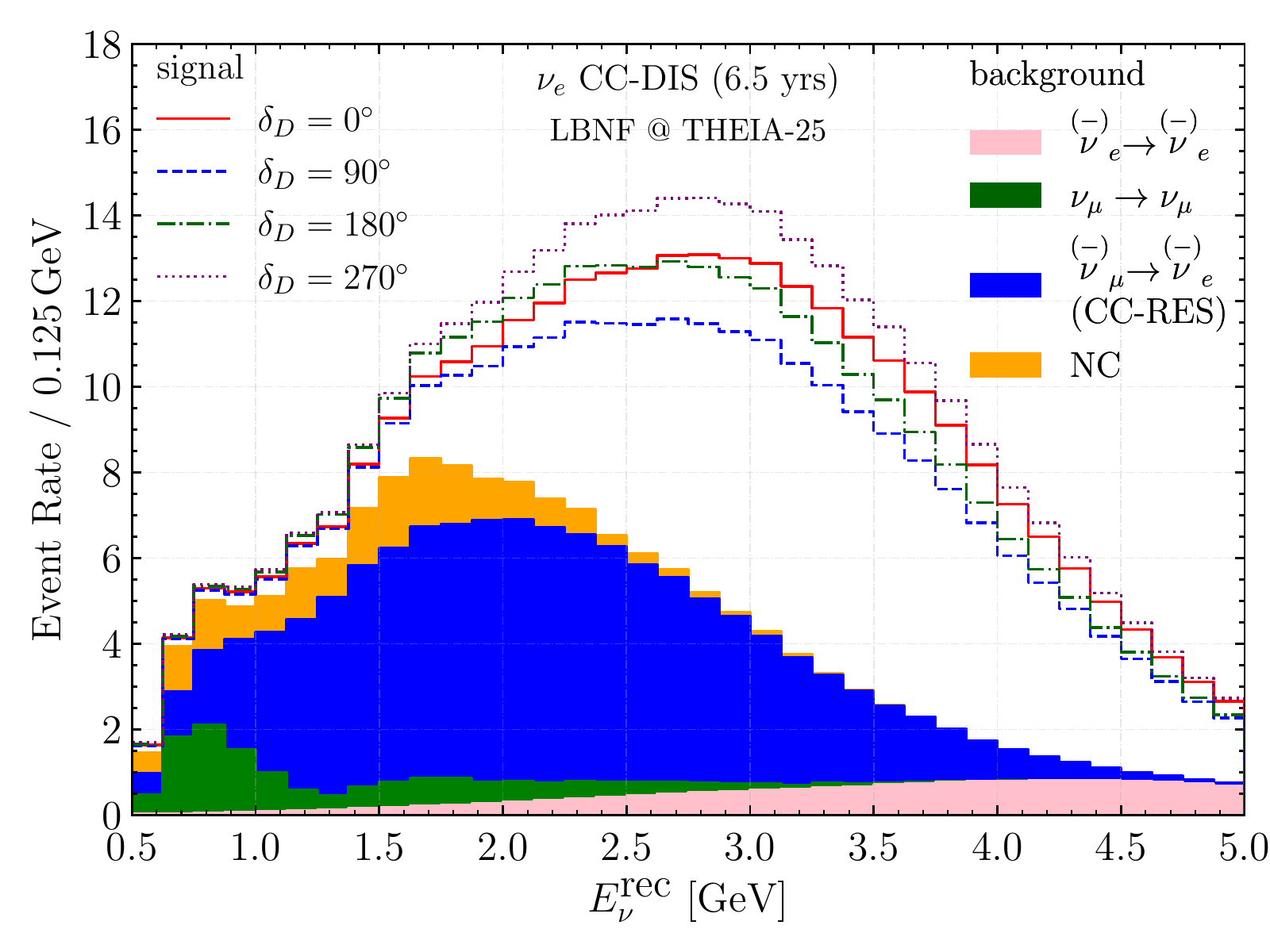}
\hfill
\includegraphics[width=0.49\linewidth]{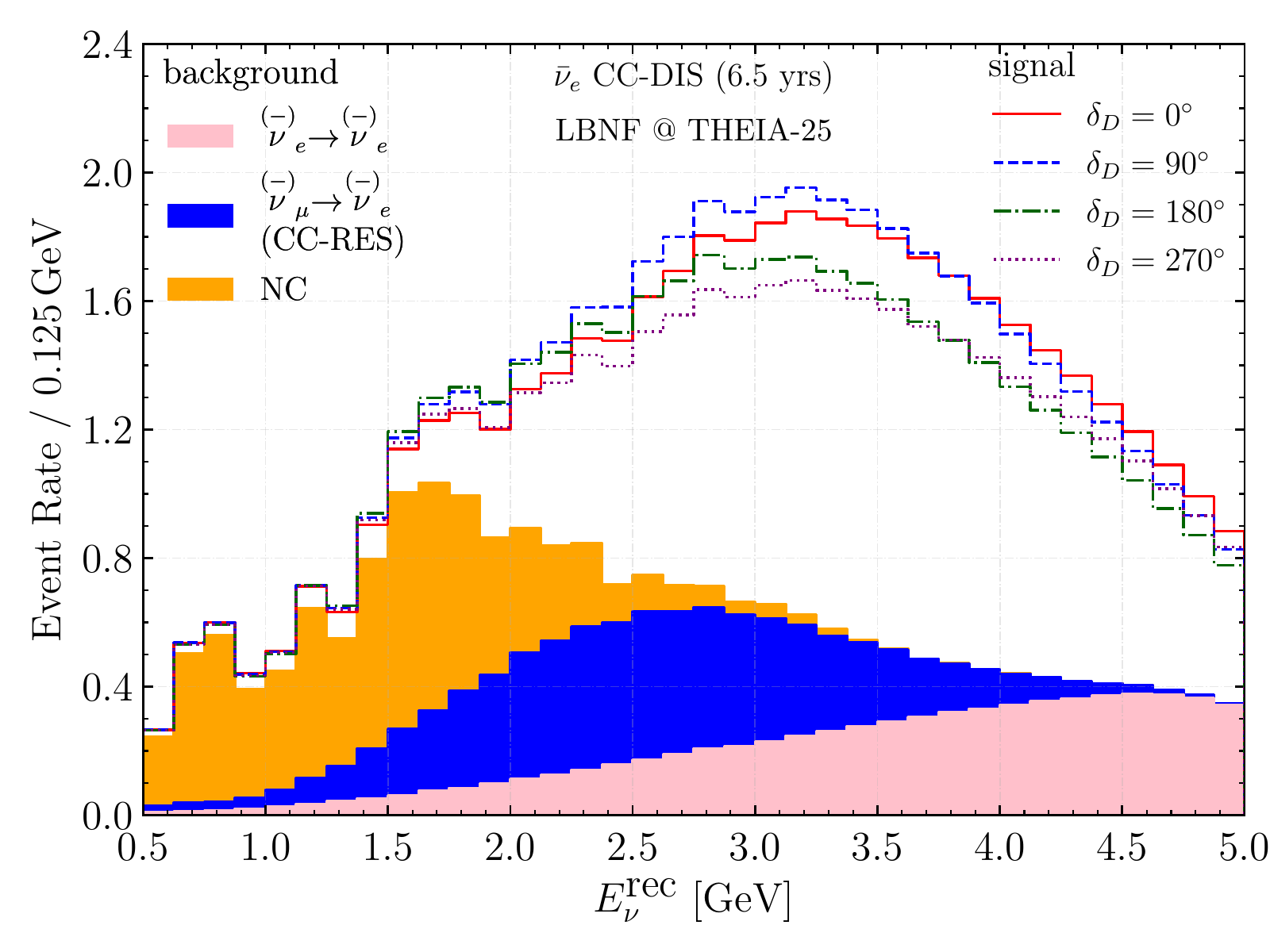}
\caption{The event spectra of CC-DIS signal and its background
at the THEIA-25 detector with LBNF neutrino (left) and anti-neutrino
(right) beams. Four solid lines are used for
$\delta_D = 0^\circ$ (red), $90^\circ$
(blue), $180^\circ$ (green), and $270^\circ$ (purple). For
comparison, the filled regions stand for the  backgrounds of the intrinsic $\nu_e / \bar \nu_e$ beam CC (pink),
the $\nu_\mu$ CC (green), the $\nu_e / \bar \nu_e$ CC-RES (blue), and the
NC multi-pion events (yellow). Both neutrino and anti-neutrino
modes have 6.5 years running.}
\label{fig:DUNEDIS}
\end{figure}  
    \end{minipage}
\end{widetext}
\noindent

The second component comes from muon neutrinos $\nu_\mu$
interacting via CC-RES or CC-DIS to produce a $\mu^-$, 
a $\pi^0$ and one or two other pions of any type.
While $\mu^-$ is misidentified as a pion,
the $\pi^0$ can be misidentified as an electron if it
has a soft photon $E_\gamma < 30\,$MeV
or the two photons are almost collinear,
$\theta_{\gamma\gamma} < 17^\circ$.
The multi-pion production of two
or three pions occurs for 4\% of the total $\nu_\mu$ CC events.
In addition, the misidentification 
of the $\pi^0$ as an electron occurs in 17\% (9\%) of the events 
for 2-pion (3-pion) final state. This $\nu_\mu$ background is
shown as green curve in the left panel of \gfig{fig:DUNEDIS}
only for the neutrino mode. Again, there is no
$\overline \nu_\mu$ CC background due to the delayed
Michel electron veto. 

The third component is the NC multi-pion background of
1$\pi^0 1\pi^x 1\pi^y$ with $x,y = 0,\pm$. The triple-pion
final state is reconstructed as background if one $\pi^0$
is misidentified as electron. This occurs for only 0.1\% of
the total NC events at the 2.5\,GeV peak energy, which
is represented as the orange region in \gfig{fig:DUNEDIS}.
It is most relevant at low energies, $E_{\nu}^{\rm rec} < 1.5$\,GeV where the CP value has small impact.

\vspace{3mm}
\textbf{Event Selection} --
We can see that the CC-DIS category has much similarity as the
CC-RES one with single-pion final state replaced by the multiple
one. Although the CC-DIS signal would not contribute much as
shown in \gfig{fig:DUNETHEIA}, this separation reveals the
feature of various channels. Below we summarize the selection
criteria for the CC-DIS signal,
\begin{itemize}
\item[(1)] Only one primary $e-$like Cherenkov ring with
$|p_{e}|> 100$ MeV. In the presence of multiple $e$-like rings,
the most energetic one is identified as the primary lepton.

\item[(2)] Reconstructed neutrino energy inside the range [0.25, 5]\,GeV.

\item[(3)] At least two pions in the final state with
    energy $\pi(|{\bf p}_{\pi}|)>200$\,MeV. Similar to
    the CC-RES category, all CC-DIS events are put into
    a single category without division according to the
    final-state pions. This conservative treatment can
    be further improved with more careful studies.
\end{itemize}

The CC-DIS signals and their backgrounds are shown in
\gfig{fig:DUNEDIS}. Different from the CC-QES in
\gfig{fig:DUNEQES} and CC-RES in \gfig{fig:DUNERES},
the CC-DIS signal can be smaller than the background,
especially for the neutrino mode. Around the peak,
$E^{\rm res}_\nu \approx 2.5$\,GeV, the dominant
background comes from the CC-RES multi pion background
which can reach more than 50\% of 
the total events at $E_\nu^{\rm rec}\approx 2$\,GeV.
The anti-neutrino case is slightly better with signal
still dominating around the peak energy. In addition,
the three background components for the $\bar \nu_e$
mode all have sizable contributions.
The NC background background dominates for 
$E_\nu^{\rm rec} < 2$\,GeV, while the CC-RES 
dominates at intermediate energies $2\,{\rm GeV} < 
E_\nu^{\rm rec} < 3.5$\,GeV and the intrinsic $\nu_e$
dominates for $E_\nu^{\rm rec} > 3.5$\,GeV. Nevertheless,
$\nu_e$ events are 7 times larger than $\bar \nu_e$.

\section{CP Sensitivity with $\mu$THEIA and DUNE}
\label{sec:CP}

As discussed in \gsec{sec:muTHEIA}, the combination of
$\mu$THEIA and DUNE is expected to improve the CP measurement.
Below we give a quantitative estimation. We first summarize
the experimental setups in \gsec{sec:exp} for completeness
and then establish
the $\chi^2$ formalism in \gsec{sec:chi2}. In order to maximize
the CP sensitivity, \gsec{sec:baseline} explores the optimal
baseline between the $\mu$DAR source and the THEIA detector
for $\mu$THEIA. Based on these, the CP sensitivity and the
influence of matter effect are studied in \gsec{sec:results}.

\subsection{Experimental Setups}
\label{sec:exp}

The $\mu$DAR neutrinos are typically produced by cyclotrons
\cite{Adelmann:2013isa}.
We adopt the configuration that the $\mu$DAR flux is generated
by a 9\,mA
proton beam with 800\,MeV protons hitting the high-$Z$ target
to deliver $1.83 \times 10^{25}$ POT in 10 years
\cite{Evslin:2015pya}. Since the $\mu$DAR
cyclotron is not a pulsed beam, it is possible to achieve almost
full duty factor. Of the charged pions produced by proton
hitting the target, $\pi^-$ is mainly absorbed by the
positively charged nuclei while $\pi^+$ first loses energy
in the thick material and then decays at rest to produce $\mu^+$.
Similar process of decay at rest also happens for $\mu^+$.
Consequently, the $\mu$DAR neutrino spectra are well defined
and predicted by the SM interactions.

For DUNE, the LBNF neutrinos are produced by 120\,GeV protons
with 1.2\,MW beam power and the flux can reach $1.1\times
10^{21}$\,POT/year.
We take 6.5 years running for each neutrino and anti-neutrino
mode \cite{DUNE:2015lol}.
With completely different energy windows [30, 55]\,MeV
for $\mu$DAR and  [0.5, 5]\,GeV for LBNF, there is no energy
overlap between the low-energy $\mu$DAR flux and the
high-energy LBNF one. So the two fluxes can run simultaneously.

For the THEIA detector we consider two possibilities, THEIA-25
with a fiducial mass of 17\,kt and THEIA-100 with a fiducial 
mass of 70\,kt \cite{Theia:2019non}. The DUNE detector has
four modules, each with 10\,kt fiducial mass \cite{Abi:2021arg}.
Both THEIA and DUNE detectors are at the same SURF site and
1289\,km away from the LBNF source.
To install the THEIA detector, one of the DUNE modules needs
to be replaced. As mentioned in \gsec{sec:muTHEIA}, the average 
matter density 2.85\,g/cm$^3$ is used for both DUNE and
$\mu$THEIA.

In the following, we consider three combinations: (1) the
full DUNE configuration with 4 far detector modules,
(2) the reduced DUNE with 3 far detector modules and $\mu$THEIA-25,
and (3) the reduced DUNE and $\mu$THEIA-100. 
The high-energy mode (HEM) can be detected by both the DUNE
and THEIA detectors while the low-energy mode (LEM) only
applies for the THEIA detector as discussed in \gsec{sec:modes}.
The energy window, fiducial
mass, baseline, and running time of each
experimental configurations are summarized in \gtab{tab:setup}.

\begin{table}[h]
\centering
\begin{tabular}{c|ccc}
  & DUNE & \multicolumn{2}{c}{DUNE + $\mu$THEIA-25/100} \\
  & (HEM) & (HEM) & (LEM) \\
\hline
\hline
\makecell[c]{Energy \\ Window\\ (GeV)} & [0.5, 5] & [0.5, 5] & [0.03, 0.055] \\
\hline
\makecell[c]{Fiducial \\ Mass \\ (kt)} & 40 & 30+17/70 & 17/70\\
\hline
\makecell[c]{Running \\ Time \\ ($\nu$ y, $\overline \nu$ y)} & (6.5, 6.5) & (6.5, 6.5) & (0, 10) \\
\hline
\makecell[c]{Baseline \\ (km)} & 1289 & 1289 & ??
\\
\hline
\makecell[c]{Density \\ (g/cm$^3$)} & 2.85 & 2.85 & 2.85
\end{tabular}
\caption{The energy window, fiducial mass, running time, baseline,
and matter density for the three experimental setups considered
in this paper.}
\label{tab:setup}
\end{table}

\subsection{Simulation and $\chi^2$ Analysis} 
\label{sec:chi2}

We then use GLoBES \cite{Huber:2004ka,Huber:2007ji} to simulate
the event rates and evaluate the CP sensitivities. A quantitative
evaluation is realized by minimizing the $\chi^2$ function that
contains three contributions,
\begin{eqnarray}
  \chi^2
\equiv
  \chi^2_{\rm stat}
+ \chi^2_{\rm sys}
+ \chi^2_{\rm para},
\end{eqnarray}
for statistical ($\chi^2_{\rm stat}$) and systematical
($\chi^2_{\rm sys}$) uncertainties in addition to the prior
constraint on the oscillation parameters ($\chi^2_{\rm para}$).

For event rate $> 10$ in a single bin, a Gaussian $\chi^2$
is much more convenient with analytical fit 
\cite{Ge:2012wj,Ge:2016zro,Ge:2022ius}. However, the event
rates considered in this paper are not large enough.
To make the sensitivity evaluation exact, the statistical
part in the first term takes the Poisson form,
\begin{eqnarray} \nonumber 
  \chi^2_{\rm stat}
& \equiv &
  2 \sum_{\rm bins} 
\Biggl[
  (1 + a_{\rm sig}) N^{\rm sig}_i 
+ (1 + a_{\rm bkg}) N^{\rm bkg}_i 
- N^{\rm data}_i 
\\ & - & 
  N^{\rm data}_i 
  \ln \left(
      \frac{
        (1 + a_{\rm sig}) N^{\rm sig}_i
     +
        (1 + a_{\rm bkg}) N^{\rm bkg}_i 
     }{N^{\rm data}_i} 
    \right) 
  \Biggr],
\qquad
\end{eqnarray}
where $N^{\rm data}_i$, $N_i^{\rm sig}$ and $N_i^{\rm bkg}$ 
are the pseudo data, signal and background event numbers
in the $i$-th bin, respectively. The coefficients $a_{\rm sig}$
and $a_{\rm bkg}$ are nuisance parameters for the 
signal and background normalizations, respectively.

The $ \chi^2_{\rm sys}$ term contains the uncorrelated
Gaussian priors of signal and background normalizations.
We take $\sigma_{\rm sig} = 5\%$ and $\sigma_{\rm bkg} 
= 10\%$ for each channel of both the low- and high-energy
modes that are observed at the THEIA detector. For the
low-energy mode, the normalization uncertainties are the
same as the configuration given in \cite{Evslin:2015pya}
while the uncertainties
are more conservative than the values (2\% for $\nu_e$ and 5\%
for $\bar \nu_e$) used in \cite{Theia:2019non}.
The systematics of the LBNF beam detection by the DUNE
detector are described by the official configuration files
\cite{Abi:2021arg}. 

Finally, $\chi^2_{\rm para}$ contains the prior information
on the oscillation parameters. Their
best fit values are obtained 
\begin{widetext}
\centering
\begin{minipage}{\linewidth}
\begin{figure}[H]
\centering
\includegraphics[width=0.49\linewidth]{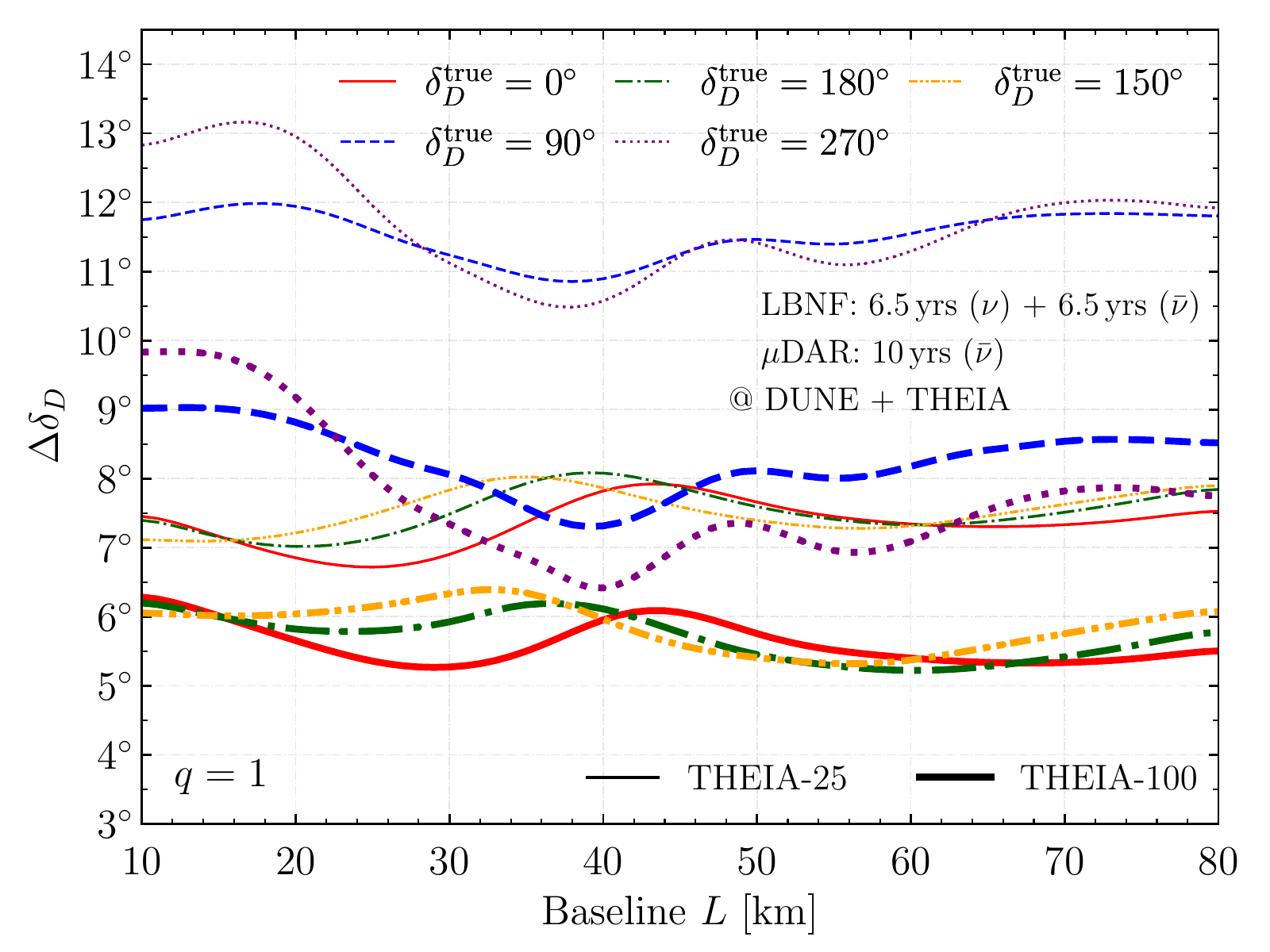}
\hfill
\includegraphics[width=0.49\linewidth]{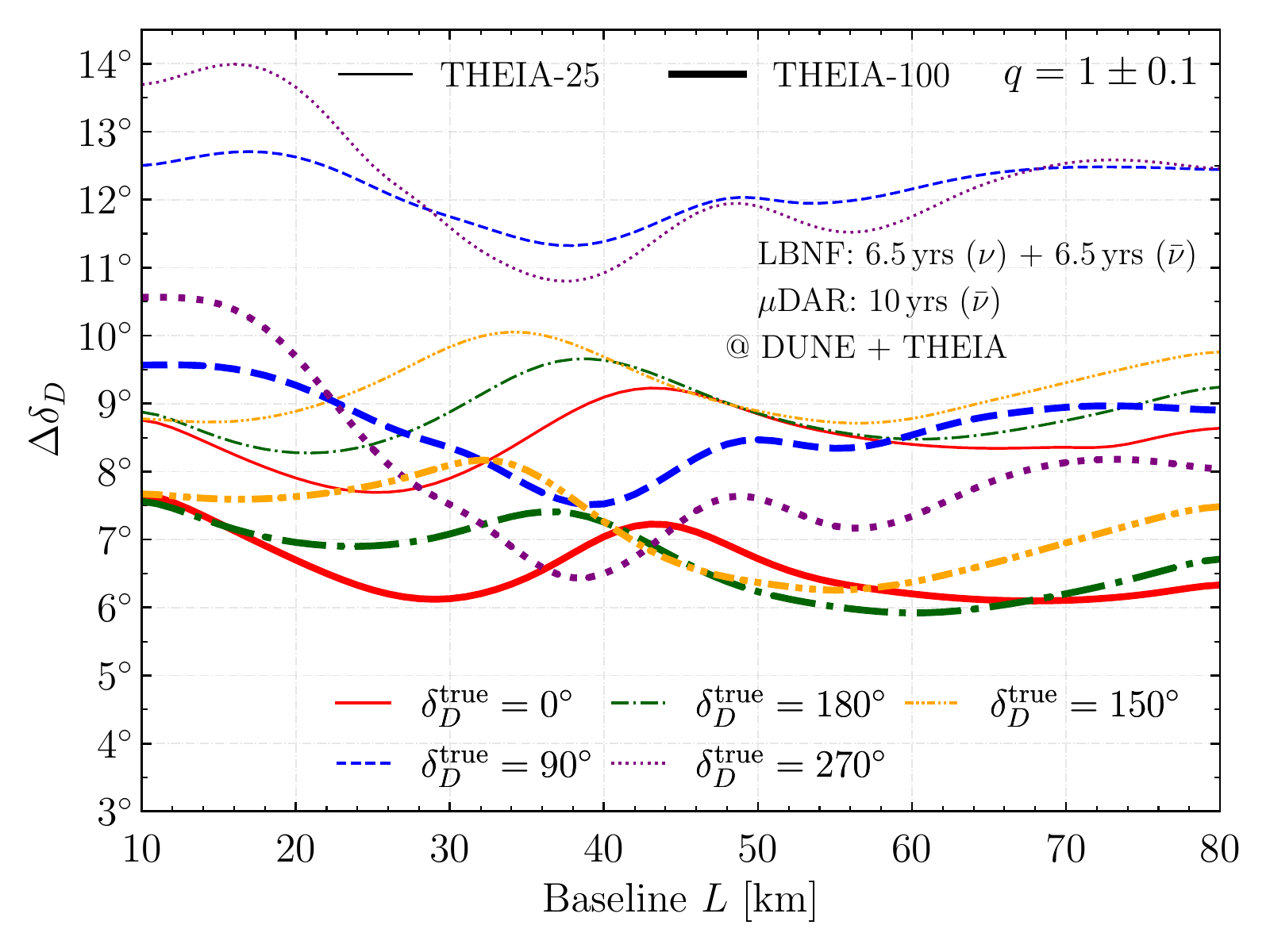}
\caption{The CP phase uncertainty, $\Delta\delta_D$, for
the combination of DUNE and THEIA as a function of the
$\mu$THEIA baseline $L$ for true values $\delta^{\rm true}_D = 0^\circ$
(red), $90^\circ$ (blue), $180^\circ$ (green), $270^\circ$
(purple), and $150^\circ$ (orange).
While the LBNF flux runs for 6.5 years each in the neutrino
and anti-neutrino modes, $\mu$DAR provides 10 years
for the anti-neutrino mode. The larger DUNE + $\mu$THEIA-100
(thick lines) has much smaller CP uncertainty than DUNE +
$\mu$THEIA-25 (thin lines).
To illustrate the matter effect on the CP uncertainty, simulations
with fixed $q = 1$ and a 10\% prior on $q$ are shown in the left 
and right panels, respectively.}
\label{fig:uncertainty}
\end{figure}
\end{minipage}
\end{widetext}
\noindent
from the global fit result
\cite{deSalas:2020pgw}, 
\begin{subequations}
\begin{eqnarray}
&&
  \sin^2 \theta_s = 0.318,
\ 
  \sin^2 \theta_a = 0.574,
\
  \sin^2 \theta_r = 0.022,
\\
&&
  \Delta m^2_s = 7.50 \times 10^{-5}\,\mbox{eV}^2,
\
  \Delta m^2_a = 2.55\times 10^{-3}\,\mbox{eV}^2,
\hspace{8mm}
\end{eqnarray}
\end{subequations}
where we take NO in our study.

Among these oscillation parameters, the solar mass 
squared difference $\Delta m_s^2$
and the solar mixing angle $\theta_s$ are kept
fixed throughout our analysis. On 
one hand, the contribution of the solar mass
squared difference $\Delta m^2_s$ enters via a
coefficient parameter $\alpha \equiv \Delta
m_s^2/\Delta m_a^2$ as shown in
\geqn{eq:Pnumu_nue_approx}. The 
parameter $\alpha$ has a small value ($\approx 0.03$)
and the error in $\Delta m_s^2$ has an even smaller
contribution. Hence it can be neglected comparing with 
the $\mathcal{O}(1)$ uncertainty on the CP phase. 
On the other hand, the solar 
mixing angle appears in the first and the
third term on the right-hand side of
\geqn{eq:Pnumu_nue_approx}. Since the
current prior on the solar mixing angle is roughly 
3\%, it can also be neglected for the study
of the large CP phase uncertainty. Moreover,
the next generation reactor neutrino experiment
like JUNO \cite{JUNO:2015zny} will provide a 
sub-percent uncertainty on $\theta_s$ that
is negligibly small.

For the other mixing parameters, the reactor 
mixing angle $\theta_r$, the atmospheric mixing
angle $\theta_a$, and the atmospheric 
mass-square difference $\Delta m^2_a$ are
treated as free parameters. We use the 
marginalized one-dimensional $\chi^2$ curves
\cite{deSalas:2020pgw} as our priors,
$\chi^2_{\rm para} \equiv
\chi^2_{\theta_r} + \chi^2_{\theta_a} +
\chi^2_{\Delta m_a^2}$.
Due to the existing tension between the T2K and 
NO$\nu$A results, which is discussed in \gsec{intro}, 
we do not include any prior on $\delta_D$. 

Since the matter effect is a natural source of fake CP,
we take its uncertainty by including a parameter $q$ to scale its 
average value defined in \gsec{sec:muTHEIA},
$A \rightarrow q A$ \cite{Bharti:2020gnu}.
The average matter density corresponds to $q = 1$
while the vacuum case takes $q=0$. Note that $q$ is
undetermined and hence treated as the fifth  
free parameter. In the following 
discussions, we consider two different 
scenarios for its uncertainty: fixed $q=1$ (no 
uncertainty) and a conservative 10\% Gaussian uncertainty.
For comparison, previous studies have used $1\% \sim 2\%$
uncertainty \cite{Roe:2017zdw,Kelly:2018kmb,DUNE:2020jqi,
DeRomeri:2016qwo}.

The CP uncertainty $\Delta \delta_D$ is defined as the
half-width of the
$\Delta \chi^2 = \chi^2(\delta_D) - \chi^2_{\rm min} = 1$
band where $\chi^2_{\rm min}$ corresponds to the best-fit
value $\delta^{\rm BF}_D$ of the CP phase. Since our
simulation uses pseudo-data, the best-fit value is the
same as the true value, $\delta^{\rm true}_D$. Note that it
is not necessary for the $\chi^2 (\delta_D)$ function to be
symmetric around the minimum. For this case, the previous
definition, $\Delta \chi^2 = 1$, gives two boundaries
below ($\delta^-_D$) and above ($\delta^-_D$) the
best-fit value $\delta^{\rm BF}_D$. Then we take the
average deviation as the CP uncertainty,
$\Delta \delta_D \equiv (\delta^+_D - \delta^-_D) / 2$.

\subsection{Baseline Options of $\mu$THEIA}
\label{sec:baseline}

The baseline between the $\mu$DAR source and the THEIA
detector can significantly affect the CP uncertainty $\Delta 
\delta_D$. \gfig{fig:uncertainty} shows $\Delta 
\delta_D$ as a function of the $\mu$THEIA baseline $L$
in the range from 10\,km to 80\,km. Since the true value
of the CP phase $\delta^{\rm true}_D$ is unknown, it needs
to be varied. Four typical CP values
$\delta^{\rm true}_D = 0^\circ, 90^\circ, 180^\circ$ and 
$270^\circ$ are chosen for illustration. In addition,
$\delta^{\rm true}_D=150^\circ$ is not
only preferred by the current result of NO$\nu$A but also
significantly affected by the matter effect uncertainty
as elaborated in \gsec{sec:results}. 
\begin{widetext}
\centering
\begin{minipage}{\linewidth}
\begin{figure}[H]
\centering
\includegraphics[width=0.49\linewidth]{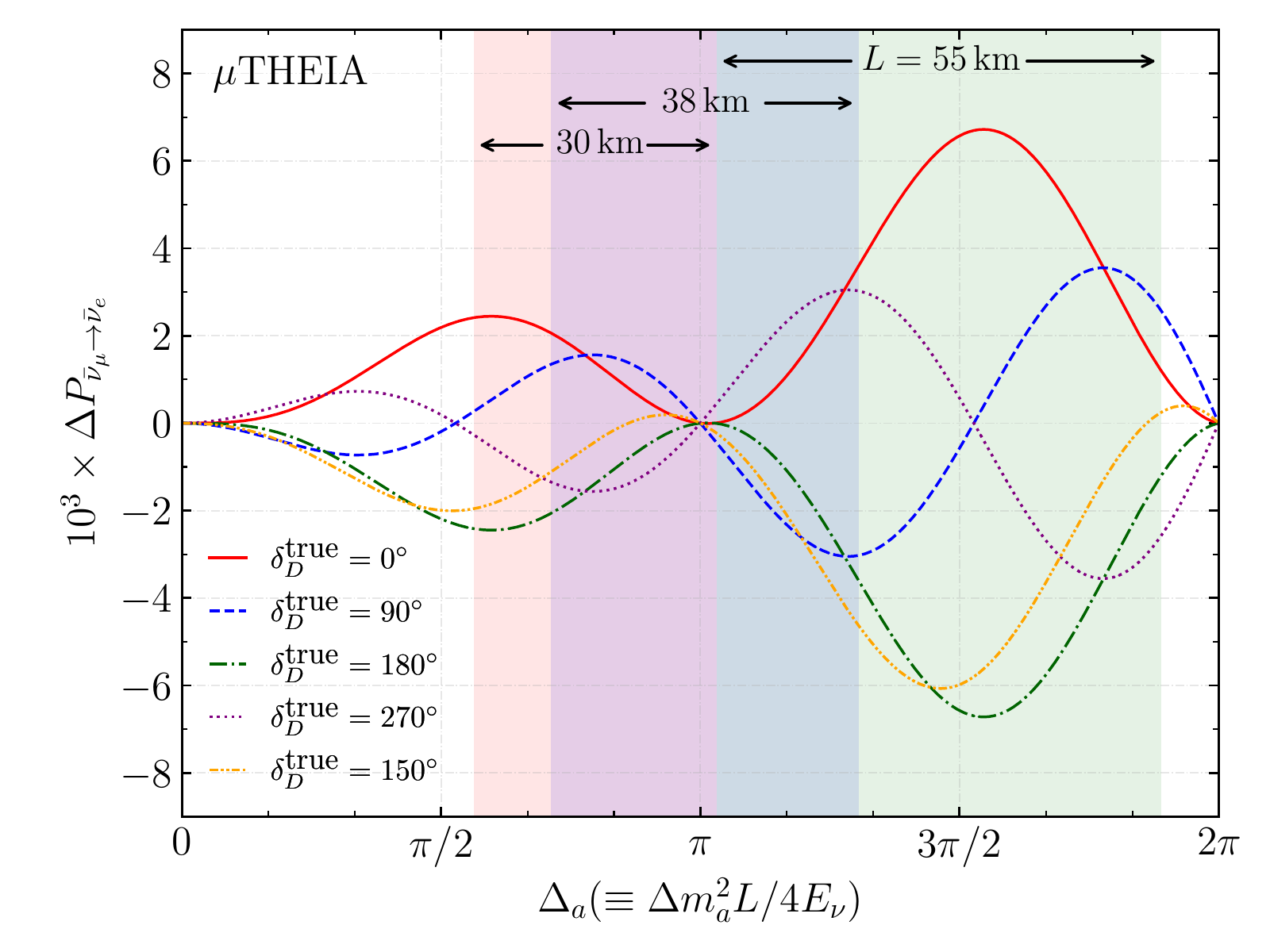}
\includegraphics[width=0.49\linewidth]{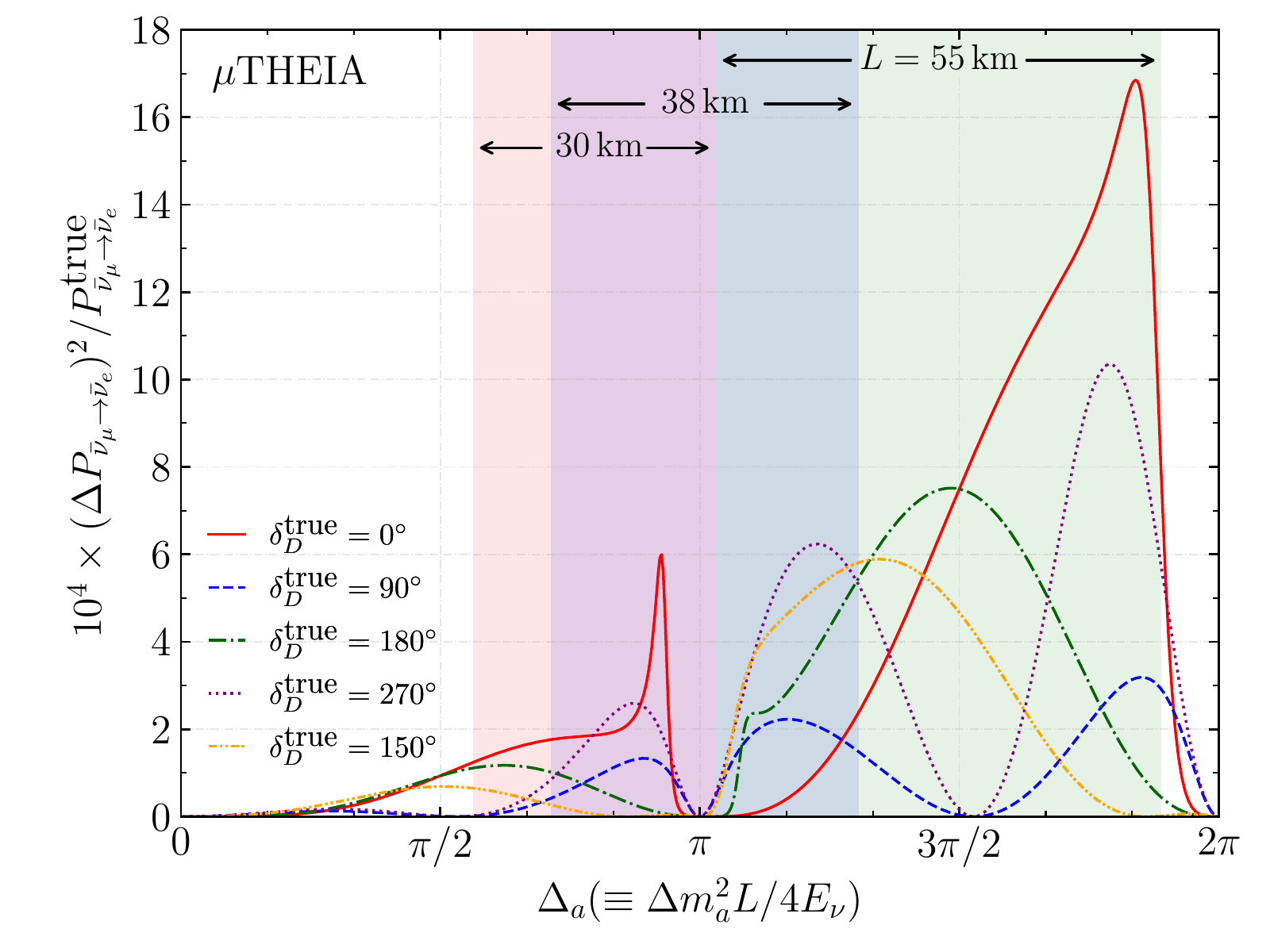}
\caption{\textbf{Left:} The oscillation probability difference
$\Delta P_{\bar \nu_\mu \rightarrow \bar \nu_e}$
between the true ($\delta^{\rm true}_D$) and fitting
($\delta^{\rm fit}_D$) CP values as a function of the
atmospheric oscillation phase $\Delta_a \equiv \Delta m^2_a L/4E$.
Five different true CP values $\delta^{\rm true}_D = 0^\circ$
(red), $90^\circ$ (blue), $180^\circ$ (green), $270^\circ$
(purple), and $150^\circ$ (orange) are shown for comparison.
The fitting values $\delta^{\rm fit}_D$ are assigned $10^\circ$
larger than the corresponding $\delta^{\rm true}_D$ values.
The neutrino energy varies within [30, 55]\,MeV while the
baseline is fixed at three different baselines, $L = 30$\,km
(red region), 38\,km (purple region), and 
55\,km (light green region).
\textbf{Right:} The relative difference
$(\Delta P_{\bar \nu_\mu \rightarrow \bar \nu_e})^2 /
        P^{\rm true}_{\bar \nu_\mu \rightarrow \bar \nu_e}$
with the same setups.}
\label{fig:diff}
\end{figure}
\end{minipage}
\end{widetext}
\noindent
So we also show the $\delta^{\rm true}_D
=150^\circ$ curve for comparison. Both THEIA-25 
(thin lines) and THEIA-100 (thick lines) are 
considered. With larger detector size, the CP 
uncertainty decreases but the baseline dependence 
follows the same trend.

To see the impact of matter effect on the optimal baseline,
the left panel of \gfig{fig:uncertainty} is obtained by fixing $q = 1$ while
the right one takes $q = 1 \pm 0.1$.
For both cases, the result shows two local minima in
the CP uncertainty for maximal CP violation.
One is around $L = 38$\,km and the other around $L = 55$\,km.
The longer one, $L = 55$\,km, is a local optimal
option for $\delta_D = 270^\circ$ that is preferred
by the T2K measurement. Although the true local
minimum for vanishing CP violation cases
$\delta_D = 0^\circ$ and $180^\circ$ actually happens
with $L \gtrsim 65$\,km, the difference in $\chi^2$
is not significant while the maximal CP violation
cases $\delta_D = 90^\circ$ and $270^\circ$
(or equivalently $-90^\circ$) become
much worse. Since the data-driven $\delta_D = -90^\circ$
is of larger interest, $L = 55$\,km is preferred
than the longer $65$\,km. For the shorter one,
the choice is more difficult. The global minimum around
$L \approx 38$\,km for $\delta_D = \pm 90^\circ$ is very
close to the global maximum for the vanishing CP violation
cases. So choosing $L = 38$\,km needs to pay too much
price and we take $L = 30$\,km to balance among various
CP values. Our simulations takes these three baselines
$L= 30$\,km, 38\,km, and 55\,km as possible options.
The final choice is up to the on-going T2K
and NO$\nu$A experiments. The comparison between the
left (fixed $q = 1$) and right (10\% uncertainty around
$q = 1$) panels of \gfig{fig:uncertainty} shows
that although the uncertain matter effect contaminates
the CP sensitivity, increasing $\Delta \delta_D$ by 3 
or 4 degrees to be exact, it does not affect the 
optimal baselines.

The optimal baseline options $L = (30, 38, 55)$\,km for
DUNE + $\mu$THEIA are all different from the TNT2HK
one $L = 23$\,km \cite{Evslin:2015pya}.
Not just the baseline length is different, but also
TNT2K/TNT2HK obtains only a single local minimum. The key
difference is the atmospheric invisible muon background.
Since both SK and HK are water Cherenkov detectors,
the atmospheric invisible muon dominates the background.
As $L$ increases, the beam flux decreases with $1/L^2$,
but the atmospheric background remains the same and eventually
dominates the statistics. So the relatively large
amount of the atmospheric background limits the optimal
baseline length. For comparison, THEIA with WbLS reduces the
invisible muons to negligible amount as shown in \gfig{fig:muDAR}.
Therefore, $\mu$THEIA can have longer baseline than
TNT2K/TNT2HK to optimize the CP sensitivity.

While the options $L = 30$\,km and 38\,km are not so far
from the 23\,km of TNT2K/TNT2HK and hence easier to understand,
the longer baseline $L = 55$\,km also
achieving comparable CP sensitivities seems
counter-intuitive. From 30\,km to 55\,km, the flux decreases
quadratically with distance and is suppressed by
a factor $(30/55)^2 \approx 0.3$. However, this flux
reduction is compensated by the increasing oscillation
amplitude. To make this feature
explicit, we show in the left panel of \gfig{fig:diff}
the oscillation probability
difference, $\Delta P_{\bar{\nu}_\mu \rightarrow 
\bar{\nu}_e} (\equiv 
P^{\rm fit}_{\bar{\nu}_\mu \rightarrow 
\bar{\nu}_e}-P^{\rm true}_{\bar{\nu}_\mu \rightarrow \bar{\nu}_e})$,
between the true ($\delta^{\rm true}_D$) and fit
($\delta^{\rm fit}_D$) CP values,
\begin{eqnarray}
  \Delta P_{\bar \nu_\mu \rightarrow \bar \nu_e}
& = &
  16 \Delta_s s_r c_s s_s c_a s_a
	\sin\Delta_{a}
	\nonumber \\
& \times &
  \sin\frac{\Delta'\delta_D} 2
  \sin\left(\frac{2 \Delta_a - 2 \delta^{\rm true}_D - \Delta'\delta_D} 2 \right).
\qquad
\label{eq:diff}
\end{eqnarray}
Note that this formula is obtained from
\geqn{eq:Pnumu_nue_approx} in the vacuum limit
($A \rightarrow 0$). The oscillation and CP phases are
defined as $\Delta_{a,s} \equiv \Delta m^2_{a,s} L/4E$
and $\Delta'\delta_D \equiv \delta^{{\rm 
fit}}_D - \delta^{\rm true}_D$. 
To further illustrate the CP sensitivity, the parameter 
$(\Delta P_{\bar{\nu}_\mu \rightarrow 
\bar{\nu}_e})^2/P^{\rm true}_{\bar{\nu}_\mu
\rightarrow \bar{\nu}_e}$ that has a similar form as the 
$\chi^2$ calculation is also calculated
in the right panel of \gfig{fig:diff}. For convenience,
we name $(\Delta P)^2 / P$ as {\it pseudo-$\chi^2$} at
the oscillation probability level. Both variables are shown
as a function of the atmospheric oscillation phase $\Delta_a$
($\equiv \Delta m^2_a L/4 E_\nu$) for five different true
CP values, $\delta^{\rm true}_D = 0^\circ$, $90^\circ$,
$180^\circ$, $270^\circ$, and $150^\circ$. The difference
between the fitting and true CP values are assigned to
have $\Delta'\delta_D = 10^\circ$ for illustration.

The left panel shows that the second peak of the oscillation
probability difference $\Delta P$ is much larger than the
first one and the relative size between the pseudo-$\chi^2$
peaks in the right panel is also significantly enhanced.
Take the $\delta^{\rm true}_D = 0^\circ$ curve in the right
panel as an example, the second peak is 3 times of the first
one which can roughly compensate the flux suppression ($\sim0.3$).
Similar feature applies also for the other true CP values.
This explains why the local minimum around $L = 55$\,km has
roughly the same value at $L = 30$\,km.

The three filled regions correspond to different baselines
$L = 30$\,km, 38\,km, and $55$\,km, respectively, while
the $\mu$DAR neutrino energy $E_\nu$ spans a wide range
of [30, 55]\,MeV. With $\Delta_a$ inversely proportional
to $E_\nu$, the left boundary of each region corresponds
to the upper energy limit 55\,MeV and the right one
to the lower limit 30\,MeV. Since the $\bar \nu_\mu$ spectrum
from $\mu$DAR source peaks at the upper limit \cite{Evslin:2015pya},
the left sides of the filled regions give the largest
contribution. This important feature can explain the
location of those local minimums in \gfig{fig:uncertainty}.
For example, the left side of the pink region for $L = 30$\,km
covering the first oscillation peak/valley of
$\delta^{\rm true}_D = 0^\circ$ in \gfig{fig:diff}
explains why this baseline corresponds to the best
sensitivity of this true CP value in \gfig{fig:uncertainty}. 
The same thing happens for $L = 38$\,km with the left
side of the purple region covering the
$\delta^{\rm true}_D = \pm 90^\circ$
peak/valley in \gfig{fig:diff} to justify the local minimum
in \gfig{fig:uncertainty}. Not to say the left side
of the light green region of $L = 55$\,km covers the second
peak/valley of $\delta^{\rm true}_D = \pm 90^\circ$ in
\gfig{fig:diff}.

\subsection{CP Sensitivity and Matter Effect}
\label{sec:results}

\begin{figure}[H]
\centering
\includegraphics[width=\linewidth]{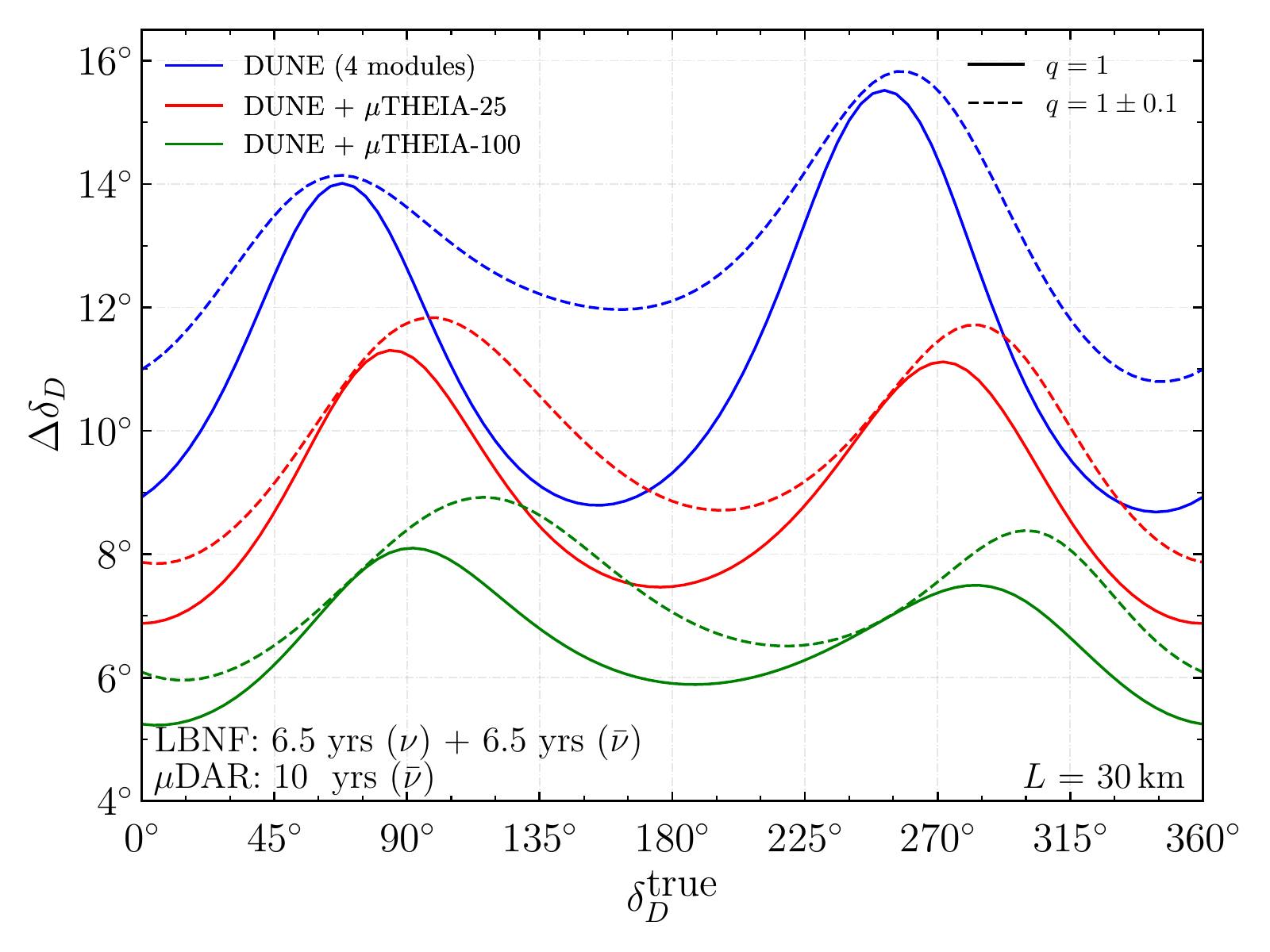}
\includegraphics[width=\linewidth]{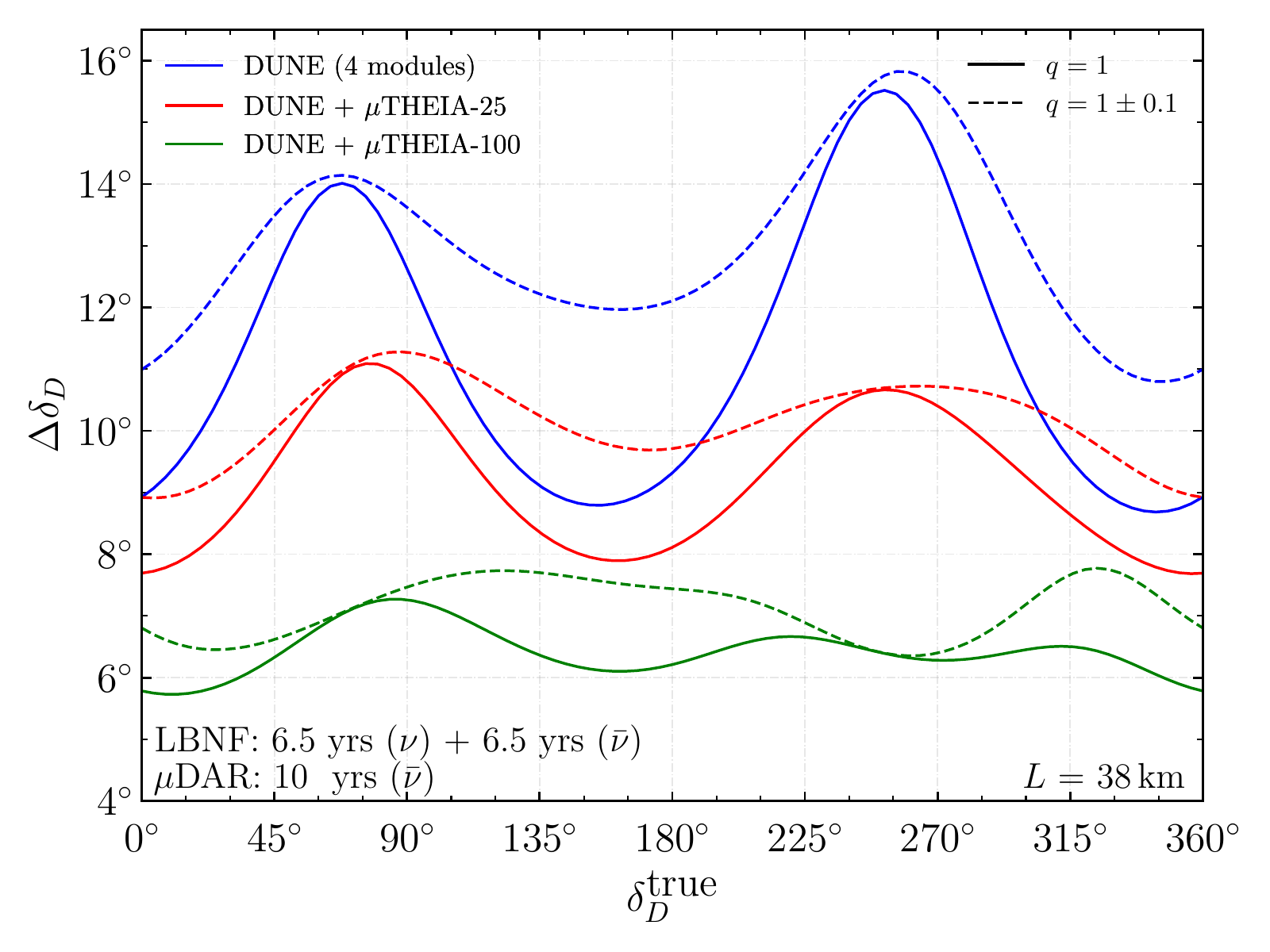}
\includegraphics[width=\linewidth]{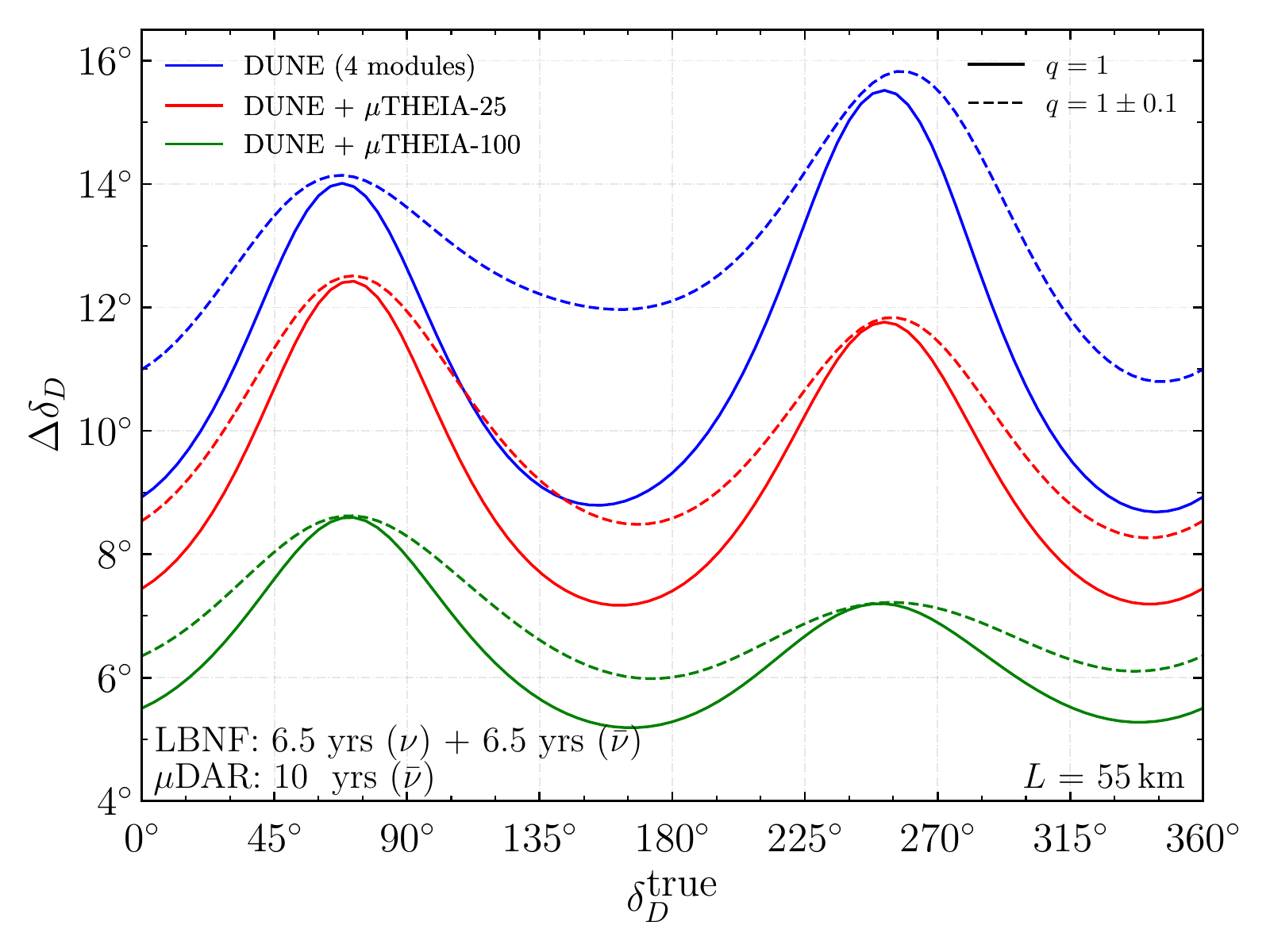}
\caption{The CP phase uncertainty $\Delta \delta_D$ as a
function of the true CP value $\delta^{\rm true}_D$.
For illustration, the three $\mu$THEIA baseline options
$L = 30$\,km (top), 38\,km (middle), and 55\,km (bottom)
are shown separately. Each panel takes
three different experimental setups, DUNE with 4 modules
(blue), DUNE with 3 modules and $\mu$THEIA-25 (red) or
$\mu$THEIA-100 (green). As for the matter effect, both
the fixed case ($q=1$, solid) and 10\% uncertainty 
($q=1\pm 0.1$, dashed) are implemented to show how it
affects the CP sensitivity.} 
\label{fig:result}
\end{figure}

As emphasized in earlier discussions, the CP sensitivity
suffers from matter effect contamination. The DUNE + $\mu$THEIA
configuration we propose in this paper can overcome this issue
to provide a clean measurement of the Dirac CP phase $\delta_D$.
\gfig{fig:result} shows the CP uncertainty $\Delta
\delta_D$ as a function of the true value
$\delta_D^{\rm true}$ for three $\mu$THEIA benchmark
baselines, $L = 30$\,km (top), 38\,km (middle), and
55\,km (bottom) found in the previous \gsec{sec:baseline}.
For each baseline, we consider several scenarios:
1) DUNE alone with 4 modules, DUNE with 3 modules and
2) $\mu$THEIA-25 or 3) $\mu$THEIA-100. More details are
summarized in \gtab{tab:setup}.

In all panels, the DUNE configuration contains the full
40\,kt detector and the blue curves do not change with
varying $L$ since it does not depend on $\mu$THEIA baseline.
For fixed matter effect ($q = 1$), the CP
uncertainty peaks around $\delta_D^{\rm true} \approx 70^\circ$
and $250^\circ$ which are consistent with 
\cite{Ballett:2016daj,Rout:2020emr,DeRomeri:2016qwo,DUNE:2020jqi}
where either fixed matter effect or just 2\% uncertainty
is adopted. Slightly shifted peaks at
$\delta_D^{\rm true} \approx 90^\circ$ and $270^\circ$
are obtained and attributed to different treatment of
systematics in \cite{Chatterjee:2021wac}.

From DUNE alone to DUNE + $\mu$THEIA-25,
the CP uncertainty $\Delta \delta_D$ significantly
reduces by roughly 1/4. This is especially true around
the maximal CP values, $\delta^{\rm true}_D = \pm 90^\circ$
since the $\mu$DAR spectrum is especially
wide to provide both $\cos \delta_D$ and $\sin \delta_D$
terms. With only $\sin \delta_D$ term, the CP uncertainty
around the maximal CP phase is intrinsically large,
$\Delta \delta_D \propto 1 / \cos \delta_D$. But a wide
spectrum can also introduce a large enough $\cos \delta_D$
term to make the CP uncertainty decrease. Previous study
shows that this feature is expected to appear when
$\mu$DAR flux is added to supplement the narrow beam
accelerator experiments, such as TNT2K/TNT2HK
\cite{Evslin:2015pya,Ge:2017qqv,Ge:2020xkm,Ge:2020ffj}.
Nevertheless, the
result turns out that this is also true for the addition
of $\mu$THEIA to DUNE, although the LBNF flux spectrum is
already quite wide. Adding a larger $\mu$THEIA-100
can even further reduce the CP uncertainty to almost
only $5^\circ$ for the best case.

\begin{figure}[t]
\centering
\includegraphics[width=1\linewidth]{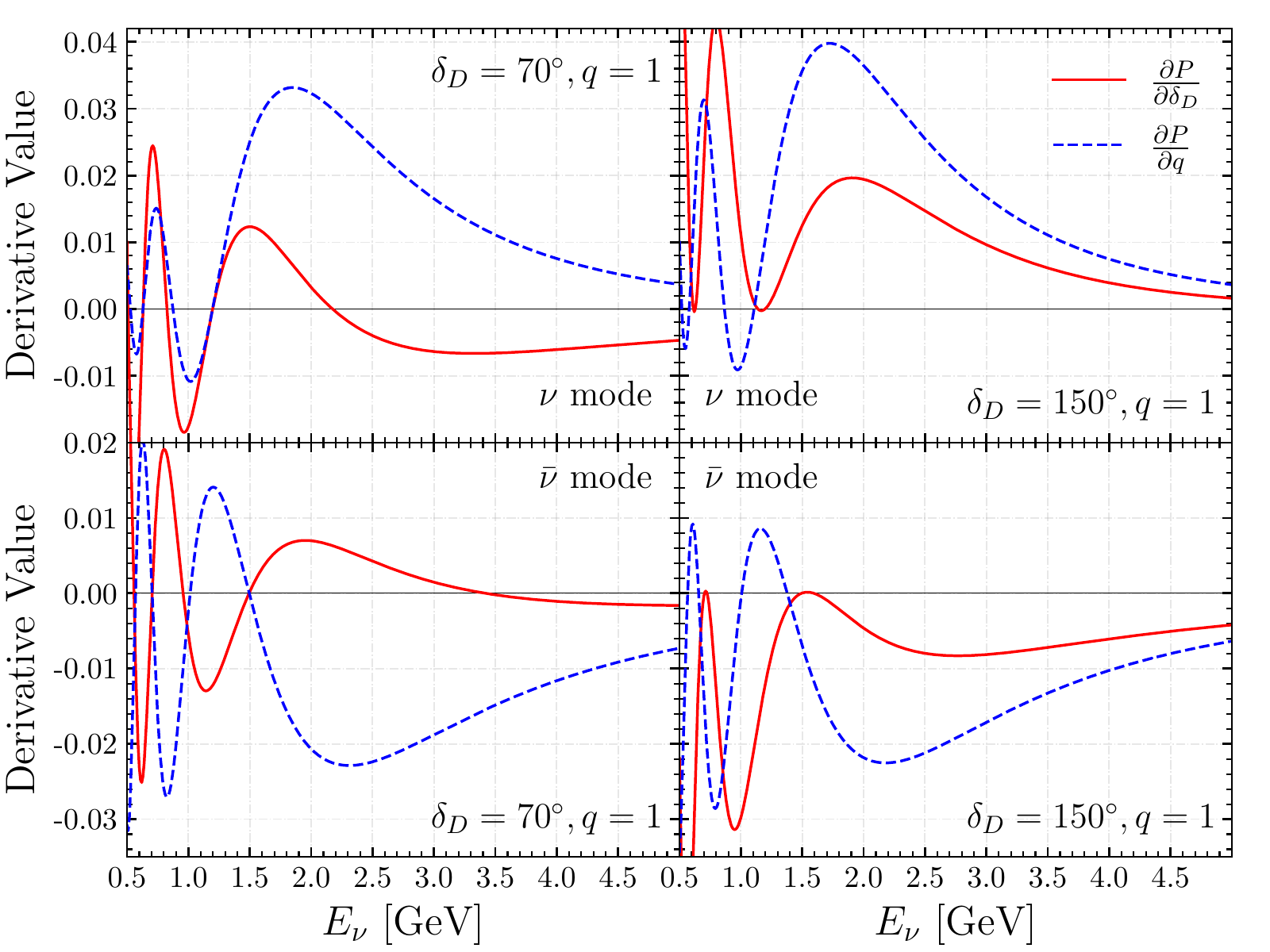}
\caption{Oscillation probability derivatives
$\partial P / \partial \delta_D$ (red solid) and
$\partial P / \partial q$ (blue dashed) with respect
to the matter effect parameter $q$ at $q = 1$ and
the Dirac CP phase $\delta_D$, respectively.
Both neutrino and anti-neutrino
modes are shown in the upper and lower rows for comparison.
The left column adopts $\delta_D = 70^\circ$ while the
right one takes $\delta_D = 150^\circ$.}
\label{fig:deriv}
\end{figure}

As expected, the matter effect can fake the CP violation
and hence its uncertainty can significantly modify the
CP uncertainty. In addition to the fixed $q = 1$ scheme
(solid lines), \gfig{fig:result} also shows the results
obtained with 10\% uncertainty in the matter effect or
equivalently $q = 1 \pm 0.1$ (dashed lines). With $q$
relaxed, the CP uncertainty becomes much worse especially
around the vanishing CP violation,
$\delta^{\rm true}_D \approx 0^\circ$ or $180^\circ$.
The most significantly affected point is around
$\delta^{\rm true}_D \approx 150^\circ$ or $330^\circ$.
With the uncertainty of matter effect taken into account,
the improvement brought by $\mu$THEIA is even more significant.
This is exactly because of the fact that matter effect
plays more important role at DUNE with much higher energy
than the low-energy $\mu$THEIA. With $\mu$THEIA added,
even switching on matter effect uncertainty would not
make the situation much worse. It also is interesting to
see that the CP uncertainty around the maximal CP violation
is almost not affected by switching on/off the matter
effect uncertainty. The $\mu$THEIA improvement is quite
stable against matter effect.

To understand the interplay between the Dirac CP phase
and the matter effect qualitatively, \gfig{fig:deriv}
shows the partial derivative $\partial
P/\partial \delta_D$ of oscillation
probability \geqn{eq:Pnumu_nue_approx} with
respect to the Dirac CP phase $\delta_D$ and $\partial P/\partial q$
to the matter effect parameter $q$ at $q = 1$ as a function
of neutrino energy $E_\nu$. The matter effect mimics
the CP effect quite well at $\delta^{\rm true}_D = 150^\circ$
with the two derivative curves having similar
shapes and peak positions. For comparison, the two
derivatives have very different features at $70^\circ$.

Note that the green
curves with DUNE and $\mu$THEIA-100 are much more flat
than the original DUNE alone. For most of the parameter
space, the CP uncertainty is better than $8^\circ$.
Among the three panels of \gfig{fig:result}, the
$L = 38$\,km one has the most flat CP uncertainty
curves for the DUNE + $\mu$THEIA-100 configuration.
Especially, the CP certainty is always better than
$8^\circ$ no matter what is the value of $\delta^{\rm true}_D$.
In this sense, $L = 38$\,km is probably the optimal
baseline for $\mu$THEIA.

\begin{figure}[t]
\centering
\includegraphics[width=\linewidth]{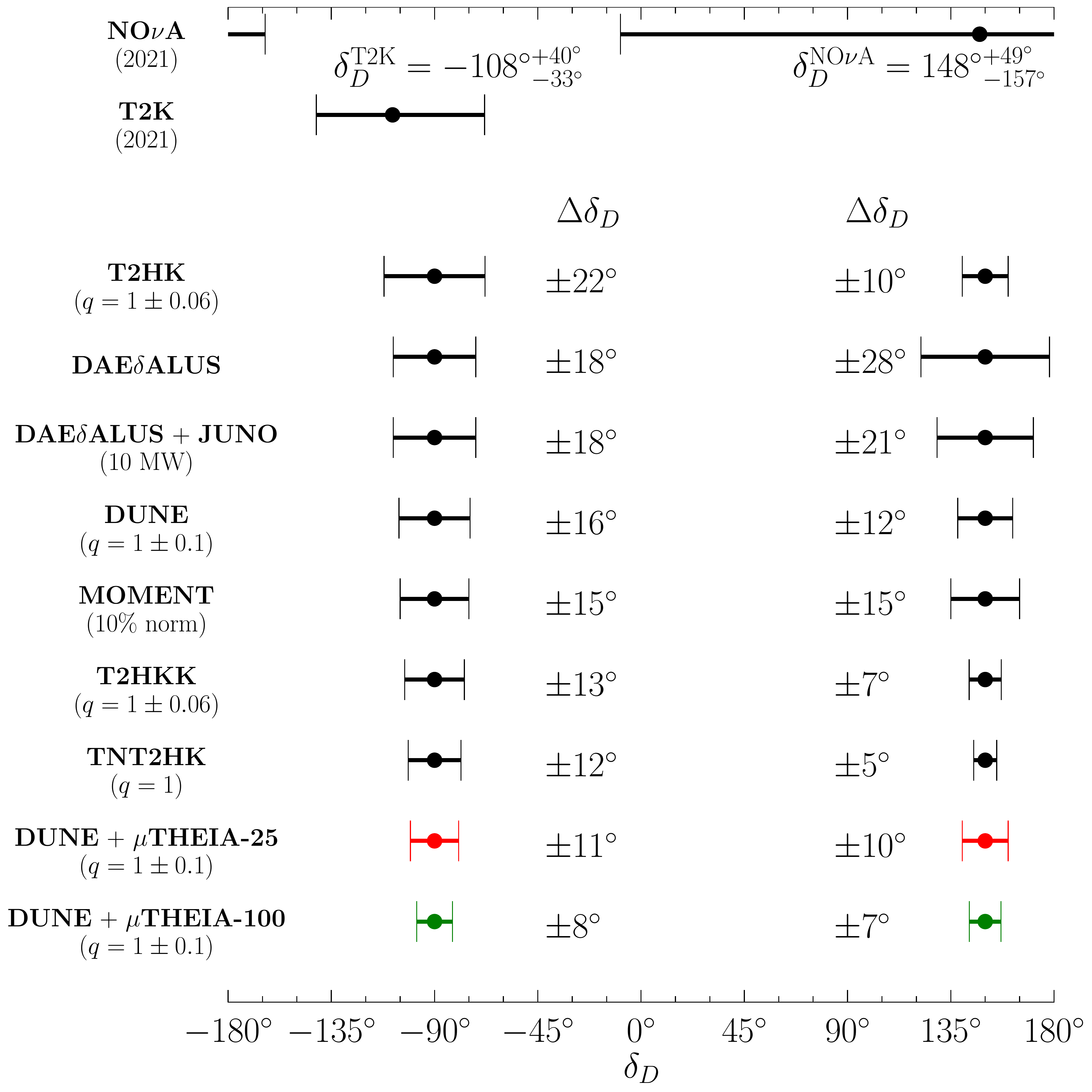}
\caption{The current $\delta_{\rm D}$ measurement at T2K
and NO$\nu$A. For comparison, the projected CP sensitivities at
various future and planned experiments are sorted by the
size of CP uncertainty. For T2K and NO$\nu$A that are already
running, their latest results from real experimental data
are shown while for those still in plan or design, we simply
quite the CP uncertainty at two typical CP phase values
$\delta_D = -90^\circ$ and $150^\circ$ for illustration.
These experiments are sorted according to their uncertainties
around $\delta_D = - 90^\circ$.
} 
\label{fig:Ddcp_compare}
\end{figure}

To further illustrate the advantages of the DUNE
+ $\mu$THEIA combination, we compare with other existing
experiments or designs in \gfig{fig:Ddcp_compare}.
The first two rows show the latest measurements
from the NO$\nu$A ($\delta_D =
{148^\circ}^{+49^\circ}_{-157^\circ}$) 
\cite{NOvA:2021nfi} and T2K ($\delta_D =
- {108^\circ}^{+40^\circ}_{-33^\circ}$)
\cite{T2K:2021xwb}. The T2K experiment has two major
upgrades: T2HK \cite{Hyper-Kamiokande:2018ofw} with
a much larger Hyper-K detector and T2HKK
\cite{Hyper-Kamiokande:2016srs} with another
detector at the second oscillation peak. 
For both of them, a matter effect uncertainty
of 6\% is considered \cite{Hyper-Kamiokande:2016srs}. Since
T2HK is in construction and T2HKK still being planned,
there is no real data yet. We take two typical values
$\delta_D = -90^\circ$ and $150^\circ$ for illustration.
At these two true values, the CP uncertainty at T2HK (T2HKK) can reach $22^\circ (13^\circ)$ and $10^\circ (7^\circ)$, respectively
\cite{Hyper-Kamiokande:2018ofw,Hyper-Kamiokande:2016srs}.

The experiments listed in \gfig{fig:Ddcp_compare}
are sorted according to their CP uncertainties.
After T2K and their upgrades, the next one is the
DAE$\delta$ALUS  experiment
that uses $\mu$DAR neutrinos. With three cyclotrons,
its CP uncertainty touches down to $18^\circ$
($28^\circ$) at the chosen typical CP phase
$\delta_D = -90^\circ$ ($150^\circ$) \cite{Alonso:2010fs}.
Another $\mu$DAR
experiment design \cite{Ciuffoli:2014ika} uses the
JUNO detector \cite{JUNO:2015zny}. The combination
of DAE$\delta$ALUS + JUNO can achieve even better
sensitivity than DAE$\delta$ALUS alone with an 
uncertainty of $18^\circ$ ($21^\circ$) at
$\delta_D = -90^\circ$ ($150^\circ$) \cite{Smirnov:2018ywm}. 

The next group is the accelerator-based DUNE
\cite{DUNE:2015lol} and MOMENT \cite{Cao:2014bea}.
The CP uncertainty at DUNE is derived from our
own simulation with a conservative 10\% uncertainty in
the matter potential, as shown in \gfig{fig:result}.
Different from the $\mu$DAR flux, the neutrino
flux from muon decay in flight is adopted by the
MOMENT experiment to $15^\circ$ \cite{Tang:2019wsv}.

Finally, TNT2HK and the two DUNE+$\mu$THEIA-25/100
configurations use $\mu$DAR neutrinos to supplement
the accelerator measurements. Since the Hyper-K detector
is going to be built, TNT2HK will always dominate
over TNT2K once a $\mu$DAR source is added around the
Kamioka site. So we only show TNT2HK in \gfig{fig:Ddcp_compare}.
Among these three options, TNT2HK has the advantage of
using the full Hyper-K detector including Super-K
for the detection of both the accelerator and $\mu$DAR
neutrinos. Nevertheless, the atmospheric invisible muon
background limits its CP uncertainty to $12^\circ$ at regions around maximum CP phase.
\cite{Evslin:2015pya}.
In addition, the J-PARC beam energy and flux \cite{T2K:2011qtm}
are also lower than the LBNF ones \cite{DUNE:2015lol}.
Although the $\mu$THEIA-25 can only use a fiducial
volume of 17\,kt, the CP uncertainty $11^\circ$ ($10^\circ$)
at DUNE + $\mu$THEIA-25 is already slightly better
than TNT2HK. With 70\,kt fiducial volume at
$\mu$THEIA-100, the CP uncertainty further reduces
to only $7^\circ \sim 8^\circ$. This clearly shows
the advantages of supplementing DUNE with $\mu$THEIA-25
or $\mu$THEIA-100.

\section{Conclusion and Outlook}
\label{sec:conclu}

The leptonic CP phase measurement at accelerator-based
neutrino oscillation experiments suffers from the
contamination of matter effect. The higher neutrino
energy, the more severe contamination. In this paper,
we put forward a possible combination of intrinsically
low-energy $\mu$DAR neutrinos and the recently proposed
THEIA detector to overcome this problem.

Our simulation shows that the THEIA detector using WbLS
has very good capability of particle identification.
This is especially useful for suppressing the atmospheric
invisible muon background, which was the major background
at the TNT2K/TNT2HK configuration, to negligible amount.
Then the $\mu$DAR
$\bar \nu_\mu \rightarrow \bar \nu_e$ oscillation 
leaves a very clear IBD signal in the detector.
In addition, the high-energy LBNF
flux can also be measured at the THEIA detector in
addition to DUNE.

With essentially a background free measurement,
the enhancement on CP sensitivity from $\mu$THEIA
is significant. The CP uncertainty around the maximal
CP violation $\delta_D = \pm 90^\circ$ reduces
up to 20\% (40\%) when compared to the 
standard DUNE configuration. Especially, the
CP uncertainty is controlled to be below $8^\circ$ 
and the best case can be as good as $6^\circ$
for the baseline $L = 38$\,km. In addition, the
dependence of CP uncertainty on the true CP phase
value is largely mitigated. If realized, either
the DUNE + $\mu$THEIA-25 or DUNE + $\mu$THEIA-100
configuration can bring the CP measurement into
a precision era.

\section*{Acknowledgements}

The authors are grateful to Constantinos Andreopoulos,
Junting Huang, Robert Svoboda,
Julia Tena Vidal, Zhe Wang, and Guang Yang for valuable discussions
and helps.
The authors are supported by the Double First Class start-up fund
(WF220442604) provided by Tsung-Dao Lee Institute \& Shanghai
Jiao Tong University, the Shanghai Pujiang Program (20PJ1407800),
and National Natural Science Foundation of China (No. 12090064).
This work is also supported in part by Chinese Academy of
Sciences Center for Excellence in Particle Physics (CCEPP).

\appendix

\section{The Wrong Scattering Angle Effect}
\label{sec:angle}

As demonstrated in \gsec{sec:IBDsignal}, a more precise energy
reconstruction for the IBD signal requires both momentum and the
scattering angle $\theta_e$ in \geqn{eq:EIBD-angle}. Reconstructing
the scattering angle is possible for the $\mu$DAR neutrinos with
fixed source location. Unfortunately, the direction of the incoming
atmospheric neutrino is unknown. So the atmospheric
neutrino background suffers from the
{\it wrong scattering angle effect} which can further blur
the neutrino energy reconstruction.

\begin{figure}[t]
\centering
\includegraphics[width=1\linewidth]{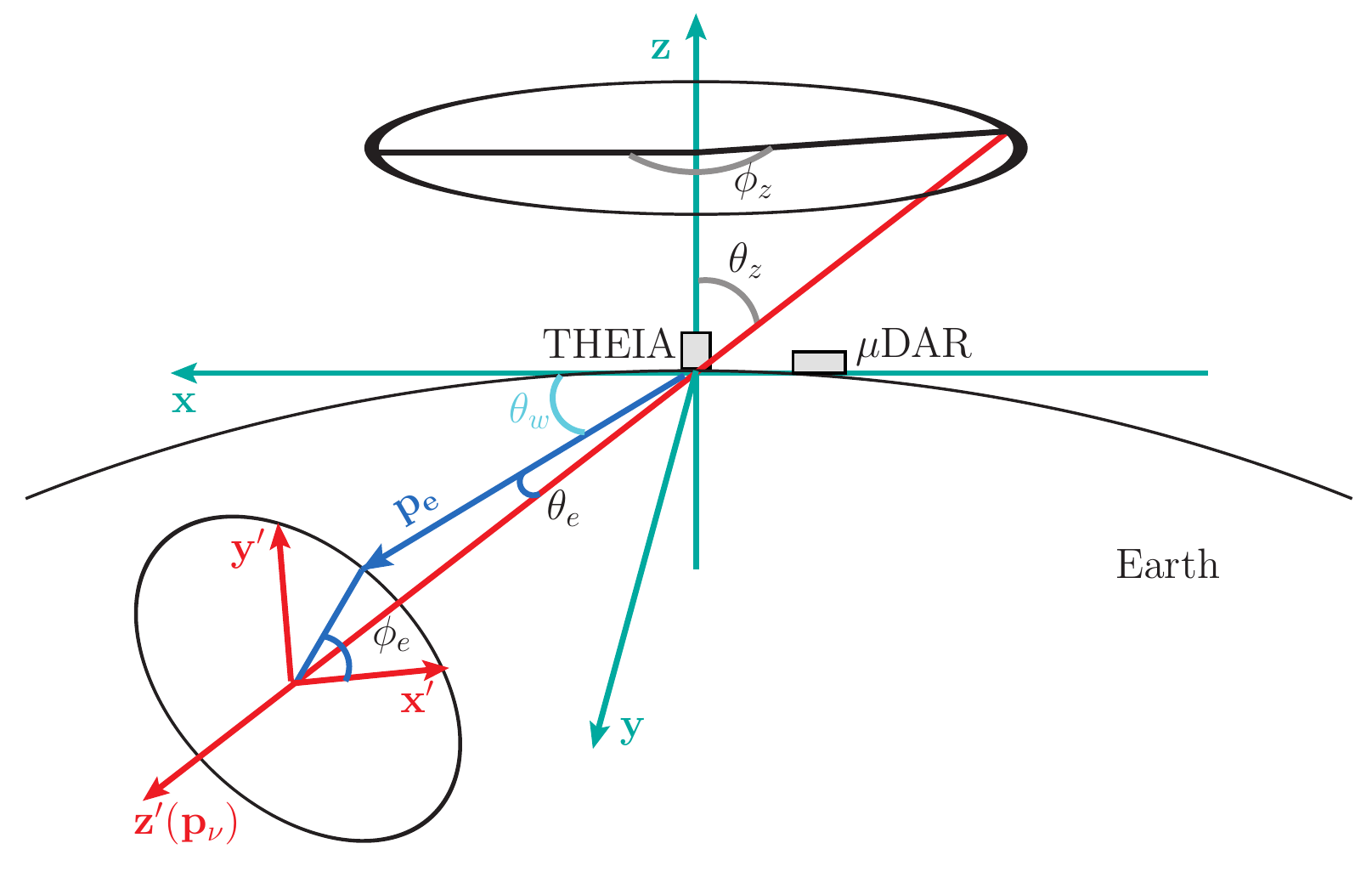}
\caption{The schematic show of the wrong scattering angle
effect for the measurement of atmospheric neutrinos.
During the scattering process, the incoming neutrino with
momentum ${\bf p}_\nu$ (long red arrow) becomes an
electron/positron with momentum ${\bf p}_e$ (blue arrow).
The neutrino momentum ($-{\bf p}_\nu$) is parameterized by
$\theta_z$ and $\phi_z$ in the lab frame (green) while
the electron/positron momentum (${\bf p}_e$) by $\theta_e$
and $\phi_e$ in the neutrino frame. Without knowing the
incoming neutrino direction, the genuine scattering angle
$\theta_e$ is wrongly reconstructed as $\theta_w$.}
\label{fig:wrong-angle}
\end{figure}

\gfig{fig:wrong-angle} shows the geometry of the atmospheric IBD background.
With $\mathcal O(10)$\,km baseline, the $\mu$DAR neutrinos essentially
travels horizontally, providing a natural definition of the $x$-axis
while the other horizontal direction is the $y$-axis of the lab frame (cyan).
Then, the zenith angle $\theta_z$ is measured from the vertical $z$-axis.
Given $\theta_z$, the incoming atmospheric neutrino direction is
parametrized by the azimuth angle $\phi_z$, measured from the $x$-axis.
The final-state lepton typically has a nonzero scattering
angle $\theta_e$ from the neutrino direction. For convenience,
a neutrino frame (blue) is established around the neutrino
momentum $\bf{p}_\nu$ with $x'$, $y'$, and $z'$-axes.

The wrong scattering angle is defined as the one between
the direction of $e^-/e^+$ (blue arrow) and the
direction of the $\mu$DAR flux (cyan $x$-axis).
For comparison, the true scattering angle $\theta_e$
and the corresponding azimuth angle $\phi_e$ are also plotted
in the figure. To calculate the wrong angle as a
function of $\theta_e$, $\theta_z$, $\phi_e$, and $\phi_z$,
one need to first establish the connection between the
horizontal and neutrino frames. 
The $z'$ direction can be easily read out from the figure,
$\vec z' = -\sin\theta_z\cos\phi_z \vec{x}-\sin\theta_z\sin\phi_z\vec{y}-\cos\theta_z\vec{z}$.
But the $x'$- and $y'$-axes can be randomly set,
as long as they satisfy
$\vec{n}_{x'}\cdot \vec{n}_{z'}=\vec{n}_{y'}\cdot \vec{n}_{z'}=\vec{n}_{x'}\cdot \vec{n}_{y'}=0$, 
$|\vec{n}_{x'}|=|\vec{n}_{y'}|=1$ and 
$\vec{n}_{x'}\times\vec{n}_{y'}=\vec{n}_{z'}$.
Accordingly, the following transformation matrix between
two coordinate systems is chosen,
\begin{eqnarray}
\left\lgroup
\begin{matrix}
  \vec{x'} \\
  \vec{y'} \\
  \vec{z'}
\end{matrix}
\right\rgroup
\hspace{-1mm} &=& \hspace{-1mm}
\left\lgroup
\begin{matrix}
  -\cos\theta_z\cos\phi_z & -\cos\theta_z\sin\phi_z & \sin\theta_z\\
 -\sin\phi_z & \cos\phi_z & 0\\
  -\sin\theta_z\cos\phi_z & -\sin\theta_z \sin\phi_z & -\cos\theta_z
\end{matrix}
\right\rgroup
\hspace{-2mm}
\left\lgroup
\begin{matrix}
  \vec x \\
  \vec y \\
  \vec z
\end{matrix}
\right\rgroup
\hspace{-1.5mm}.
\qquad
\label{eq:frame-trans}
\end{eqnarray}

The $e^- / e^+$ direction in the neutrino frame is
($\sin\theta_e\cos\phi_e$, $\sin \theta_e \sin\phi_e$, $\cos\theta_e$).
Using the frame transformation \geqn{eq:frame-trans}, the
$e^- / e^+$ direction in the lab frame is
\begin{eqnarray}
\vec{n}_e=
\left\lgroup
\begin{matrix}
  -c^\phi_z s^\theta_z c^\theta_e
- s^\theta_e (c^\theta_z c^\phi_e c^\phi_z + s^\phi_e s^\phi_z)
\\
  -s^\phi_z s^\theta_z c^\theta_e
+ s^\theta_e (-c^\theta_z c^\phi_e s^\phi_z + c^\phi_z s^\phi_e)
\\
  -c^\theta_z c^\theta_e + c^\phi_e s^\theta_z s^\theta_e
\end{matrix}
\right\rgroup^T,
\end{eqnarray}
where we have used shorthand notations,
$(c^\phi_{e,z}, s^\phi_{e,z}) \equiv (\cos \phi_{e,z}, \sin \phi_{e,z})$ and
$(c^\theta_{e,z}, s^\theta_{e,z}) \equiv (\cos \theta_{e,z}, \sin \theta_{e,z})$.

With $\mu$DAR flux in the direction of
$\vec{n}_{\mu\mbox{\tiny{DAR}}}=(1,0,0)$, the cosine term of
the wrong scattering angle $\theta_w$ is given by
\begin{eqnarray}
  \cos\theta_w
& \equiv & 
\vec{n}_e\cdot\vec{n}_{\mu\mbox{\tiny{DAR}}}/|\vec{n}_e||\vec{n}_{\mu\mbox{\tiny{DAR}}}|
\\
& = &
  -\cos \phi_z \sin \theta_z \cos\theta_e
\nonumber \\
& &
- \sin \theta_e
 (\cos\theta_z\cos\phi_e\cos\phi_z + \sin\phi_e\sin\phi_z).
\nonumber
\end{eqnarray}

To see the effect of a wrong scattering angle, we use Monte
Carlo method to randomly generate the scattering process.
As seen from \gfig{fig:wrong-angle}, there are 
four different angles in the whole scattering process and each
of them has specific probability distribution.
First, the zenith angle of the atmospheric neutrino
flux $\theta_z$ is defined as the angle between the local
zenith and the direction of atmospheric neutrino flux
\cite{Honda:2015fha}.
Its probability distribution is sampled according to
the low energy Gran Sasso flux downloaded from the Honda
website \cite{Honda}. Given a zenith angle $\theta_z$,
the azimuth angle $\theta_e$ is isotropically sampled
since the location of the $\mu$DAR source is not known yet.

From the atmospheric neutrino scattering with target,
the scattering angle $\theta_e$ of the final-state charged
lepton distributes according to the GENIE simulation.
One important feature is that the $\cos\theta_e$ distribution
depends on the neutrino energy $E_\nu$. Moreover,
the lepton azimuth angle $\phi_e$ is isotropic.

In addition, the atmospheric neutrino flux is affected
by the neutrino oscillation through the Earth which is
a function of the propagation length $2 R \cos\theta_z$
where $R = 6371$\,km is the Earth radius. We use
the PREM Earth model \cite{Dziewonski:1981xy} that
is implemented in GLoBES to calculate the oscillation
probability to calculate the modified atmospheric neutrino flux.

As expected, the wrong scattering angle significantly
affects the energy reconstruction since the scattering
angle in the energy reconstruction formula \geqn{eq:EIBD-angle}
plays an important role as we discussed in \gsec{sec:IBDsignal}. 
The reconstructed energy spectra at several different typical
energies are shown in \gfig{fig:atm-erec}.

\addcontentsline{toc}{section}{References}
\bibliographystyle{plain}

\end{document}